\pdfoutput=1
\documentclass[twocolumn]{aastex61}
\usepackage{graphicx}
\usepackage{natbib}
\usepackage{comment}

\begin{document}

\title{Resolved H \textsc{I} observations of local analogs to $\rm z \sim 1$ luminous compact blue galaxies: evidence for rotation-supported disks}

\author{Katie Rabidoux}
\affiliation{Engineering Physics Department,
 University of Wisconsin-Platteville, 1 University Plaza, Platteville,
  WI 53818, USA}

\author{D.J. Pisano}
\altaffiliation{Adjunct
  Astronomer at Green Bank Observatory, P.O. Box 2, Rt. 28/92, Green
    Bank, WV 24944, USA}
\affiliation{Department of Physics and
    Astronomy and Center for Gravitational Waves and Cosmology, West
    Virginia University, 135 Willey St., P.O. Box 6315 Morgantown, WV
    26506, USA}

\author{C. A. Garland}
\altaffiliation{Math for America (MfA) Early Career Teacher}
\affiliation{Uncommon Charter High
  School, 1485 Pacific St., Brooklyn, NY 11216, USA}

\author{Rafael Guzm{\'a}n}
\affiliation{Department of Astronomy, University
  of Florida, 211 Bryant Space Science Center, P.O. Box 112055,
  Gainesville, FL 32611, USA}

\author{Francisco J. Castander}
\affiliation{Institut de Ciencies de
  l'Espai, (ICE, IEEC/CSIC), E-08193, Bellaterra (Barcelona), Spain }

\author{Spencer A. Wolfe}
\affiliation{Department of Physics and Astronomy
  and Center for Gravitational Waves and Cosmology, West Virginia
  University, 135 Willey St., P.O. Box 6315 Morgantown, WV 26506, USA}

\begin{abstract}

While bright, blue, compact galaxies are common at $\rm z \sim 1$,
they are relatively rare in the local universe, and their evolutionary
paths are uncertain. We have obtained resolved H \textsc{I}
observations of nine $\rm z \sim 0$ luminous compact blue galaxies
(LCBGs) using the Giant Metrewave Radio Telescope and Very Large Array
in order to measure their kinematic and dynamical properties and
better constrain their evolutionary possibilities. We find that the
LCBGs in our sample are rotating galaxies that tend to have nearby
companions, relatively high central velocity dispersions, and can have
disturbed velocity fields. We calculate rotation velocities for each
galaxy by measuring half of the velocity gradient along their major
axes and correcting for inclination using axis ratios derived from
SDSS images of each galaxy. We compare our measurements to those
previously made with single dishes and find that single dish
measurements tend to overestimate LCBGs' rotation velocities and H
\textsc{I} masses. We also compare the ratio of LCBGs' rotation
velocities and velocity dispersions to those of other types of
galaxies and find that LCBGs are strongly rotationally supported at
large radii, similar to other disk galaxies, though within their
half-light radii the $\rm V_{rot}/ \sigma$ values of their H
\textsc{I} are comparable to stellar $\rm V_{rot}/ \sigma$ values of
dwarf elliptical galaxies. We find that LCBGs' disks on average are
gravitationally stable, though conditions may be conducive to local
gravitational instabilities at the largest radii. Such instabilities
could lead to the formation of star-forming gas clumps in the disk,
resulting eventually in a small central bulge or bar.
\end{abstract}


\section{Introduction}

\subsection{LCBGs: Analogs to $\rm z \sim 1$ star-forming galaxies}
Luminous compact blue galaxies (LCBGs) are a morphologically
heterogeneous class of star-forming galaxies that are defined by their
blue colors, high luminosities in the optical B band, compact sizes,
and high surface brightnesses \citep{W04}. Their strong optical
emission lines and blue continua suggest that LCBGs harbor diverse
stellar populations, with a current starburst involving approximately
a tenth of the galaxy's mass coexisting with older cohorts of stars of
approximately solar metallicity \citep{H01, G03, H07}.

LCBGs are common in galaxy surveys at intermediate
redshifts. \citet{K94} found that $30\%$ of compact sources at $\rm z
\sim 0.1-0.7$ show strong, narrow emission lines characteristic of
star formation. \citet{G97} found that LCBGs compose $20\%$ of the
general field population of galaxies and contribute $45\%$ of the
total star formation rate density at $\rm 0.4 < z < 1$. \citet{T10}
found that LCBGs comprise $\sim 10\%$ of the total galaxy population
with $\rm M_{B} < -17$ and $\sim 5\%$ of the galaxies with $\rm M_{B}
< -16$ at a median redshift of $\rm z = 0.49$, which they note is
lower than the \citet{G97} value likely due to the rapid evolution of
LCBGs after $\rm z \sim 1$. \citet{LucasThesis} found that LCBGs
contribute $42\%$ of the luminosity density of galaxies with $\rm
M_{B} < -18.5$ at $\rm z = 0.9$, and that they comprise $30\%$ of the
galaxy population with $\rm M_{B} < -15$ and $60\%$ of the galaxy
population with $\rm M_{B} < -18.5$ at that redshift. LCBGs at $\rm z
< 1$ are selected to have optical properties that are consistent with
the small, bright, blue galaxies that appear in deep field
observations \citep{K94, P97, W04}.

In contrast to their abundance at intermediate redshifts, LCBGs are a
factor of ten rarer in number density in the local universe
\citep{G01}. \citet{LucasThesis} found that LCBGs comprise less than
$2\%$ of galaxies at $\rm z \sim 0$. This discrepancy suggests that
LCBGs are a progenitor population for one or more of the galaxy types
prevalent at $\rm z \sim 0$. \citet{CGletter} recently confirmed that
local LCBGs have similar morphologies, gas fractions, and specific
star formation rates to higher-redshift star-forming
galaxies. Following the definitions compiled by \citet{W04} to select
for local analogs of intermediate-redshift LCBGs, these galaxies have
$\rm{B-V < 0.6\ mag,\ SBe(B) < 21.0 \ mag \ arcsec^{-2}}$, and
$\rm{M_{B} < -18.5\ mag}$.

The properties of both local and $\rm z \sim 1$ LCBGs overlap with
many similar types of galaxies that have been described in the
literature. Galaxies with properties similar to those of LCBGs include
Compact Galaxies \citep{P97, G97}, Luminous Compact Galaxies
\citep{H01}, and Blue Compact Galaxies \citep{K94, G99, BvZ01,
  P01}. While they are blue and compact, LCBGs are too massive
\citep[$\rm M_{*} \sim 10^{9}\ M_{\sun}$,][]{G03,CG04,T10}, luminous
\citep[$\rm L_{B} \sim 10^{9}\ L_{\sun}$,][]{CG04}, and have
metallicities too high \citep[$\rm 12 + log(O/H) \sim 8.5$,][]{T10} to
be classified as Blue Compact Dwarfs. \citet{C09} found that LCBGs
overlap in blue luminosity, morphology, stellar mass, and metallicity
with the Green Pea galaxies detected by Galaxy Zoo at z$\sim 0.1 -
0.4$. \citet{H05} found that the lower-mass examples of compact
Ultraviolet Luminous Galaxies (UVLGs), which they identify as
low-redshift analogs of high-redshift Lyman Break Galaxies (LBGs),
overlap in mass with the higher-mass examples of compact galaxies
discussed in \citet{P97}. Similarly, \citet{G03} and \citet{H04} point
out that some LCBGs could be low-mass, lower-redshift counterparts to
LBGs, and \citet{F10} have detected fine-structure emission lines of
Si \textsc{II} that have been previously observed in $\rm z \sim 3$
LBGs in a $\rm z \sim 0.04$ LCBG, which they interpret as an
indication that star formation processes may be related in both types
of galaxies. It is useful to study $\rm z \sim 0$ LCBGs, then, to
better understand the properties of the types of galaxies that exist
at higher redshifts.

LCBGs have heterogeneous morphologies. Many LCBGs appear to be the
products of mergers, especially at intermediate redshift where the
spatial density of galaxies was larger and mergers were more common
\citep{AO01}. In particular, irregular morphologies are more common in
LCBGs than in other blue compact galaxies \citep{O01}. Many LCBGs also
have companions \citep{CG04, PG10, CGletter}. At $\rm z \sim 0.2-1.3$,
$60\%$ of LCBGs appear to have similar properties to local H II
galaxies, while $40\%$ of LCBGs resemble local starburst disk
galaxies, and $90\%$ seem to be small galaxies with some extension,
but lacking large, faint disks \citep{N06}. \citet{CGletter} find that
$40\%$ of local LCBGs are ``clumpy'', which they define to mean three
or more optical clumps. \citet{W04} point out that LCBGs are also not
a distinct class of galaxies in parameter space. They exist at the
extreme blue, bright, and compact ends of the optical properties that
serve to identify them, but they are not outliers along the continuum
of observed properties for field galaxies at the redshifts at which
they appear \citep[see Figure 1 in][]{CG04}.

As LCBGs at $\rm z \sim 0$ are rare compared to their number density
at $\rm z \sim 1$ \citep{G01}, it is likely that they evolve quickly
once their current episodes of star formation end, though it is not
known what types of galaxies LCBGs will subsequently become. It has
been suggested that LCBGs could be undergoing their final phase of
star formation, and will continue to passively evolve and fade to
become today's spheroidal or dwarf elliptical galaxies \citep{G97,B04}
or faint, low-mass spiral galaxies \citep{P97}. Other authors have
suggested that LCBGs could be spiral galaxies undergoing a burst of
star formation as they form their bulges \citep{BvZ01,H01,B06}. It has
also been asserted that LCBGs are galaxies that only appear similar in
unresolved optical images at intermediate redshift and are actually a
diverse enough population that their evolutionary paths and end
products are widely varied \citep{T10}. As LCBGs are visible across a
large range of redshifts, they are excellent candidates for studying
galaxy evolution \citep{H07}.

\subsection{Goals}
In order to determine possible evolutionary paths for LCBGs, it is
necessary to have knowledge of their H \textsc{I}
properties. Measuring the H \textsc{I} mass gives an estimate of the
fuel available for star formation and constrains the duration of the
current starburst. The internal kinematics of the H \textsc{I} and
evidence of past interactions give clues regarding the starburst
triggering and quenching mechanisms \citep{P01}, and can support or
rule out disk or spheroid models of LCBGs' morphology. To investigate
the nature of these galaxies, we have studied a selection of local
analogs to intermediate-redshift LCBGs. Previously, \citet{CG04, CG05,
  CG07} surveyed the optical, H \textsc{I}, and CO properties of Sloan
Digital Sky Survey (SDSS)- and Markarian- selected LCBGs. They took H
\textsc{I} and CO spectra of a large sample of LCBGs using single
pointings \citep{CG04, CG05}. For their study, they selected local
LCBG analogs having the same optical properties as
intermediate-redshift LCBGs as outlined by \citet{W04}. \citet{CG07}
also initiated follow-up mapping observations of four Markarian
galaxies and one SDSS galaxy with the Very Large Array (VLA). In this
paper, we follow up the previous Garland et al. studies with Giant
Metrewave Radio Telescope (GMRT) and VLA H \textsc{I} observations of
galaxies selected from the \citet{CG04} sample, plus one additional
local LCBG.

An overarching goal of this paper is to compare the H \textsc{I}
properties we derive from resolved observations of nearby LCBGs to the
properties derived from unresolved single pointings. Since LCBGs are
most common at redshifts where resolved H \textsc{I} studies are not
possible, it is important for us to understand what information is
lost in unresolved observations of these galaxies. To accurately
predict their evolution, we must first know what available
observations can definitively tell us.

Previous H \textsc{I} studies of local LCBGs have not had the spatial
resolution to distinguish the target sources from their nearby
companions. Therefore, another goal of our study is to identify H
\textsc{I}-rich companions and signatures of interacting galaxies that
were not resolved in the single-dish H \textsc{I} observations from
\citet{CG04}. Since H \textsc{I} gas traces a galaxy's gravitational
potential at a much larger radius than light from stars, our resolved
H \textsc{I} observations could indicate locations conducive to
interaction-driven star formation where it may not be obvious from
optical observations.  We therefore take advantage of the GMRT and
VLA's angular resolution to measure the extent of H \textsc{I}
emission and identify signatures of rotation in order to calculate
dynamical masses ($\rm M_{dyn}$) for these LCBGs from measurements of
rotation velocities (as opposed to estimating $\rm M_{dyn}$ from
linewidths that could potentially be biased by the inclusion of nearby
companions, tidal features, or non-rotation components). Coupled with
the H I mass ($\rm M_{HI}$), these measurements give us an estimate of
how much gas is available for the continuation of the starburst. This
constrains the evolutionary scenarios for LCBGs, as the bulge
formation scenario would imply that LCBGs have higher $\rm M_{dyn}$
than have been sampled from the central bright cores of LCBGs at
intermediate redshifts \citep[e.g.][]{P01}, and the spheroidal/dwarf
elliptical progenitor scenario requires LCBGs to undergo passive
evolution after their current starburst \citep{G99}, which would limit
their possible rotation velocities.

An additional goal of this paper is to determine whether LCBGs are
rotationally-supported disk galaxies or dispersion-dominated bulges to
better understand their likely future morphologies once their star
formation has been quenched. Since our resolved study can also
distinguish velocity dispersions from rotation velocities, we can
compare their rotation velocities to their velocity dispersions and
look for evidence of disklike or bulgelike behavior both at their
outermost regions and their centers. We can also use the ratio of
ordered to random motions and the gas fractions that we have measured
to look for evidence of disk instabilities that could trigger star
formation in these galaxies. These measurements will better constrain
the future evolution of local LCBGs, and have strong implications for
the possible evolutionary products of their $\rm z \sim 1$
counterparts.

In this paper, we describe our sample of LCBGs in Section 2. We
describe our results in Section 3, and discuss their physical
implications in Section 4. We give our conclusions in Section 5. We
briefly address the properties of each LCBG in the Appendix. We assume
$\rm H_0=70\ km\ s^{-1}Mpc^{-1}$ throughout this paper.

\section{Sample Selection, Observations, and Data Reduction}

\subsection{Sample selection}
We chose nine galaxies from the SDSS- and Markarian-selected
single-dish sample of LCBGs that \citet{CG04} observed with the Green
Bank Telescope (GBT). We also included an additional SDSS galaxy
(SDSS0125+0110) in our sample, selected from the single-dish sample of
LCBGs that Garland et al. (in prep) observed with Arecibo. The Garland
et al. samples were chosen for their blue colors, high luminosities,
and compact appearances similar to properties outlined by \citet[][see
  \citet{CG04} for a more detailed discussion of the selection
  criteria]{W04}. We selected sources that had not been previously
observed in H \textsc{I} emission at high resolution with
interferometers. The galaxies we observed span the full range of
colors of the GBT sample, but do not include the very brightest or
most compact galaxies in the GBT sample that \citet{CG04} observed. We
made sure to include galaxies with and without known optical
companions. We show the optical properties of the galaxies in our
sample in Table \ref{fig:obssummary} calculated using SDSS Data
Release 9 magnitudes and radii \citep[DR9,][]{DR9} using the equations
in Section 2.1.2 of \citet{CG04}. The galaxies in our sample were
strong detections in single-dish H \textsc{I} observations
\citep{CG04}, which makes them good candidates for interferometer
observations. The galaxies in our sample have heterogeneous
morphologies, including isolated spiral galaxies, galaxies with tidal
tails, multiple galaxies in a common H \textsc{I} envelope, galaxies
with distant companions, and galaxies with disturbed gas
components. One source (SDSS1319+5253) contains three galaxies in a
common H \textsc{I} envelope, one of which (SBS 1317+523B) is an
LCBG. The galaxy that was not included in the original \citet{CG04}
sample (SDSS0125+0110) is not consistent with the LCBG optical
parameters defined by \citet{W04} when using the photometry of DR9,
though it is consistent with these optical parameters when using the
photometry of SDSS Data Release 4 (DR4), which \citet{CG04} used to
select the original local LCBG sample.  As its optical properties
remain close to the LCBG optical cuts described by \citet{W04}, are
within the defined LCBG optical parameters when using the photometry
of earlier SDSS data releases, and remain within optical properties of
LCBGs as described by other authors \citep[for example,][]{G97}, we
include it in our analysis.  We discuss each galaxy individually in
the Appendix.

\subsection{GMRT observations and reduction}
We observed eight galaxies with the GMRT near Pune, India. The GMRT is
comprised of 30 antennas in a fixed Y-configuration with 14 antennas
within 1 km and a maximum baseline of 25 km. We observed five galaxies
(SDSS0728+3532, SDSS0934+0014, SDSS0936+0106, SDSS1319+5253, and
SDSS1402+0955) in January 2006, along with three additional galaxies
(SDSS0119+1452, SDSS0125+0110, and SDSS1507+5511) in January 2007. We
observed each galaxy during a separate session with measurements of
flux calibrators 3C48, 3C147, and/or 3C286 at the beginning and end of
the observing run. We interspersed observations of a bright,
unresolved, nearby phase calibrator every $\sim 40$ minutes that we
selected from the VLA calibrator manual for a typical observing
session of $\sim 9$ hours. We flagged and calibrated the data using
the Astronomical Image Processing System
(AIPS)\footnote{aips.nrao.edu} data reduction package using the
standard procedures. For the GMRT this requires doing an initial
calibration for a single, RFI-free channel before flagging and
calibrating the full observing band. We made data cubes from the inner
50 channels (out of an original 128 channels with a channel width of
$\rm \sim 13.7\ km\ s^{-1}$) using the AIPS task IMAGR. For each
galaxy, we made two different cubes: a ``low-resolution'' cube
(typically $\sim 50''-60''$) made from baselines shorter than $\rm 5\
k\lambda$, and a ``high-resolution'' cube (typically $5''-20''$) made
with a larger UV range (made from baselines out to $\rm 50\ k\lambda -
120\ k\lambda$). When making the high-resolution cubes, we chose
robustness parameters, UV tapers, and UV ranges for each galaxy in
order to maximize the resolution while maintaining a high level of
signal to noise. We cleaned the data cubes using the number of
iterations necessary for the total flux of the clean components in a
central channel to reach a plateau so as not to incorporate too many
negative clean components. See Table
\ref{fig:cubes} for the imaging parameters used for each galaxy.

\begin{figure*}[htb!]
\gridline{\leftfig{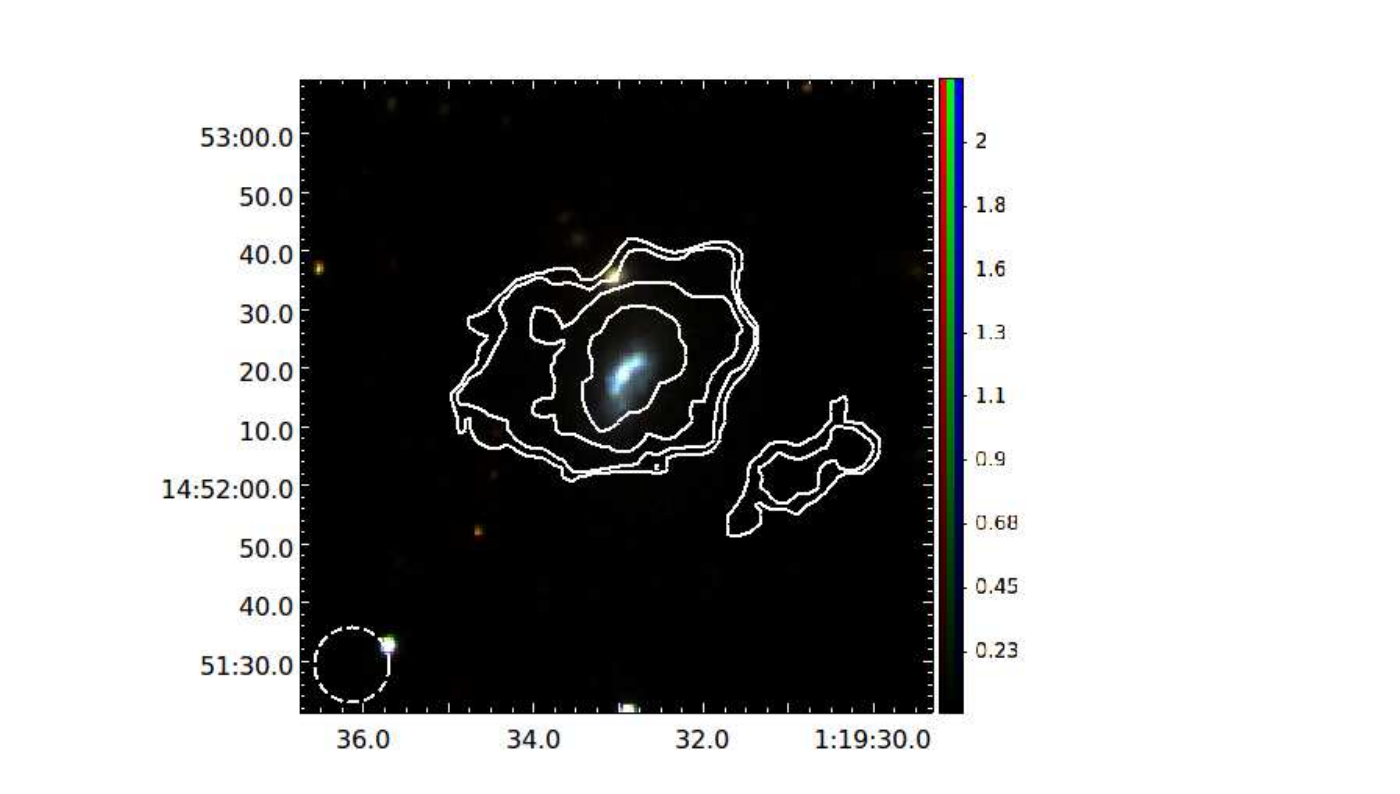}{0.57\textwidth}{(a)}
  \leftfig{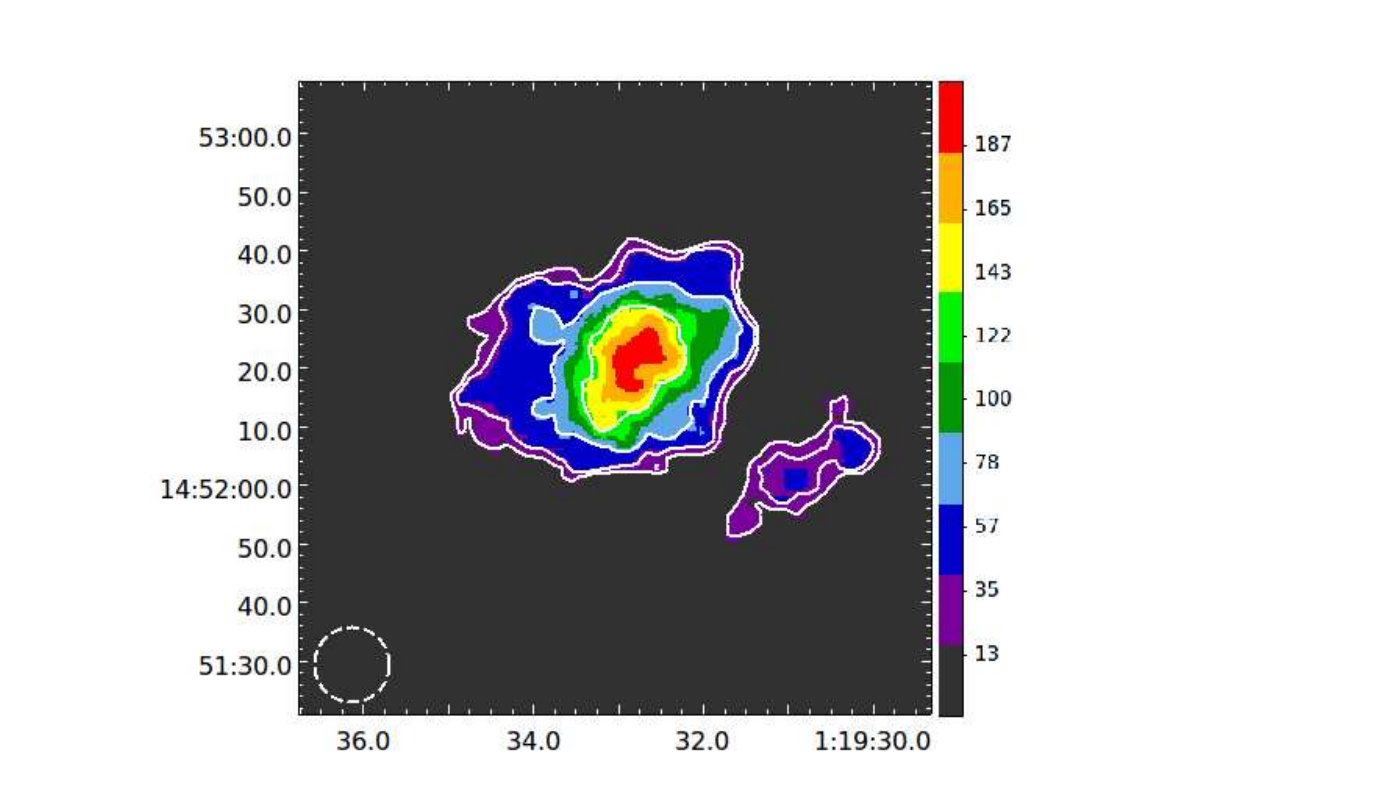}{0.57\textwidth}{(b)}}
\gridline{\leftfig{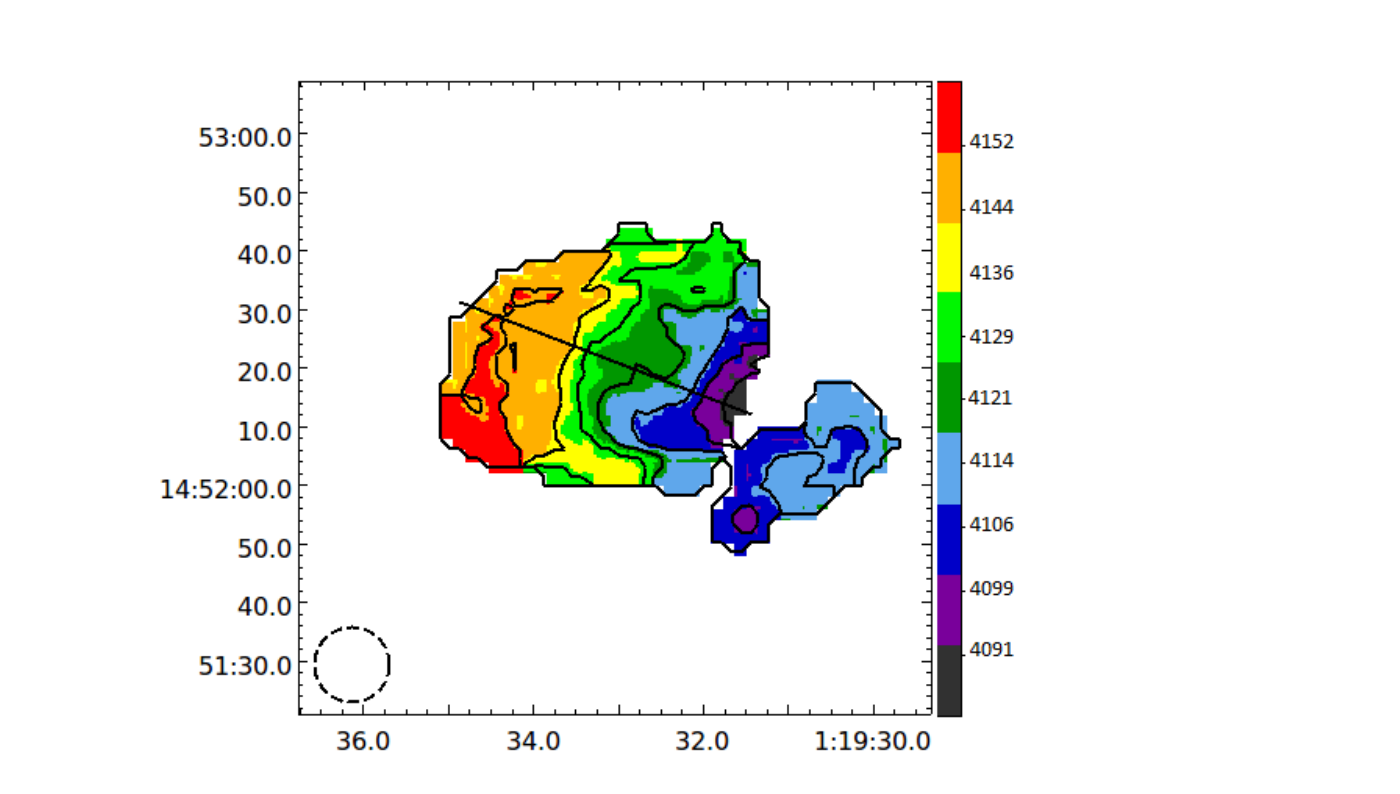}{0.57\textwidth}{(c)}
  \leftfig{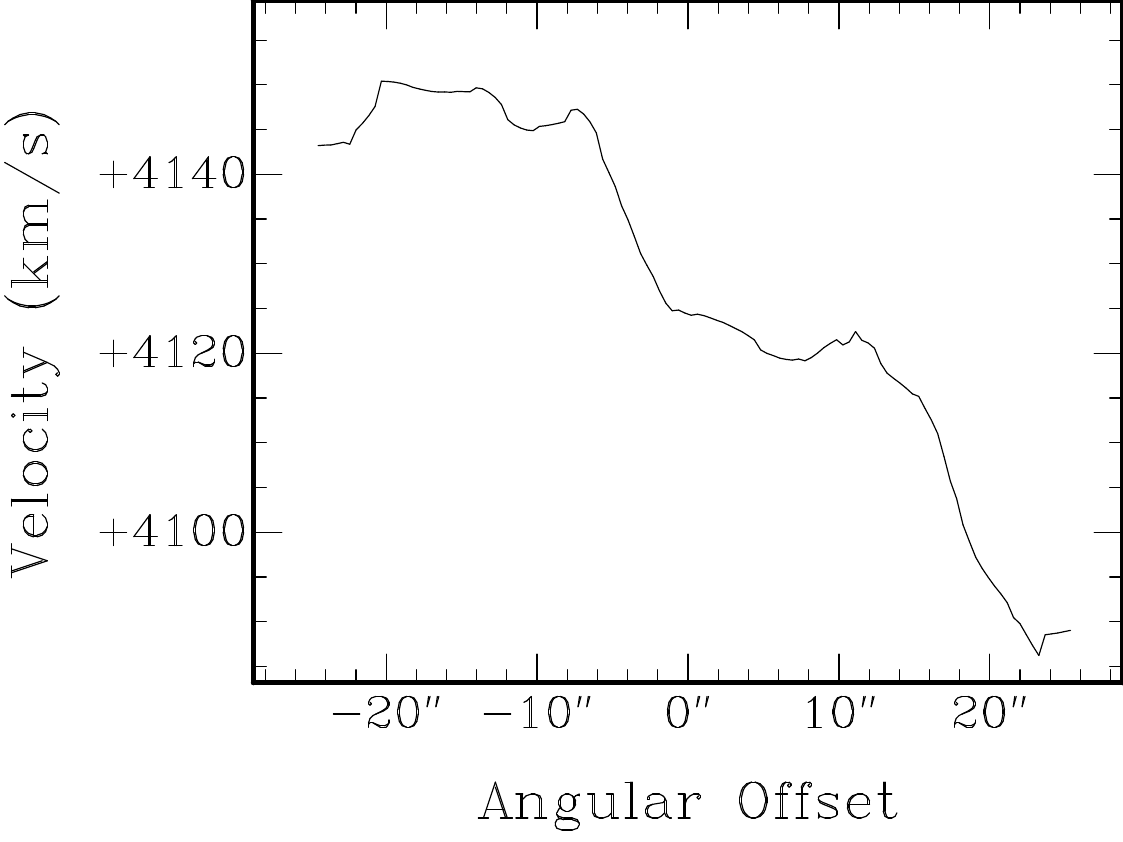}{0.37\textwidth}{(d)}}
\gridline{\leftfig{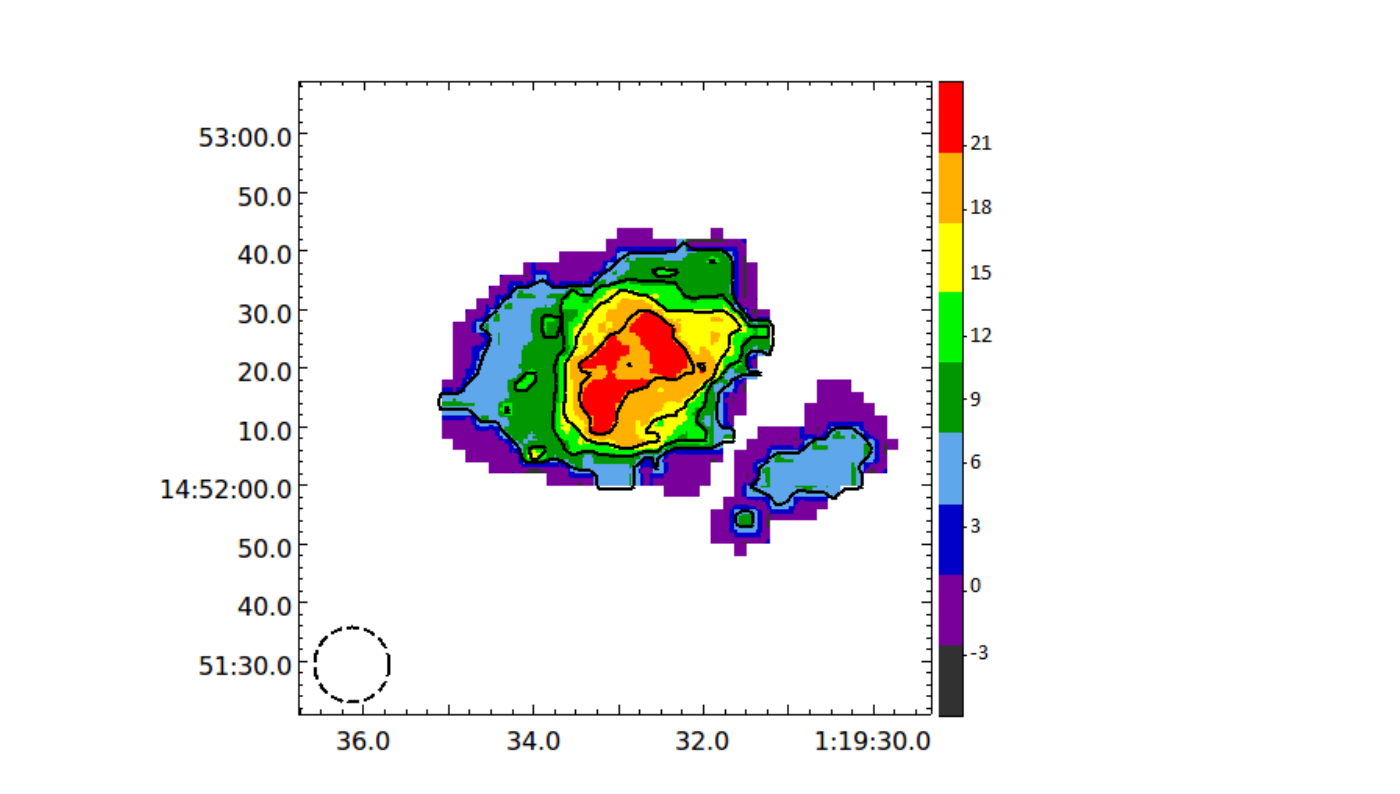}{0.57\textwidth}{(e)}
  \leftfig{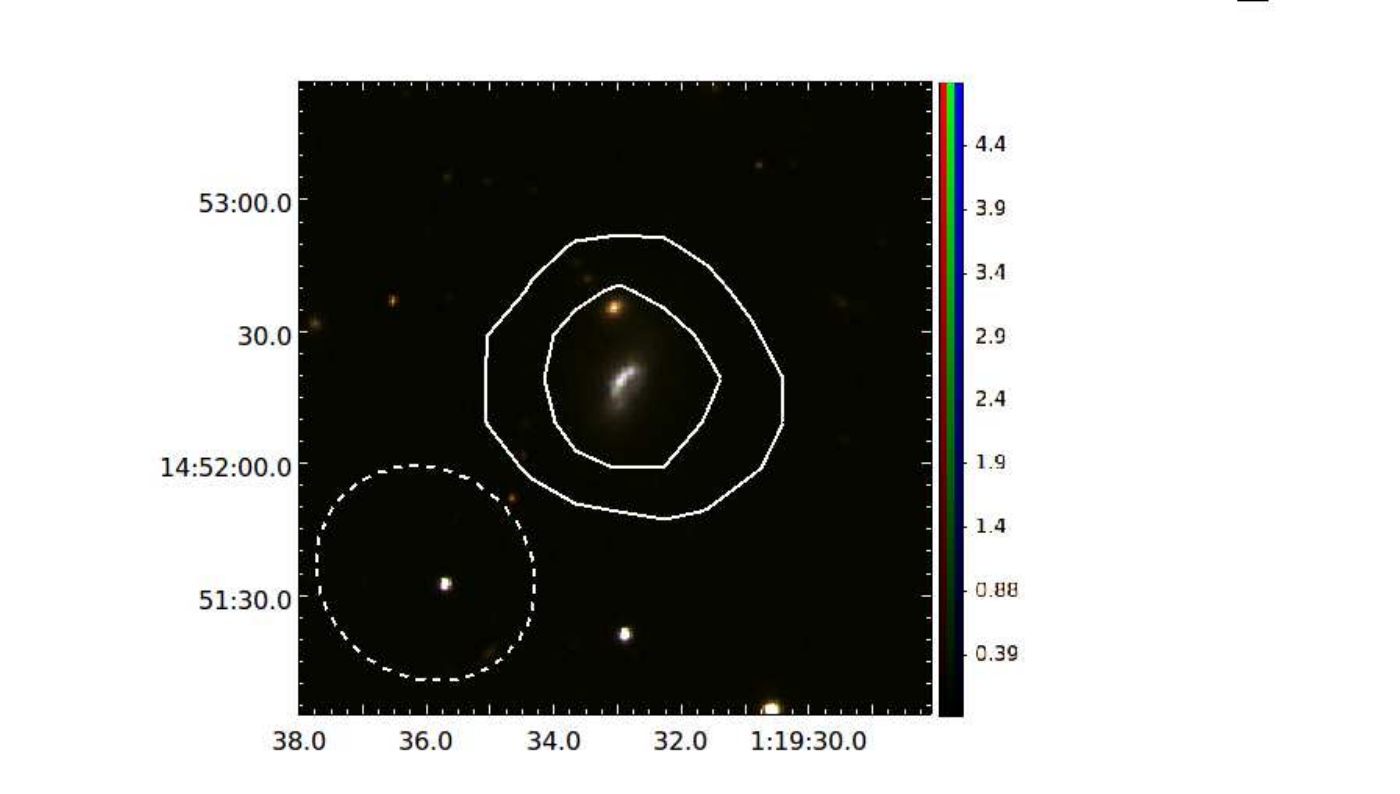}{0.57\textwidth}{(f)}}

\caption{\footnotesize SDSS0119+1452 (NGC 469): (a) Moment 0 contours made
  with a $13\arcsec \times 13\arcsec$ beam overlaid on a SDSS DR9
  \textit{gri} image. Contours represent H \textsc{I} intensities
  equivalent to column densities of $\rm 2^{n}\times10^{20}\ cm^{-2}$
  for n = 0, 1, 2, 3. Image units are analog to digital units
  (ADU). (b) Moment 0 map with the same contours as in (a). Map units
  are $\rm Jy\ Beam^{-1}\ m\ s^{-1}$. (c) Moment 1 map with a thick
  line showing the major axis. Contours are $10\ \rm km\ s^{-1}$. Map
  units are $\rm km\ s^{-1}$. (d) Velocities along the major axis
  slice shown in (c). (e) Moment 2 map with $5\ \rm km\ s^{-1}$
  contours. Map units are $\rm km\ s^{-1}$. (f) Low-resolution Moment
  0 contours made with a $52\arcsec \times 47\arcsec$ beam overlaid on
  SDSS DR9 \textit{gri} image. Contours represent H \textsc{I}
  intensities equivalent to column densities of $\rm
  2^{n}\times10^{20}\ cm^{-2}$ for n = 0, 1. Image units are the same
  as in (a). The horizontal and vertical axes of each map are right
  ascention and declination (J2000). Beam sizes are shown in the lower
  left corner of each map. \label{fig:0119}}

\end{figure*}
\begin{figure*}[htb!]

\gridline{\leftfig{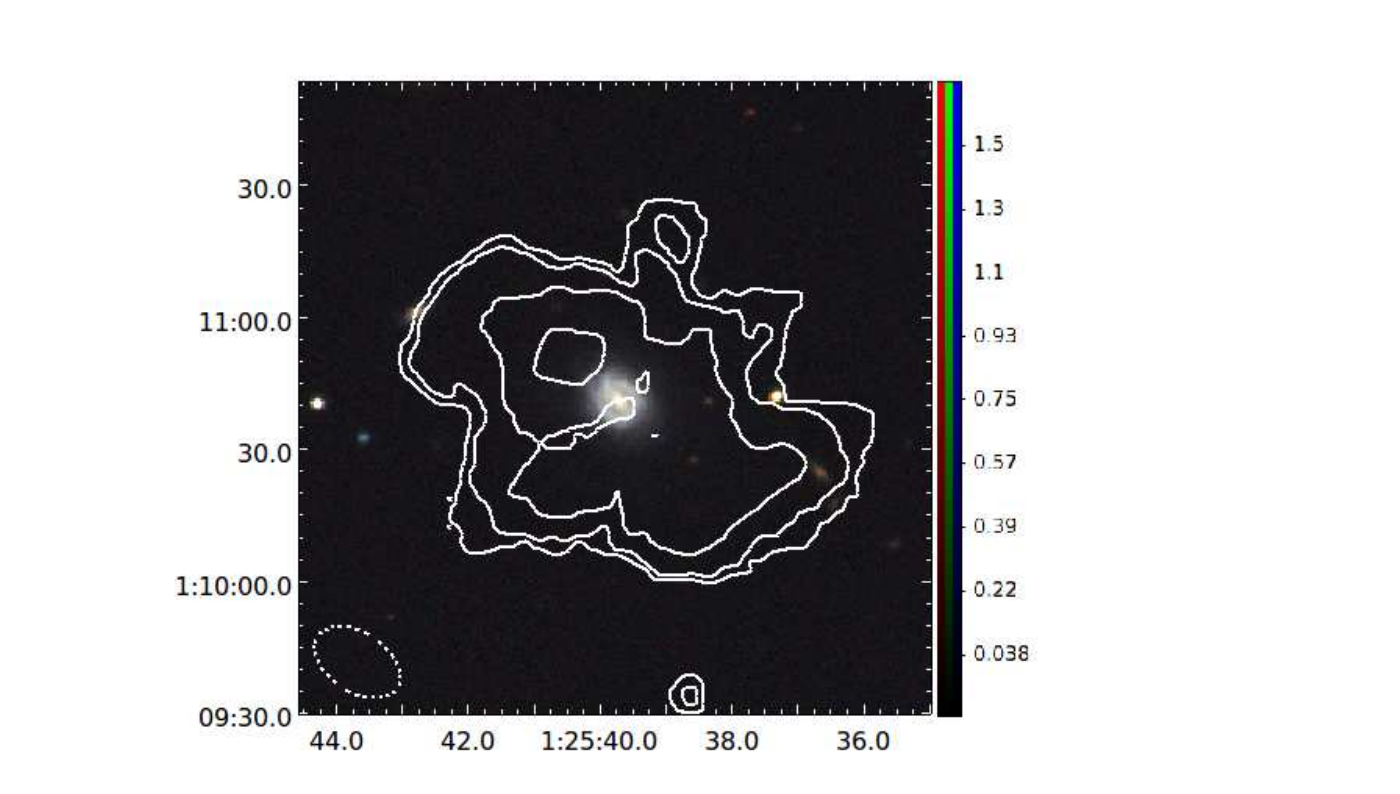}{0.57\textwidth}{(a)}
  \leftfig{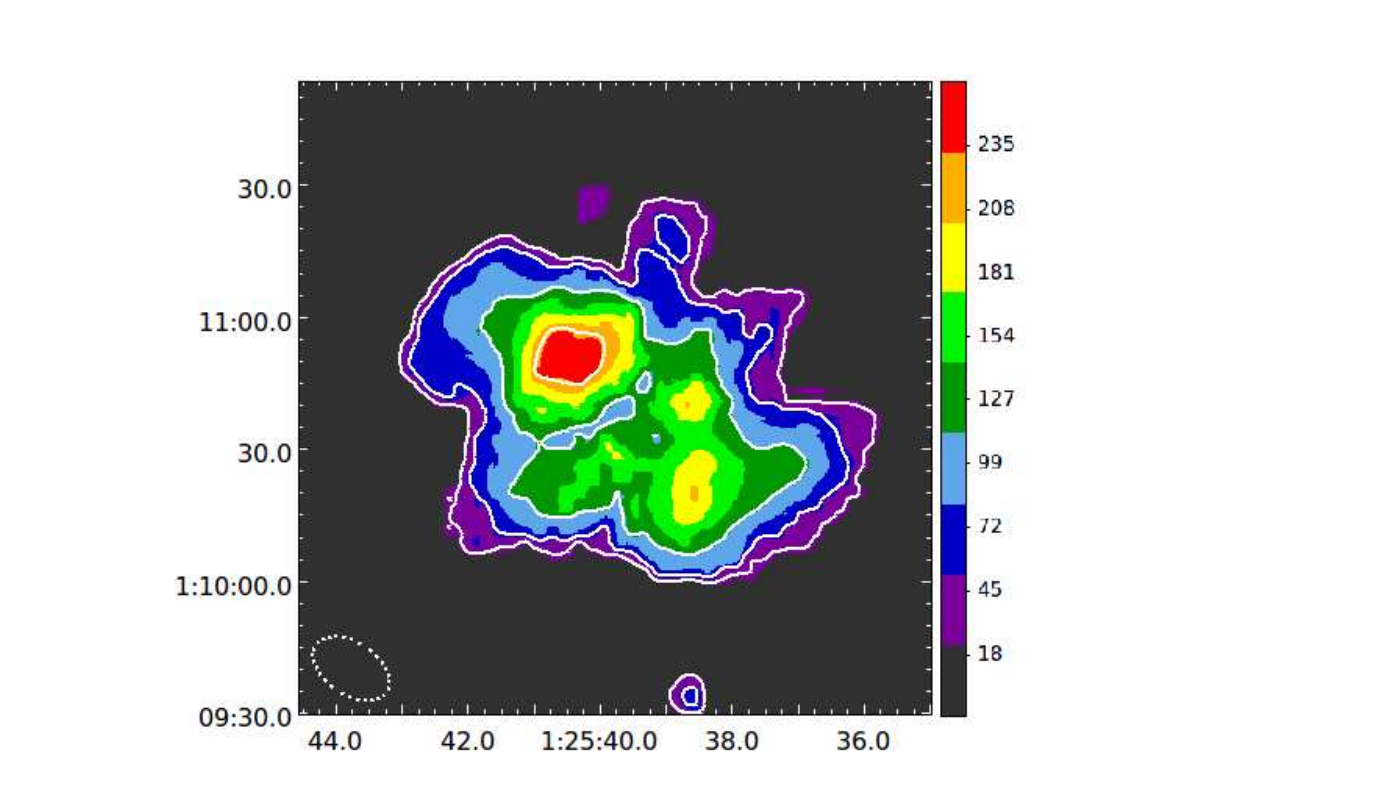}{0.57\textwidth}{(b)}}
\gridline{\leftfig{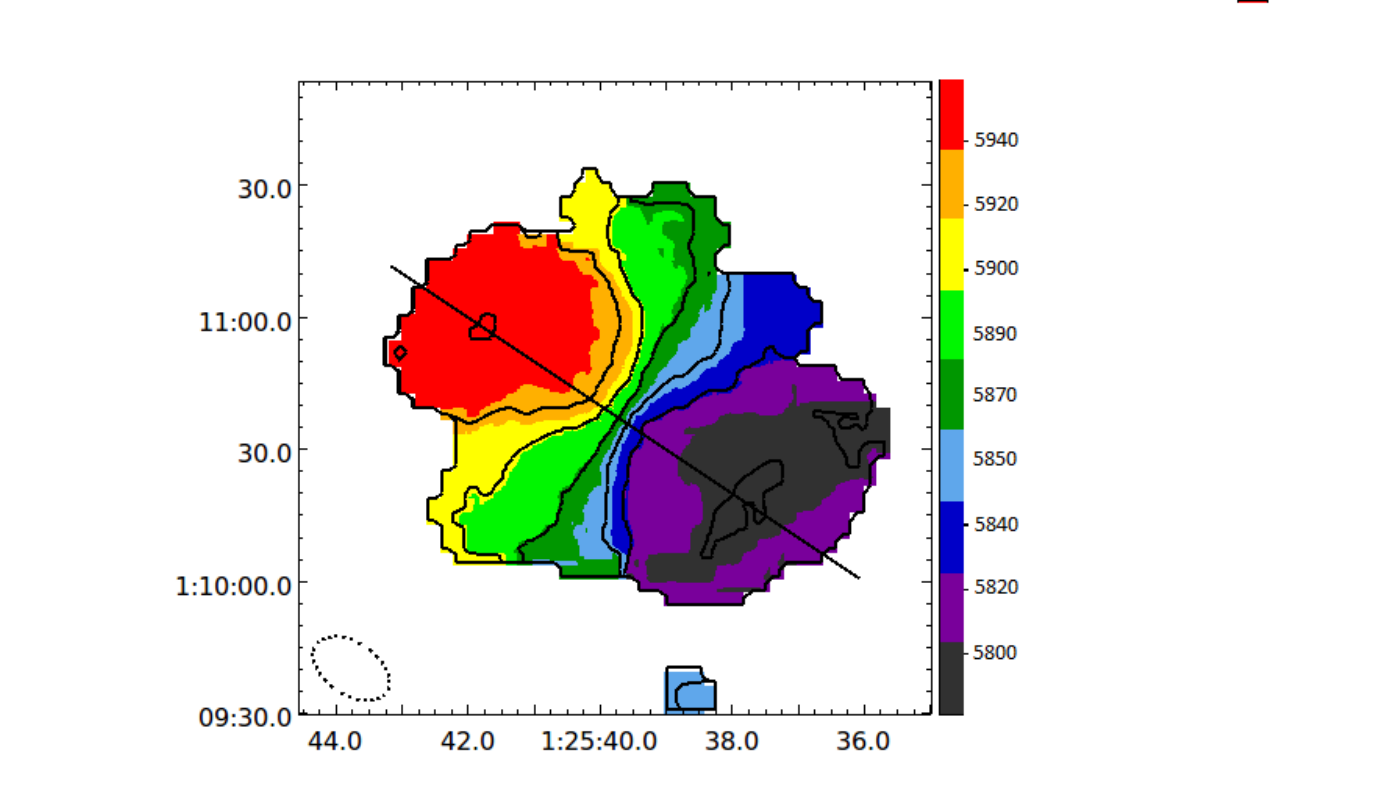}{0.57\textwidth}{(c)}
  \leftfig{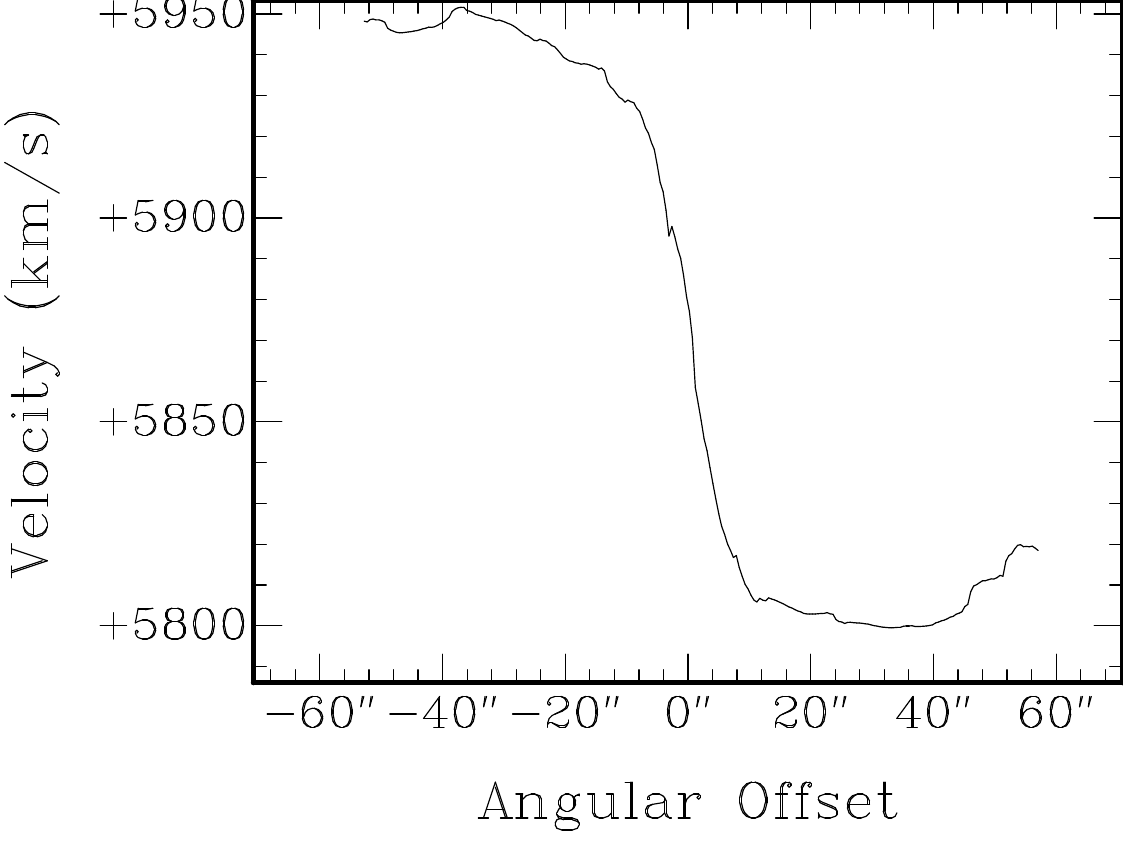}{0.37\textwidth}{(d)}}
\gridline{\leftfig{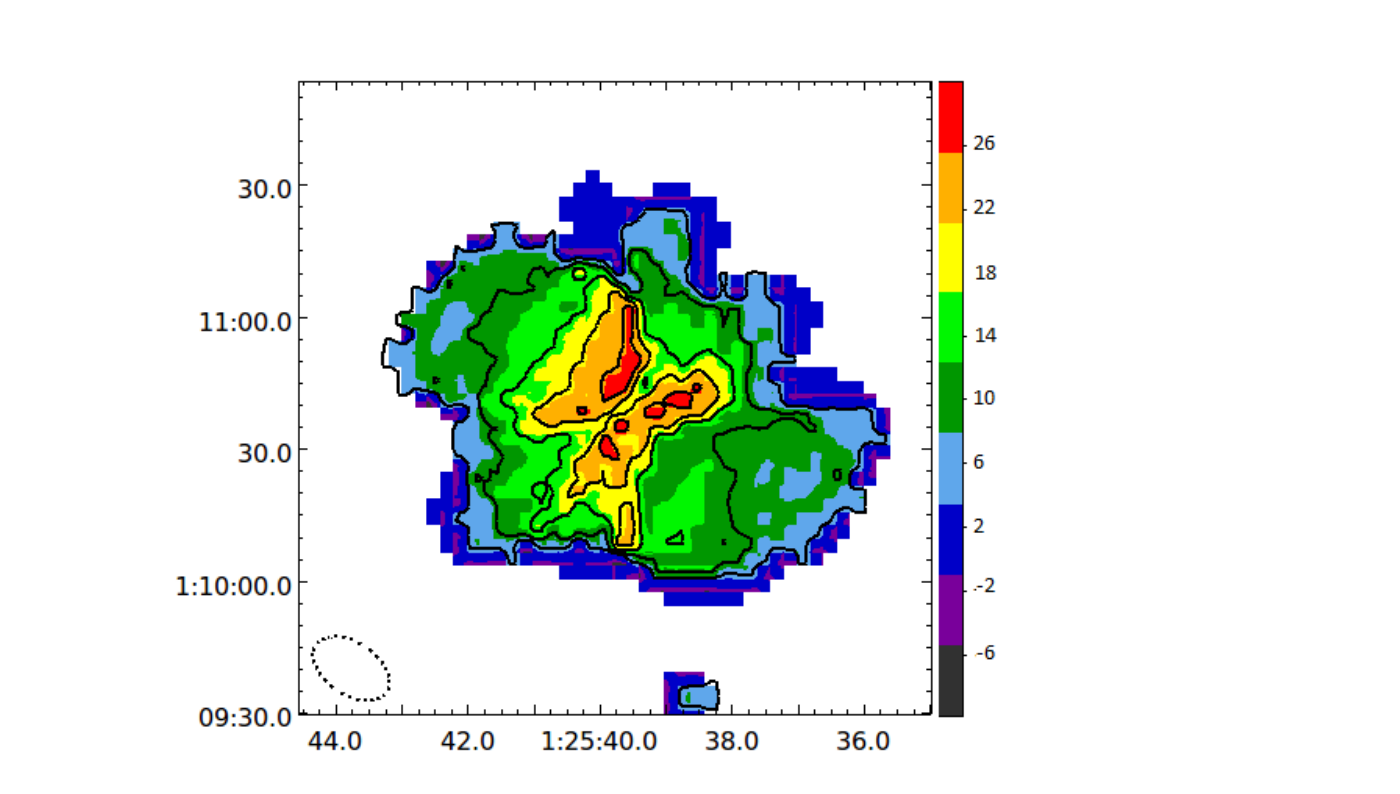}{0.57\textwidth}{(e)}
  \leftfig{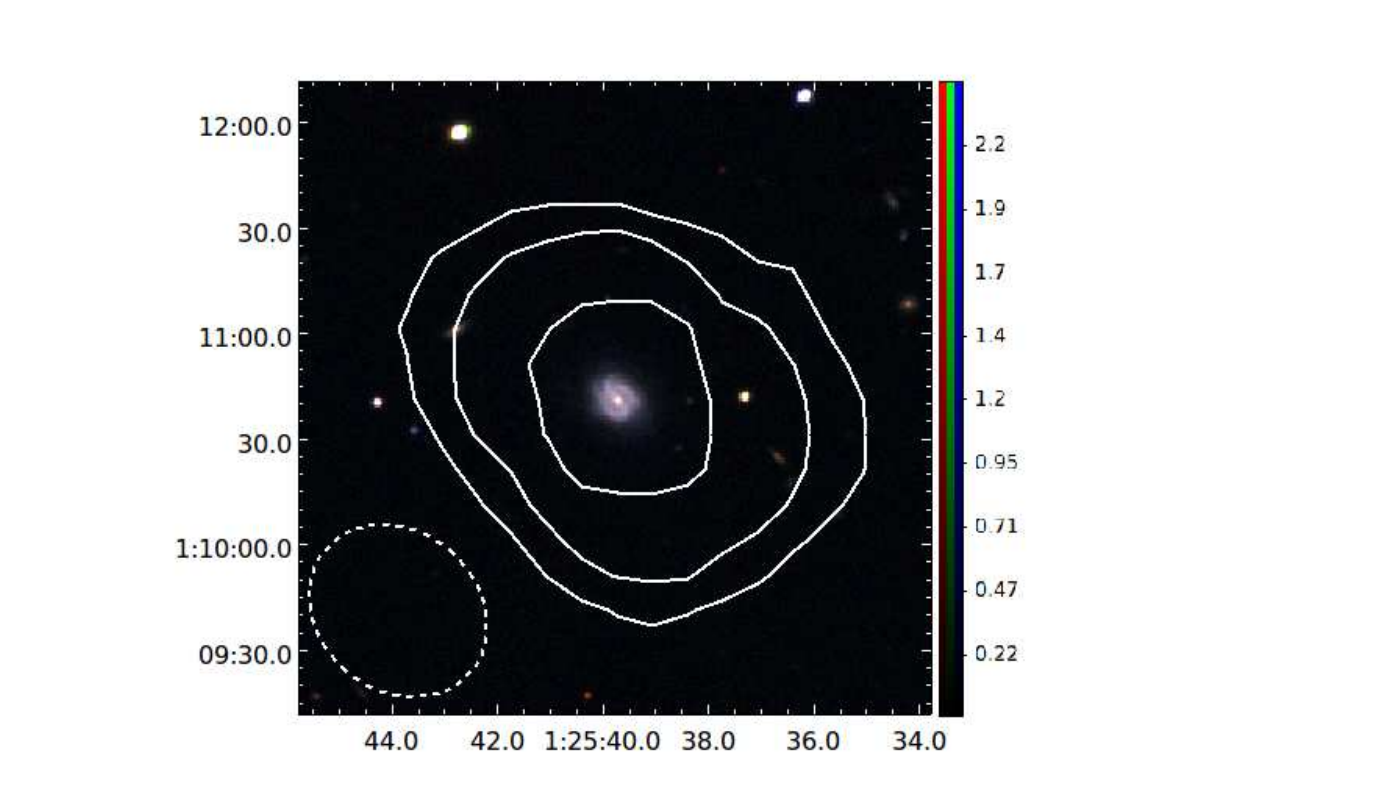}{0.57\textwidth}{(f)}}
  
\caption{\footnotesize SDSS0125+0110 (ARK 044): (a) Moment 0 contours made
  with a $22\arcsec \times 13\arcsec$ beam overlaid on a SDSS DR9
  \textit{gri} image. Contours represent H \textsc{I} intensities
  equivalent to column densities of $\rm 2^{n}\times10^{20}\ cm^{-2}$
  for n = 0, 1, 2, 3. Image units are analog to digital units
  (ADU). (b) Moment 0 map with the same contours as in (a). Map units
  are $\rm Jy\ Beam^{-1}\ m\ s^{-1}$. (c) Moment 1 map with a thick
  line showing the major axis. Contours are $25\ \rm km\ s^{-1}$. Map
  units are $\rm km\ s^{-1}$. (d) Velocities along the major axis
  slice shown in (c). (e) Moment 2 map with $5\ \rm km\ s^{-1}$
  contours. Map units are $\rm km\ s^{-1}$. (f) Low-resolution Moment
  0 contours made with a $54\arcsec \times 45\arcsec$ beam overlaid on
  SDSS DR9 \textit{gri} image. Contours represent H \textsc{I}
  intensities equivalent to column densities of $\rm
  2^{n}\times10^{20}\ cm^{-2}$ for n = 0, 1, 2. Image units are the
  same as in (a). The horizontal and vertical axes of each map are
  right ascention and declination (J2000). Beam sizes are shown in the
  lower left corner of each map. \label{fig:0125} }

\end{figure*}
\begin{figure*}[htb!]

  \gridline{\leftfig{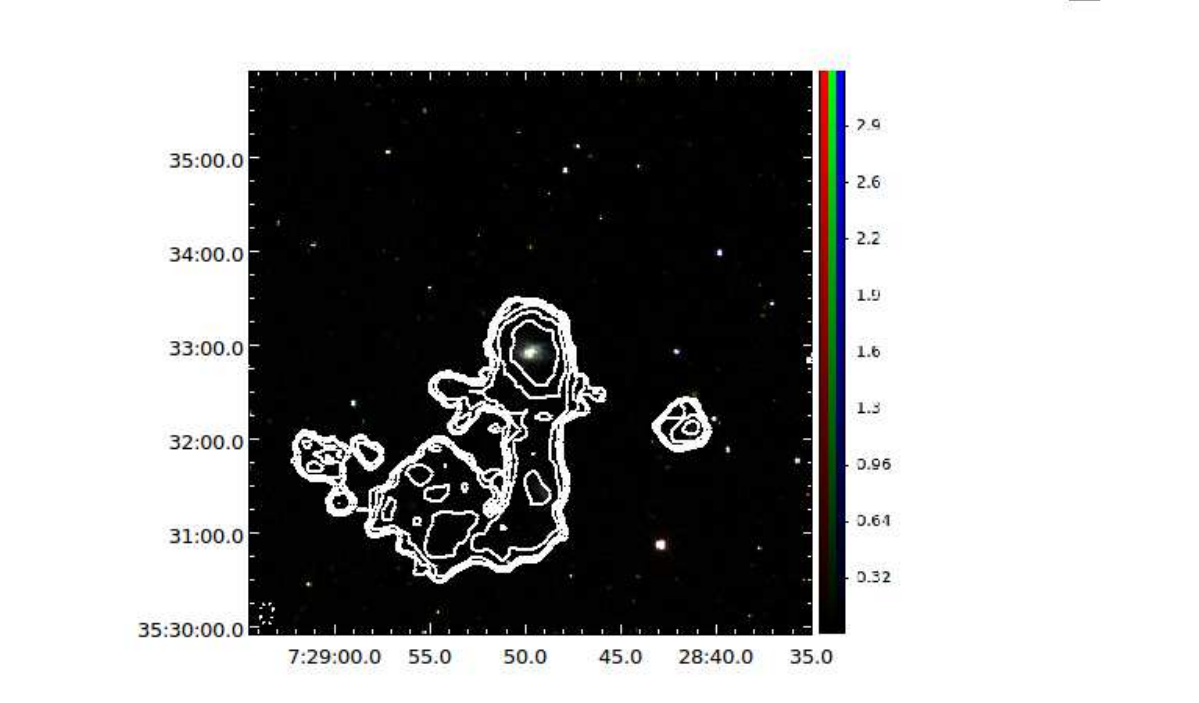}{0.57\textwidth}{(a)}
    \leftfig{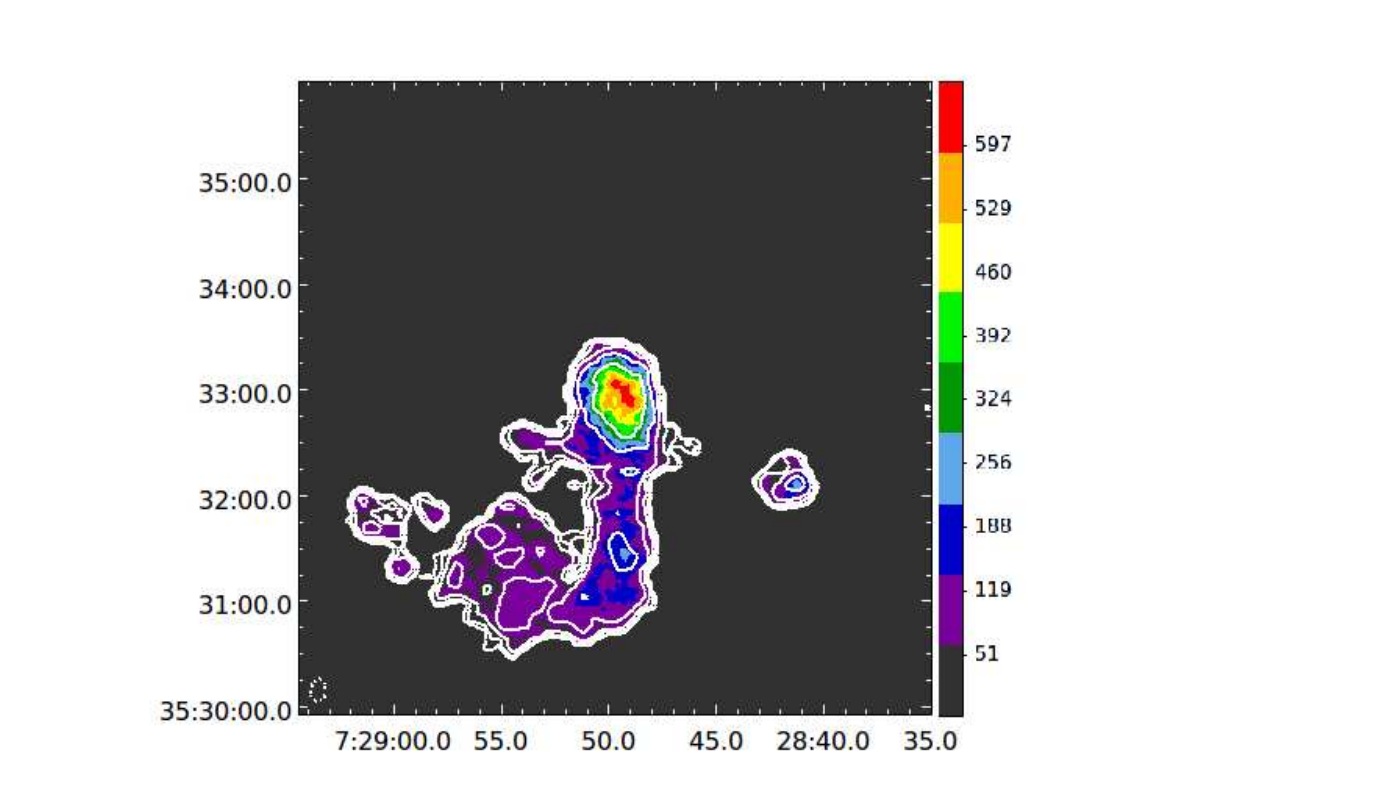}{0.57\textwidth}{(b)}}
  \gridline{\leftfig{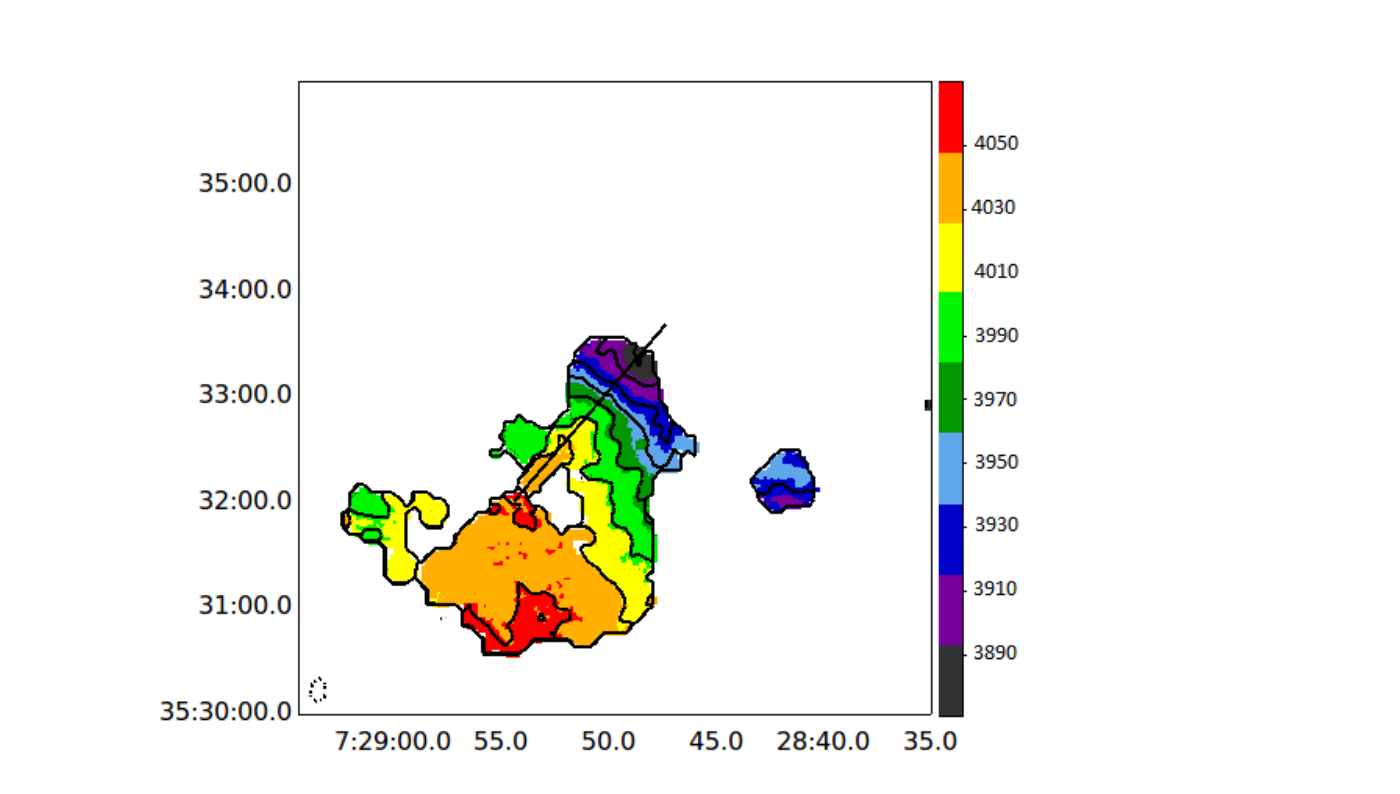}{0.57\textwidth}{(c)}
    \leftfig{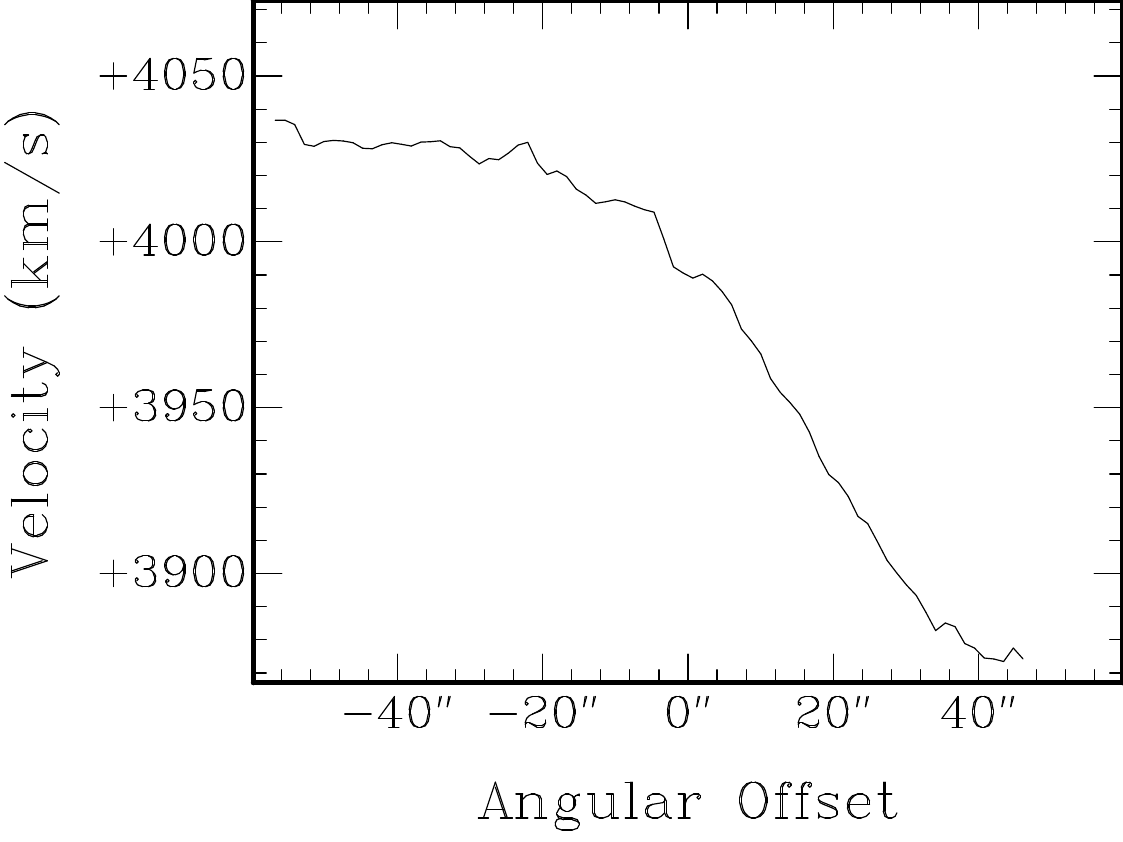}{0.37\textwidth}{(d)}}
  \gridline{\leftfig{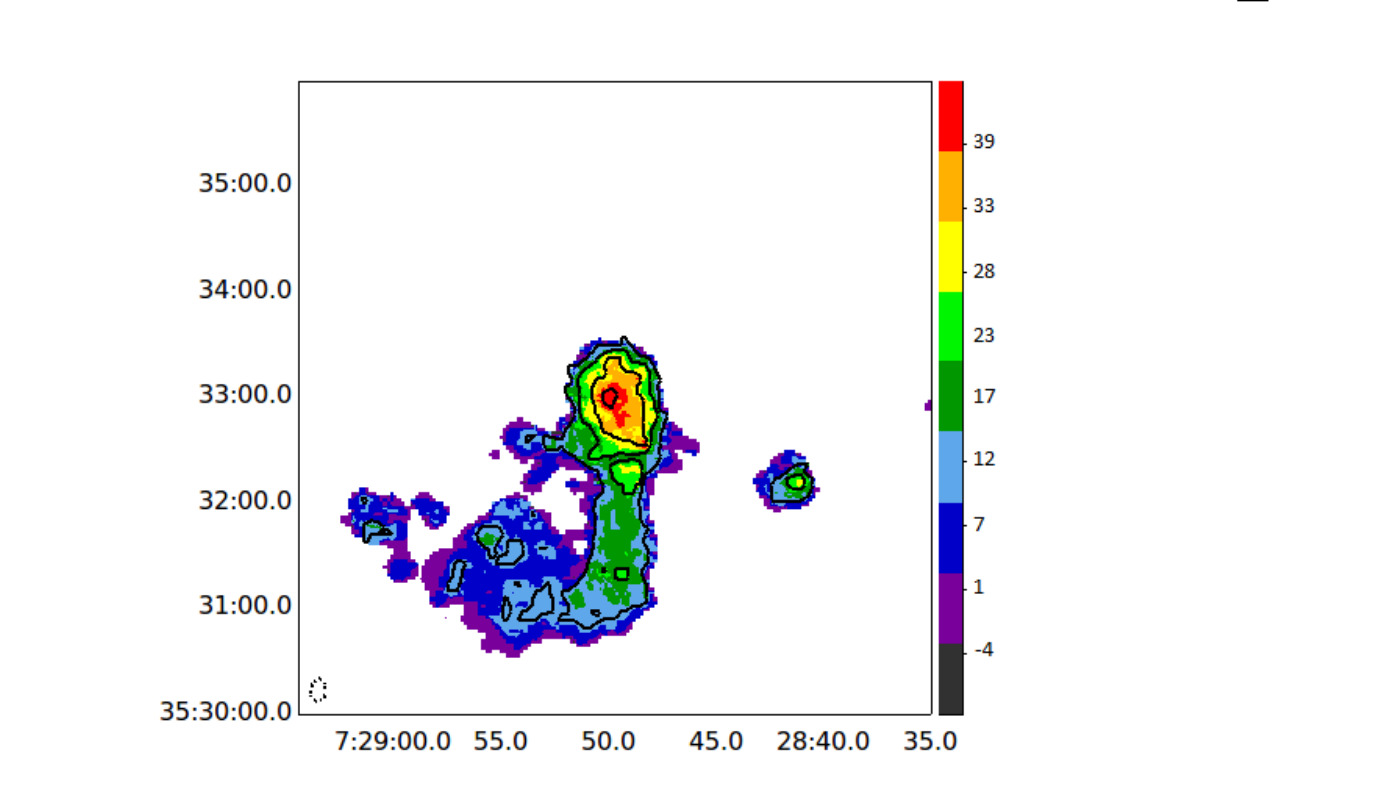}{0.57\textwidth}{(e)}
    \leftfig{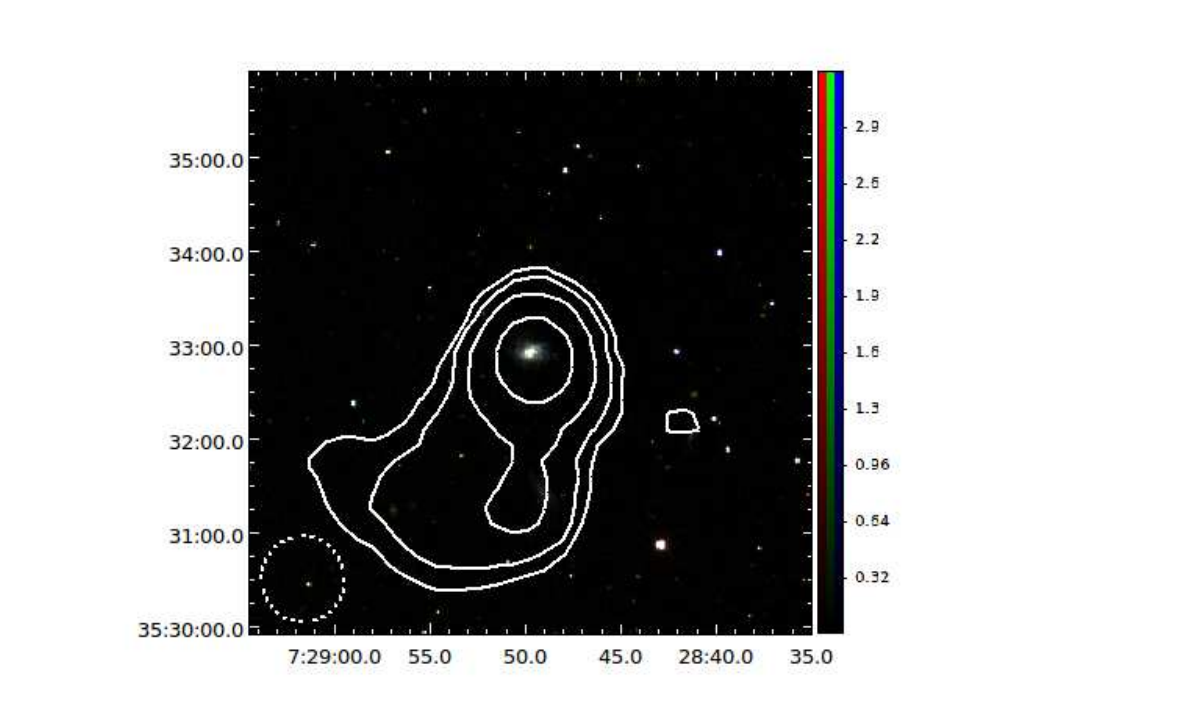}{0.57\textwidth}{(f)}}
  
\caption{\footnotesize SDSS0728+3532 (ARK 134): (a) Moment 0 contours made
  with a $13\arcsec \times 8\arcsec$ beam overlaid on a SDSS DR9
  \textit{gri} image. Contours represent H \textsc{I} intensities
  equivalent to column densities of $\rm 2^{n}\times10^{20}\ cm^{-2}$
  for n = 0, 1, 2, 3, 4, 5. Image units are analog to digital units
  (ADU). (b) Moment 0 map with the same contours as in (a). Map units
  are $\rm Jy\ Beam^{-1}\ m\ s^{-1}$. (c) Moment 1 map with a thick
  line showing the major axis. Contours are $25\ \rm km\ s^{-1}$. Map
  units are $\rm km\ s^{-1}$. (d) Velocities along the major axis
  slice shown in (c). (e) Moment 2 map with $10\ \rm km\ s^{-1}$
  contours. Map units are $\rm km\ s^{-1}$. (f) Low-resolution Moment
  0 contours made with a $55\arcsec \times 53\arcsec$ beam overlaid on
  SDSS DR9 \textit{gri} image. Contours represent H \textsc{I}
  intensities equivalent to column densities of $\rm
  2^{n}\times10^{20}\ cm^{-2}$ for n = 0, 1, 2, 3. Image units are the
  same as in (a). The horizontal and vertical axes of each map are
  right ascention and declination (J2000). Beam sizes are shown in the
  lower left corner of each map. \label{fig:0728}}

\end{figure*}
\begin{figure*}[htb!]
  
  \gridline{\leftfig{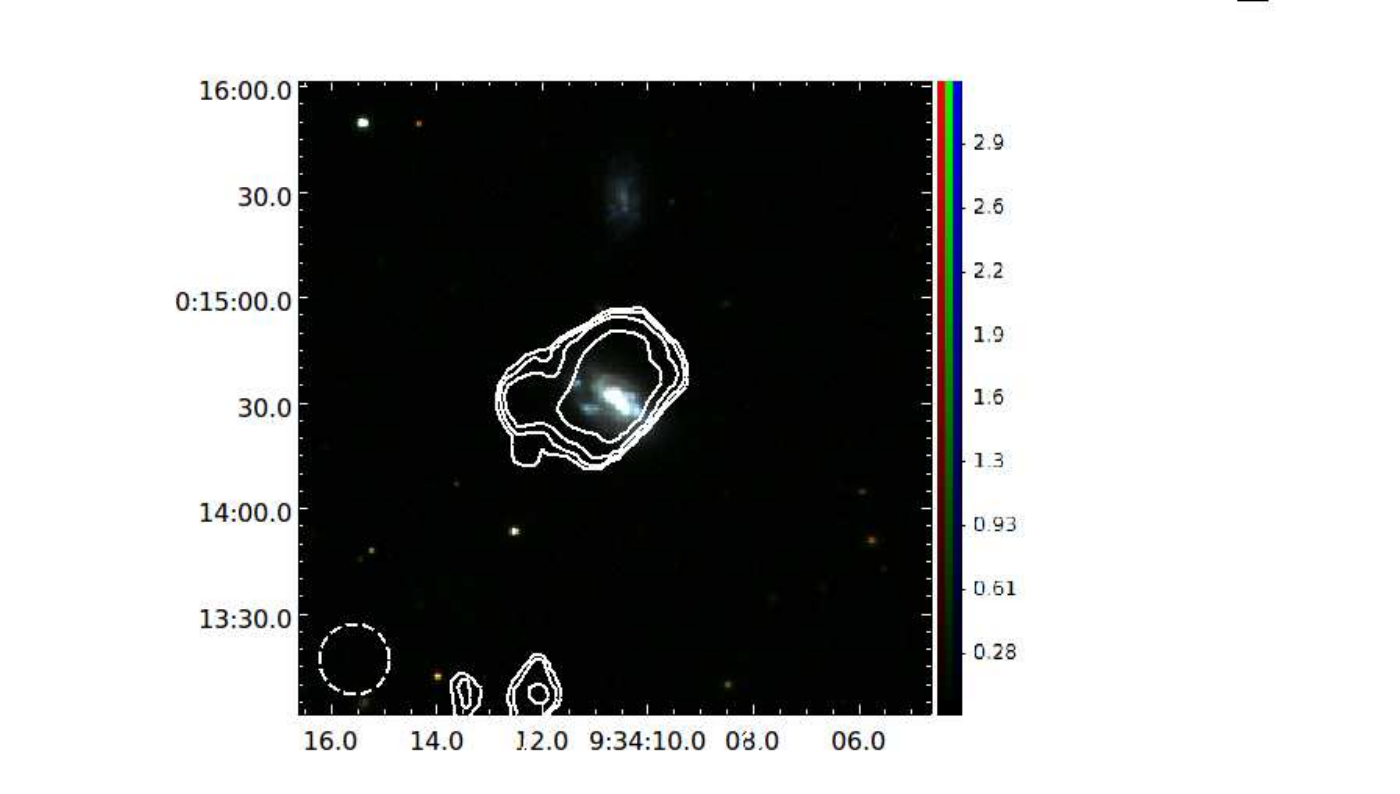}{0.57\textwidth}{(a)}
    \leftfig{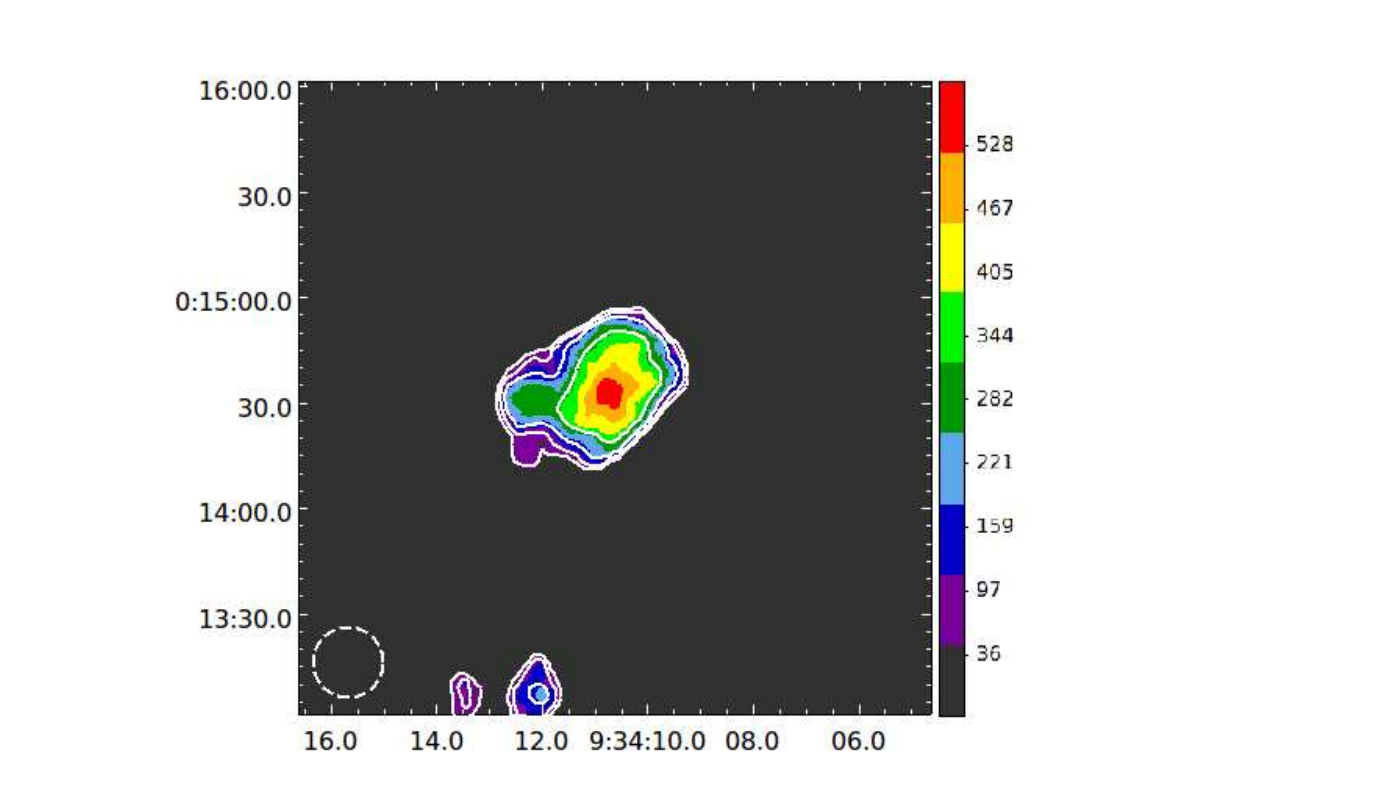}{0.57\textwidth}{(b)}}
  \gridline{\leftfig{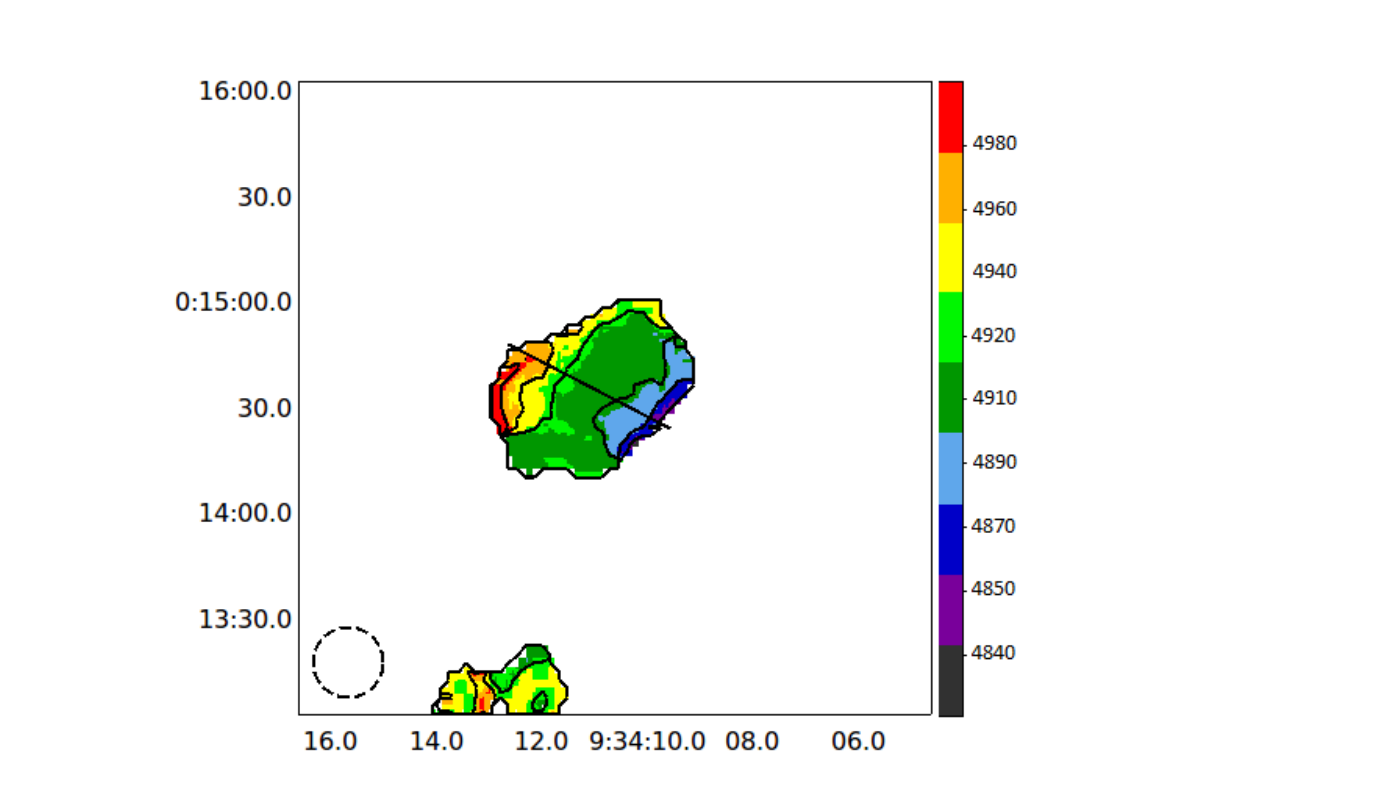}{0.57\textwidth}{(c)}
    \leftfig{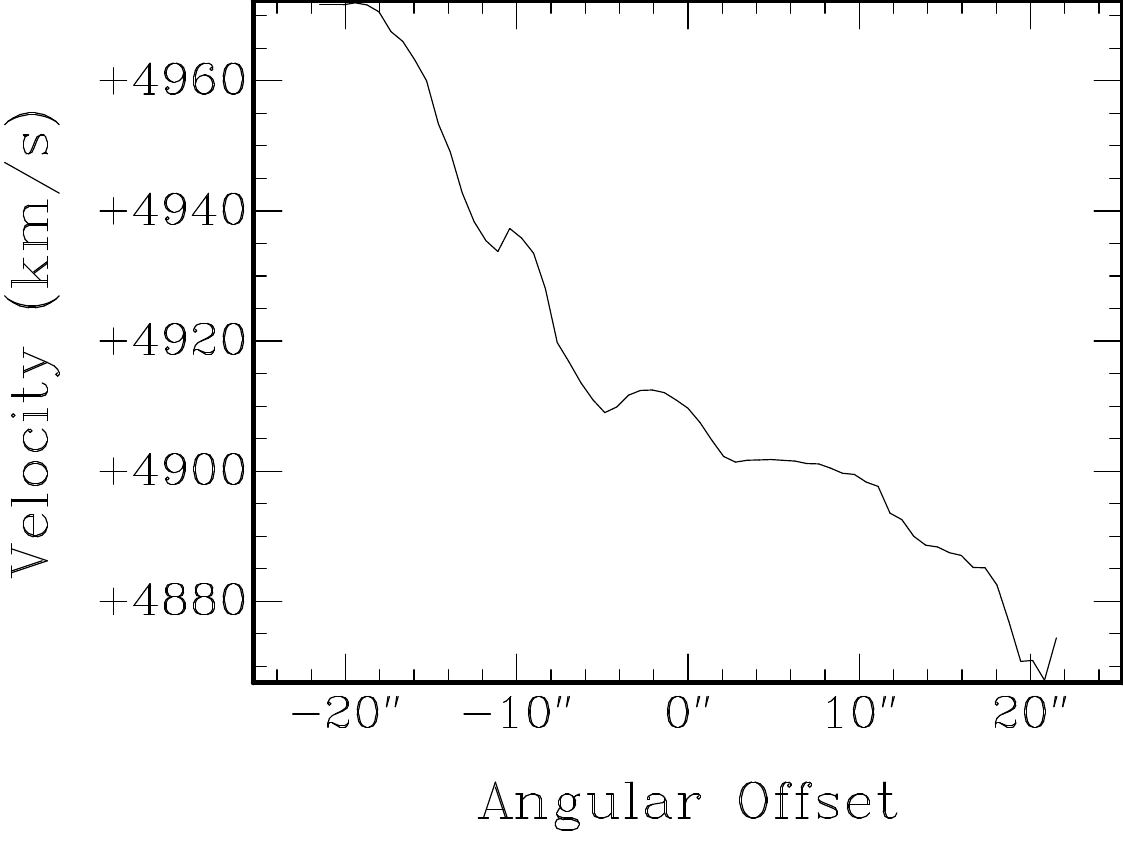}{0.37\textwidth}{(d)}}
  \gridline{\leftfig{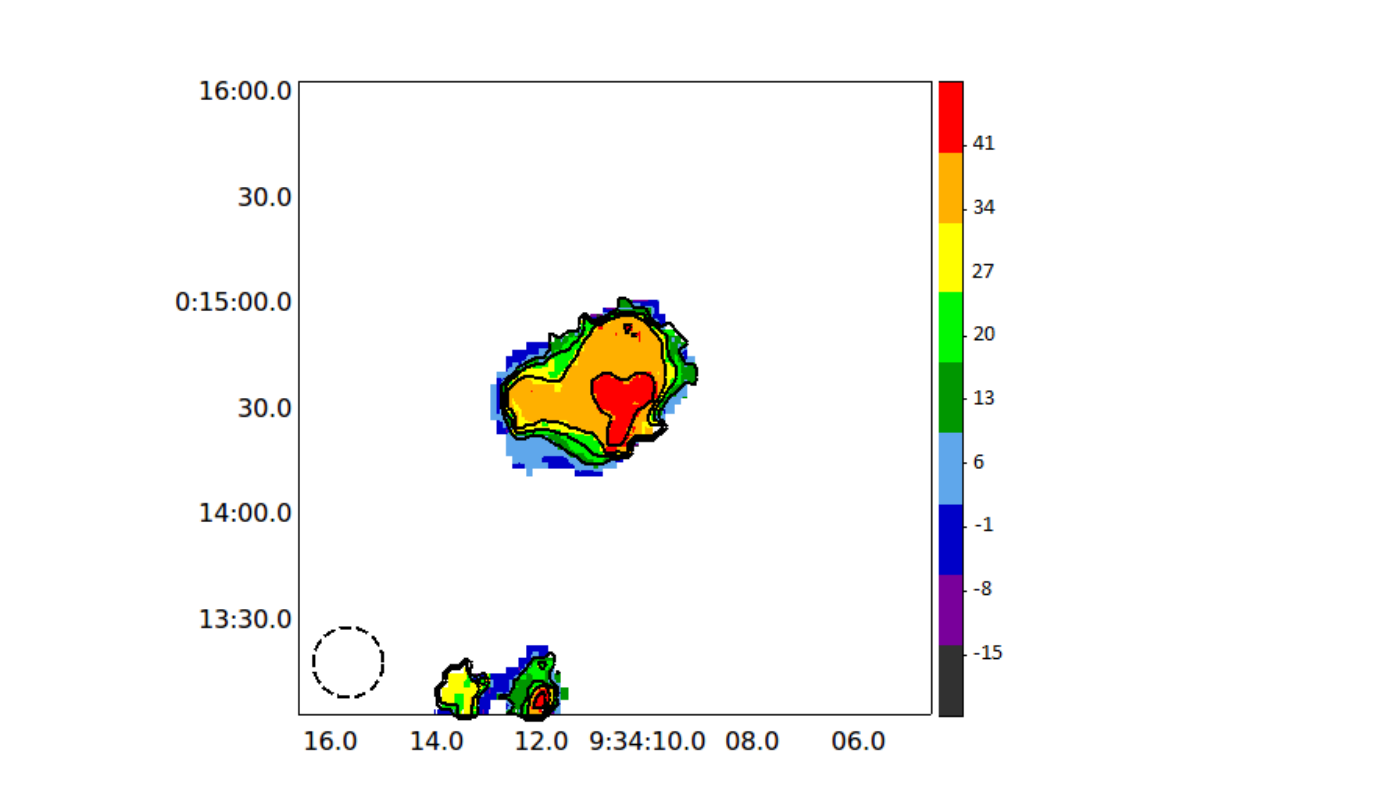}{0.57\textwidth}{(e)}
    \leftfig{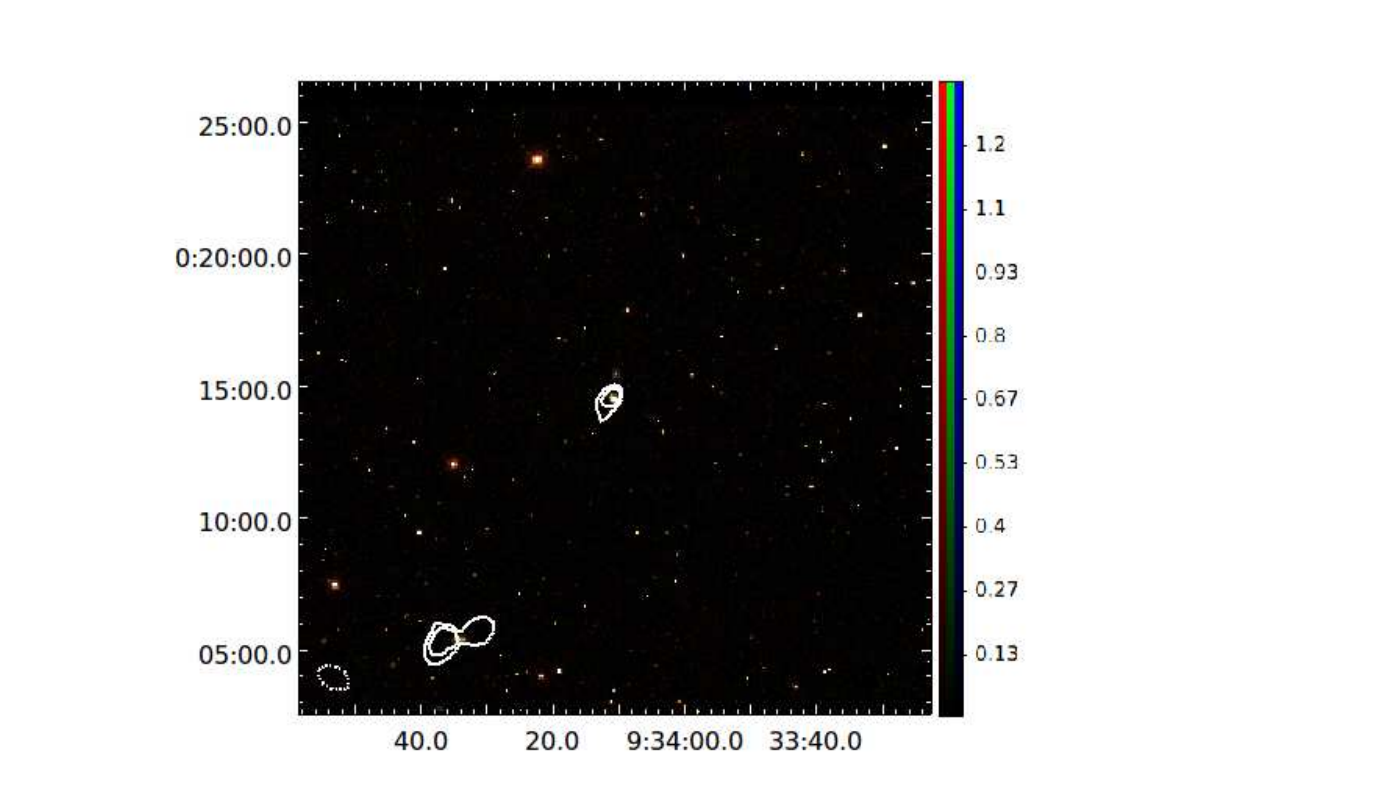}{0.57\textwidth}{(f)}}
  
\caption{\footnotesize SDSS0934+0014 (UGC 05097): (a) Moment 0 contours made
  with a $20\arcsec \times 20\arcsec$ beam overlaid on a SDSS DR9
  \textit{gri} image. Contours represent H \textsc{I} intensities
  equivalent to column densities of $\rm 2^{n}\times10^{20}\ cm^{-2}$
  for n = 0, 1, 2, 3. Image units are analog to digital units
  (ADU). (b) Moment 0 map with the same contours as in (a). Map units
  are $\rm Jy\ Beam^{-1}\ m\ s^{-1}$. (c) Moment 1 map with a thick
  line showing the major axis. Contours are $25\ \rm km\ s^{-1}$. Map
  units are $\rm km\ s^{-1}$. (d) Velocities along the major axis
  slice shown in (c). (e) Moment 2 map with $10\ \rm km\ s^{-1}$
  contours. Map units are $\rm km\ s^{-1}$. (f) Low-resolution Moment
  0 contours made with a $75\arcsec \times 49\arcsec$ beam overlaid on
  SDSS DR9 \textit{gri} image. Contours represent H \textsc{I}
  intensities equivalent to column densities of $\rm
  2^{n}\times10^{20}\ cm^{-2}$ for n = 0, 1, 2. Image units are the
  same as in (a). The horizontal and vertical axes of each map are
  right ascention and declination (J2000). Beam sizes are shown in the
  lower left corner of each map. \label{fig:0934}}

\end{figure*}
\begin{figure*}[htb!]

  \gridline{\leftfig{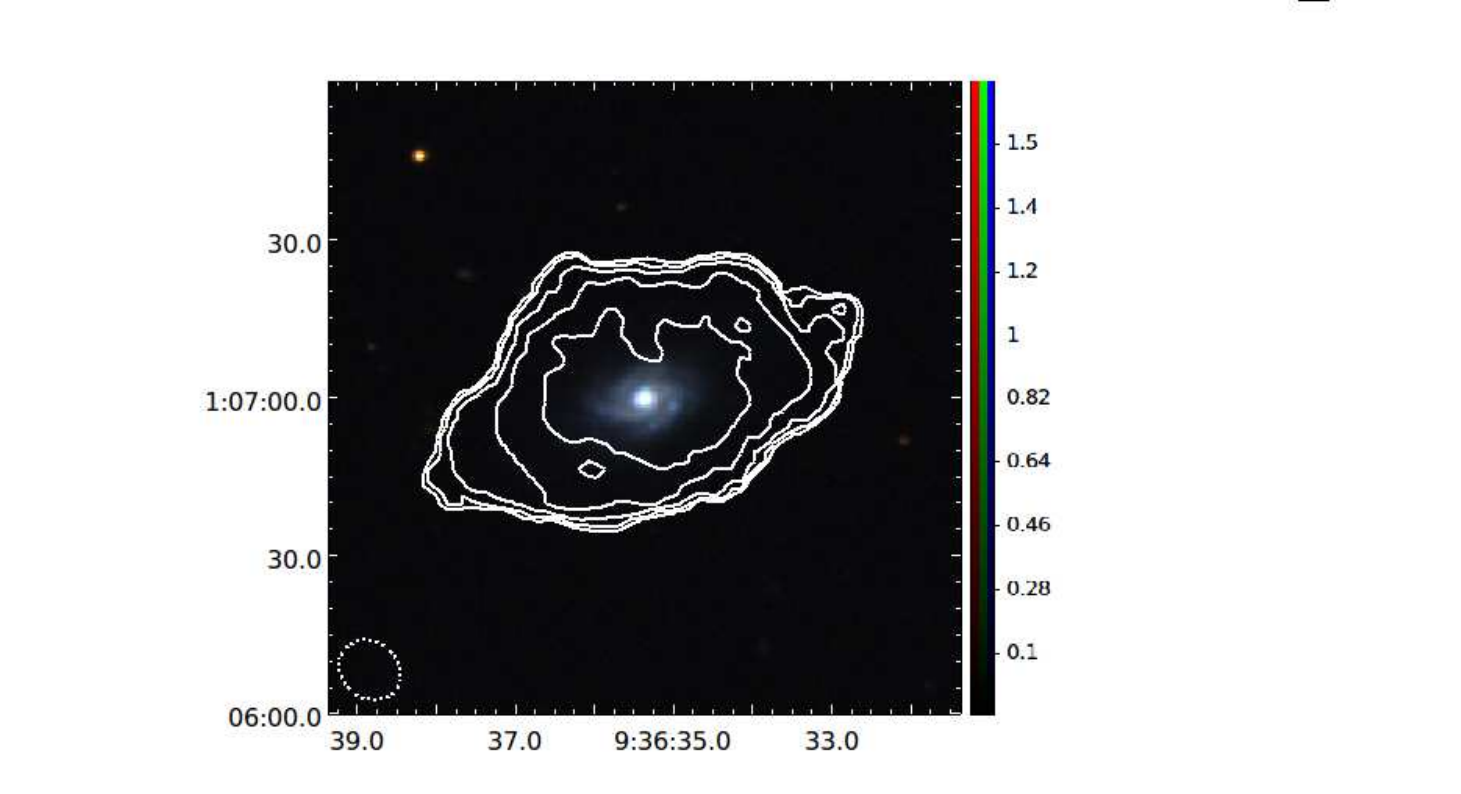}{0.57\textwidth}{(a)}
    \leftfig{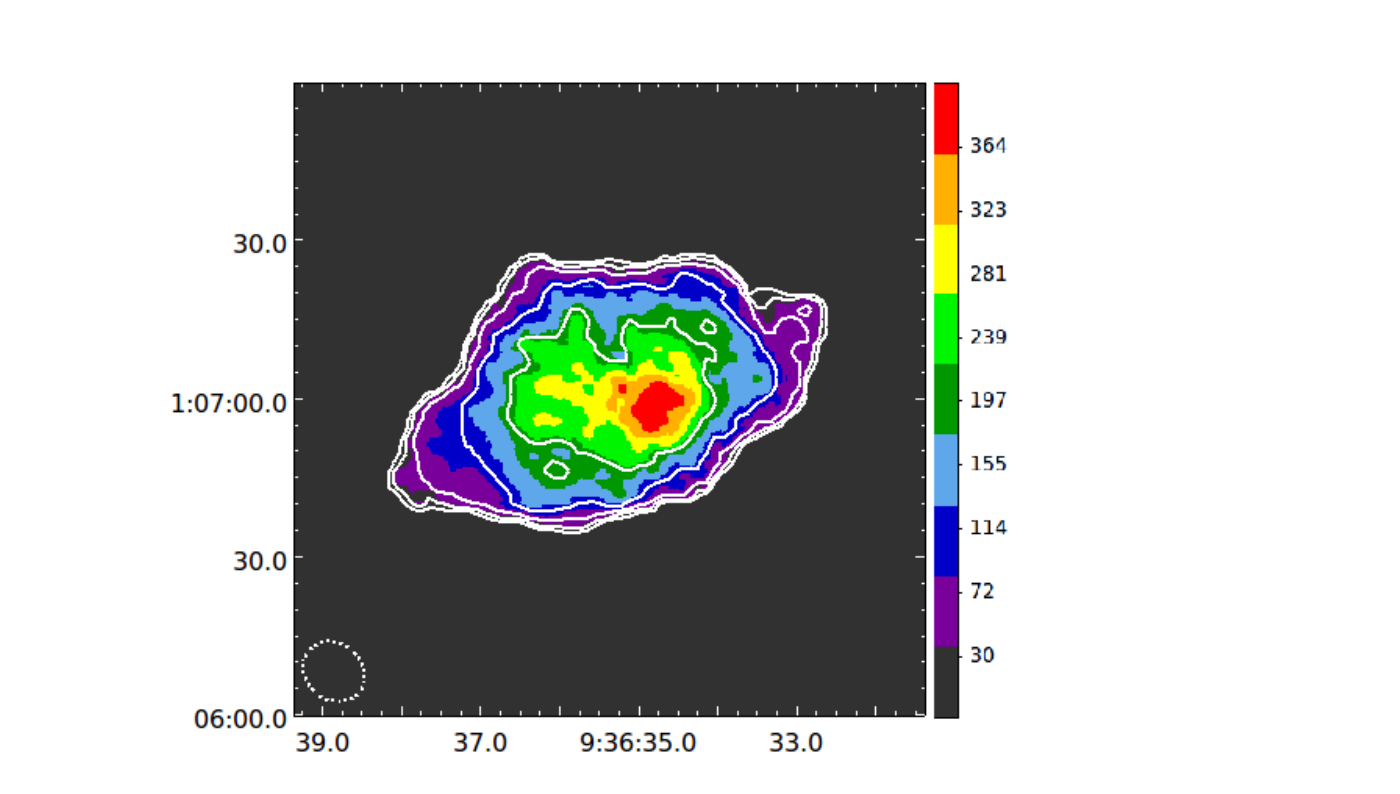}{0.57\textwidth}{(b)}}
  \gridline{\leftfig{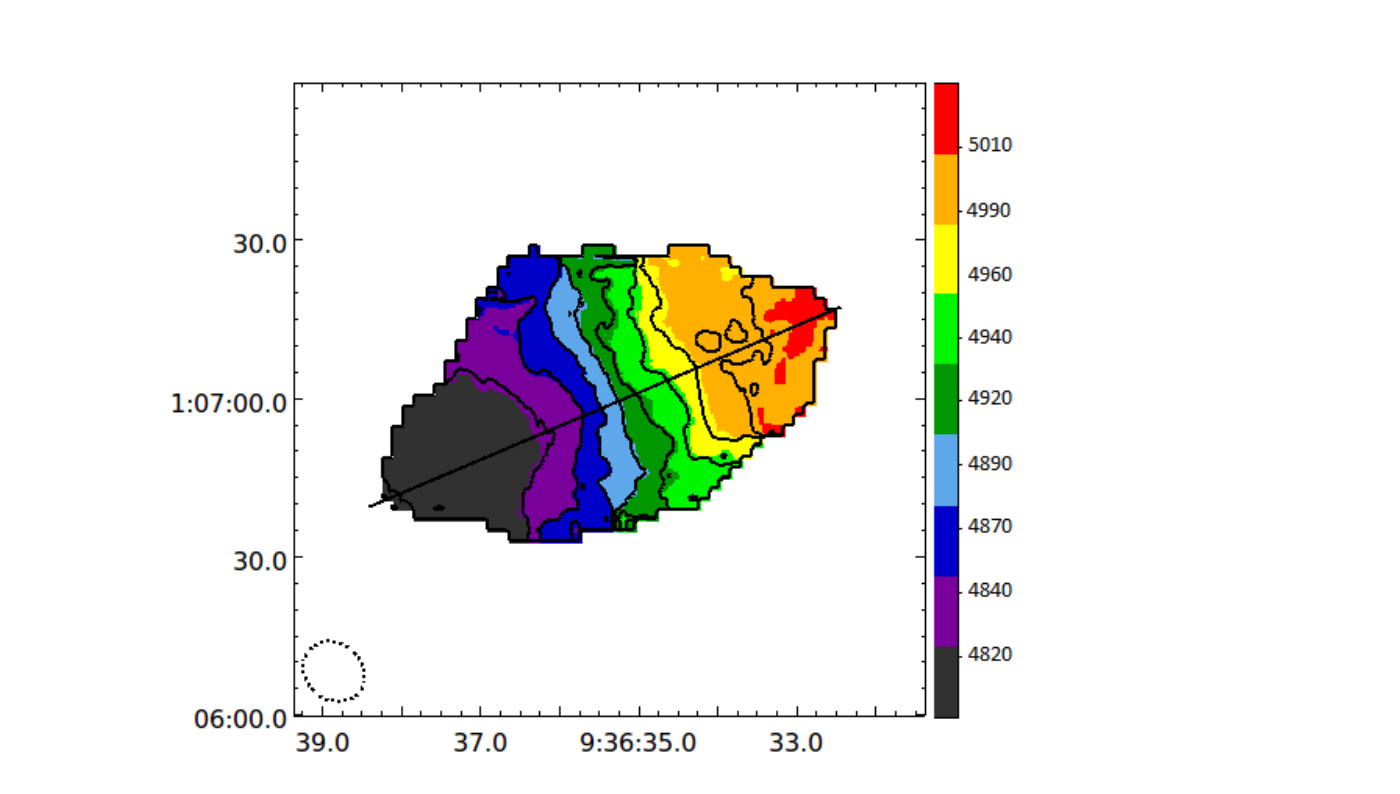}{0.57\textwidth}{(c)}
    \leftfig{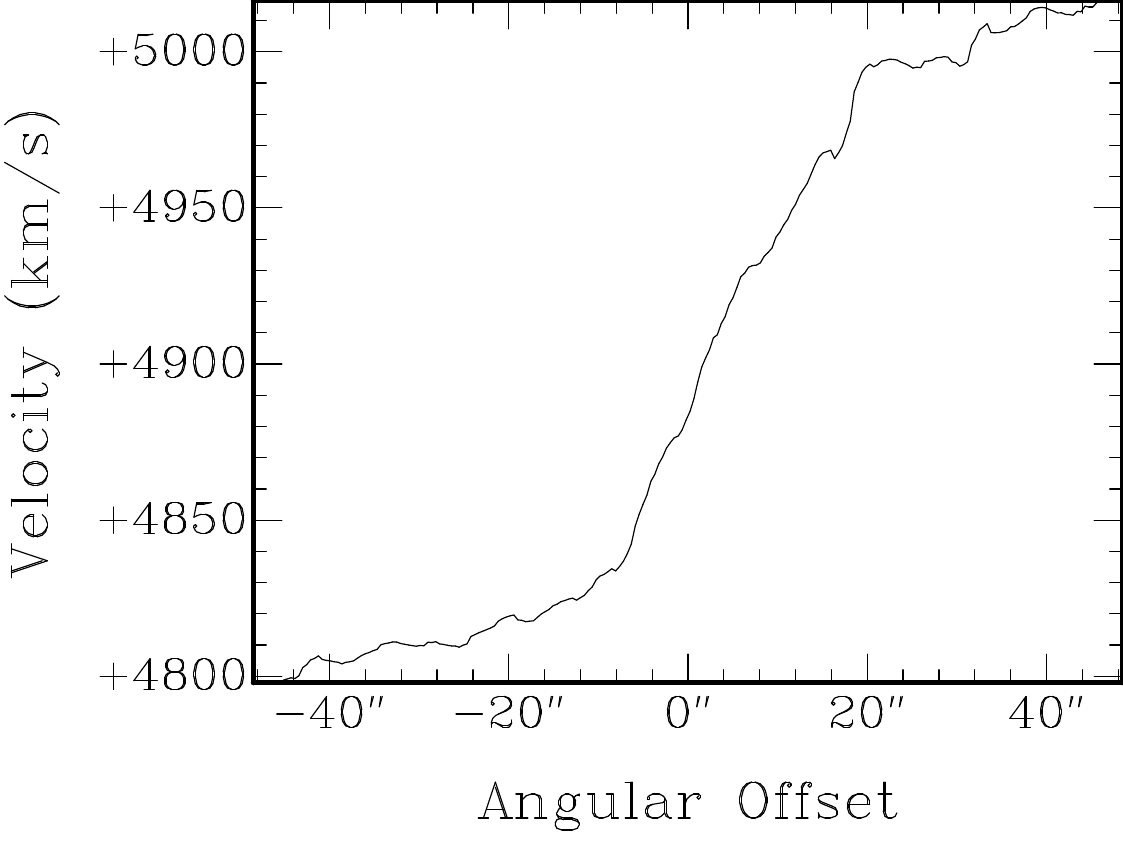}{0.37\textwidth}{(d)}}
  \gridline{\leftfig{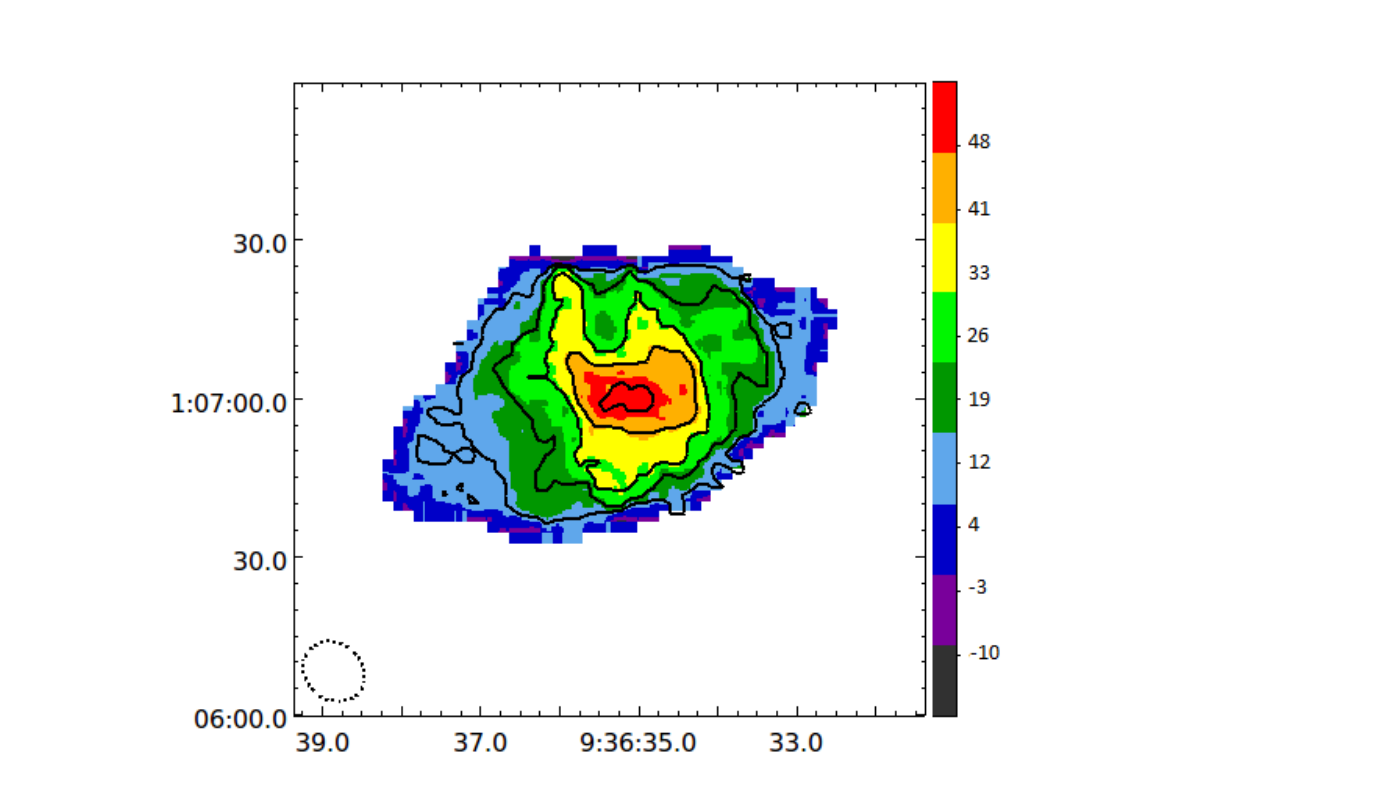}{0.57\textwidth}{(e)}
    \leftfig{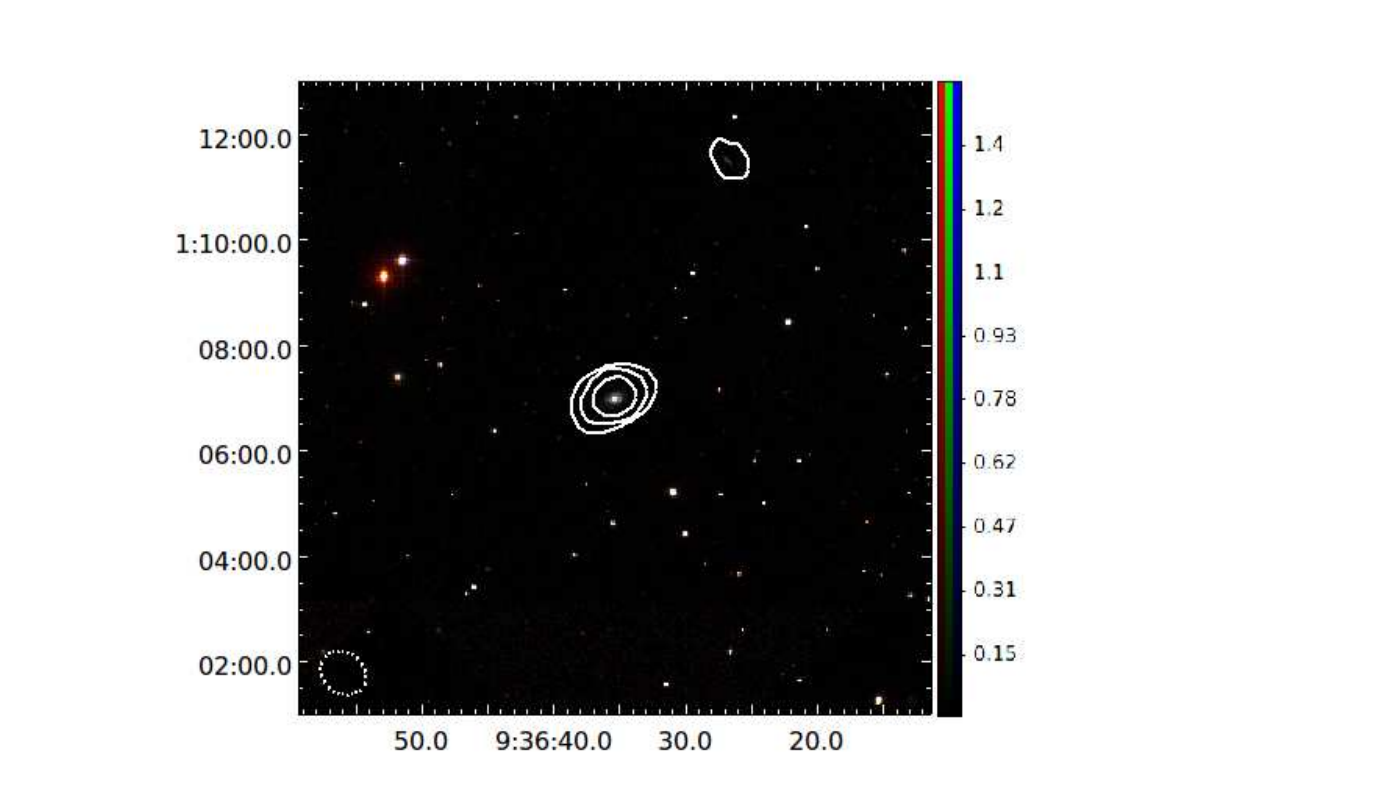}{0.57\textwidth}{(f)}}
  
\caption{\footnotesize SDSS0936+0106 (CGCG 007-009): (a) Moment 0 contours
  made with a $12\arcsec \times 11\arcsec$ beam overlaid on a SDSS DR9
  \textit{gri} image. Contours represent H \textsc{I} intensities
  equivalent to column densities of $\rm 2^{n}\times10^{20}\ cm^{-2}$
  for n = 0, 1, 2, 3, 4, 5. Image units are analog to digital units
  (ADU). (b) Moment 0 map with the same contours as in (a). Map units
  are $\rm Jy\ Beam^{-1}\ m\ s^{-1}$. (c) Moment 1 map with a thick
  line showing the major axis. Contours are $25\ \rm km\ s^{-1}$. Map
  units are $\rm km\ s^{-1}$. (d) Velocities along the major axis
  slice shown in (c). (e) Moment 2 map with $10\ \rm km\ s^{-1}$
  contours. Map units are $\rm km\ s^{-1}$. (f) Low-resolution Moment
  0 contours made with a $55\arcsec \times 46\arcsec$ beam overlaid on
  SDSS DR9 \textit{gri} image. Contours represent H \textsc{I}
  intensities equivalent to column densities of $\rm
  2^{n}\times10^{20}\ cm^{-2}$ for n = 0, 1, 2. Image units are the
  same as in (a). The horizontal and vertical axes of each map are
  right ascention and declination (J2000). Beam sizes are shown in the
  lower left corner of each map. \label{fig:0936}}

\end{figure*}
\begin{figure*}[htb!]

  \gridline{\leftfig{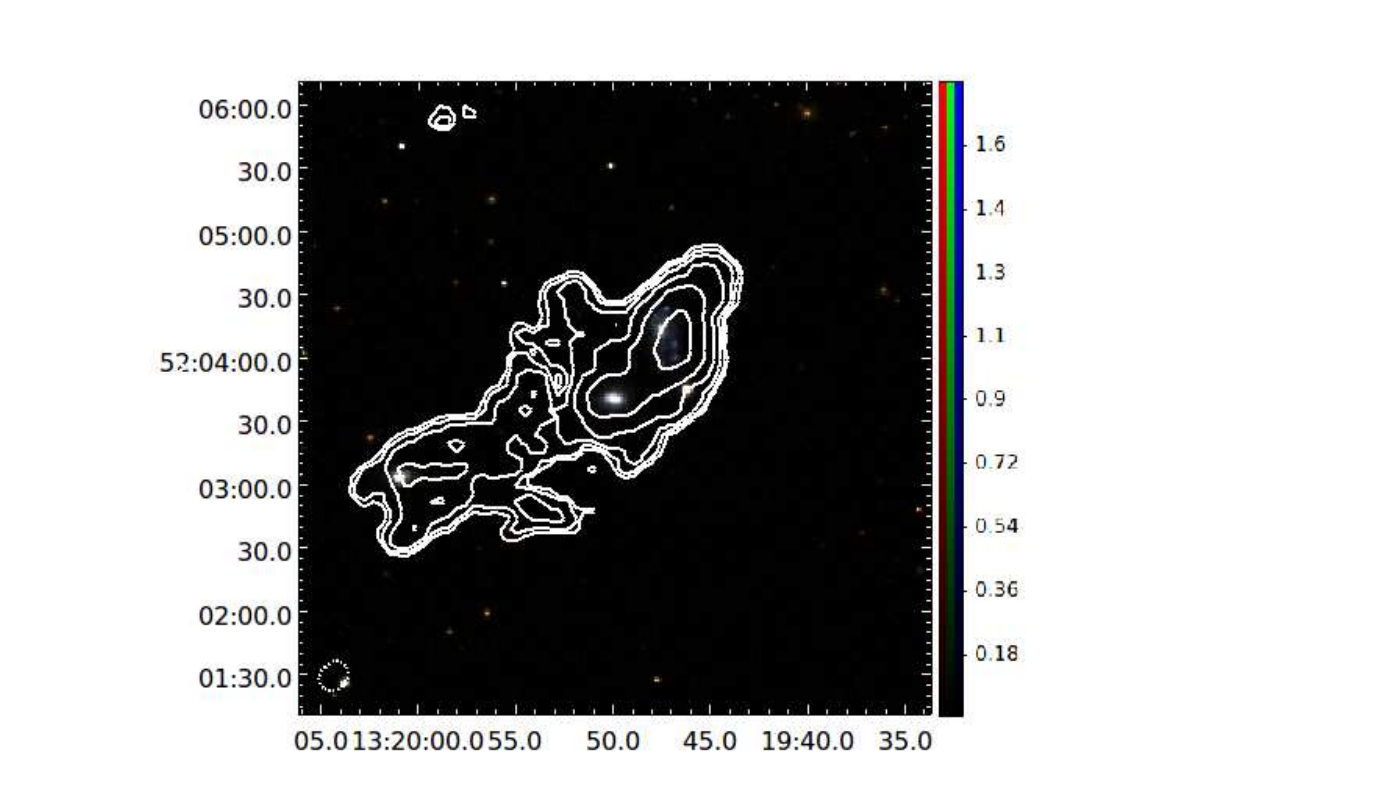}{0.57\textwidth}{(a)}
    \leftfig{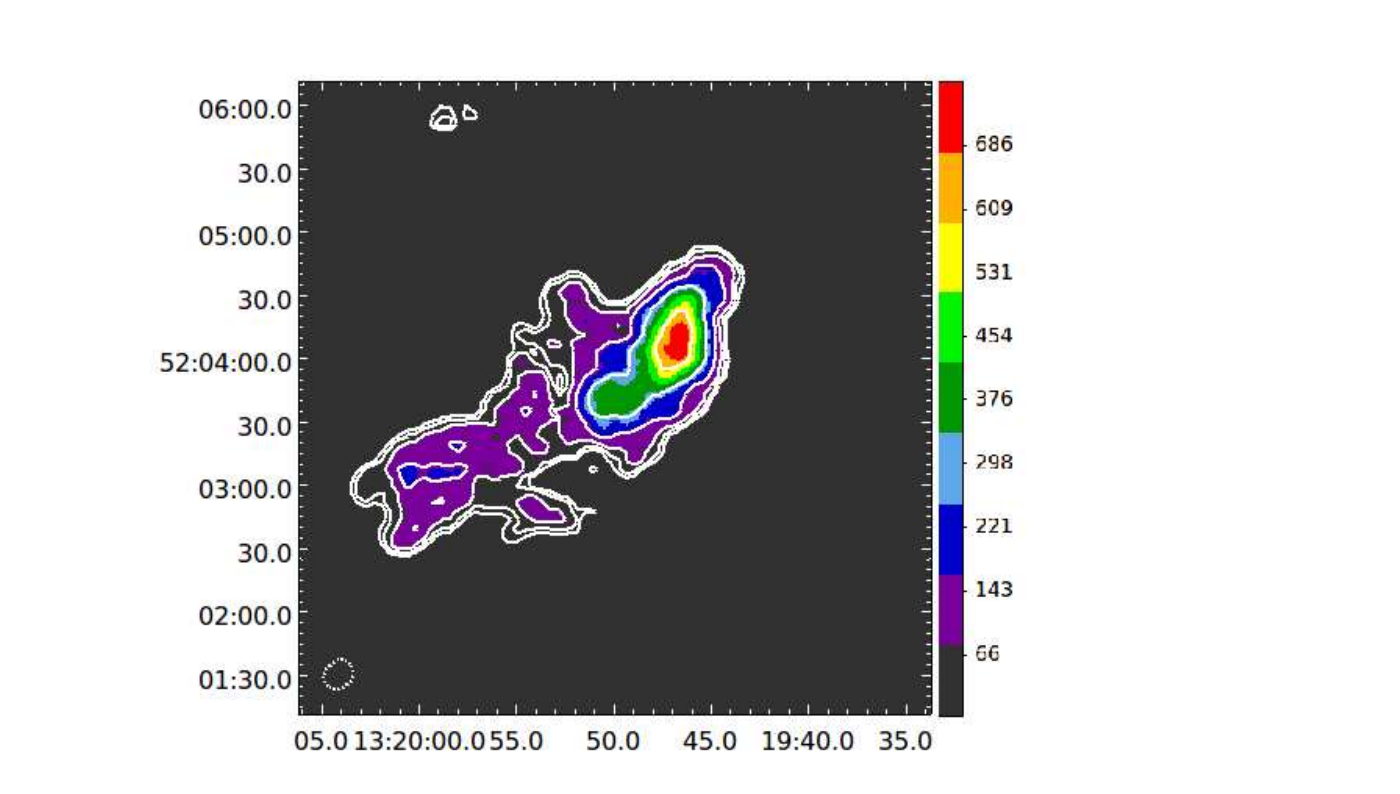}{0.57\textwidth}{(b)}}
  \gridline{\leftfig{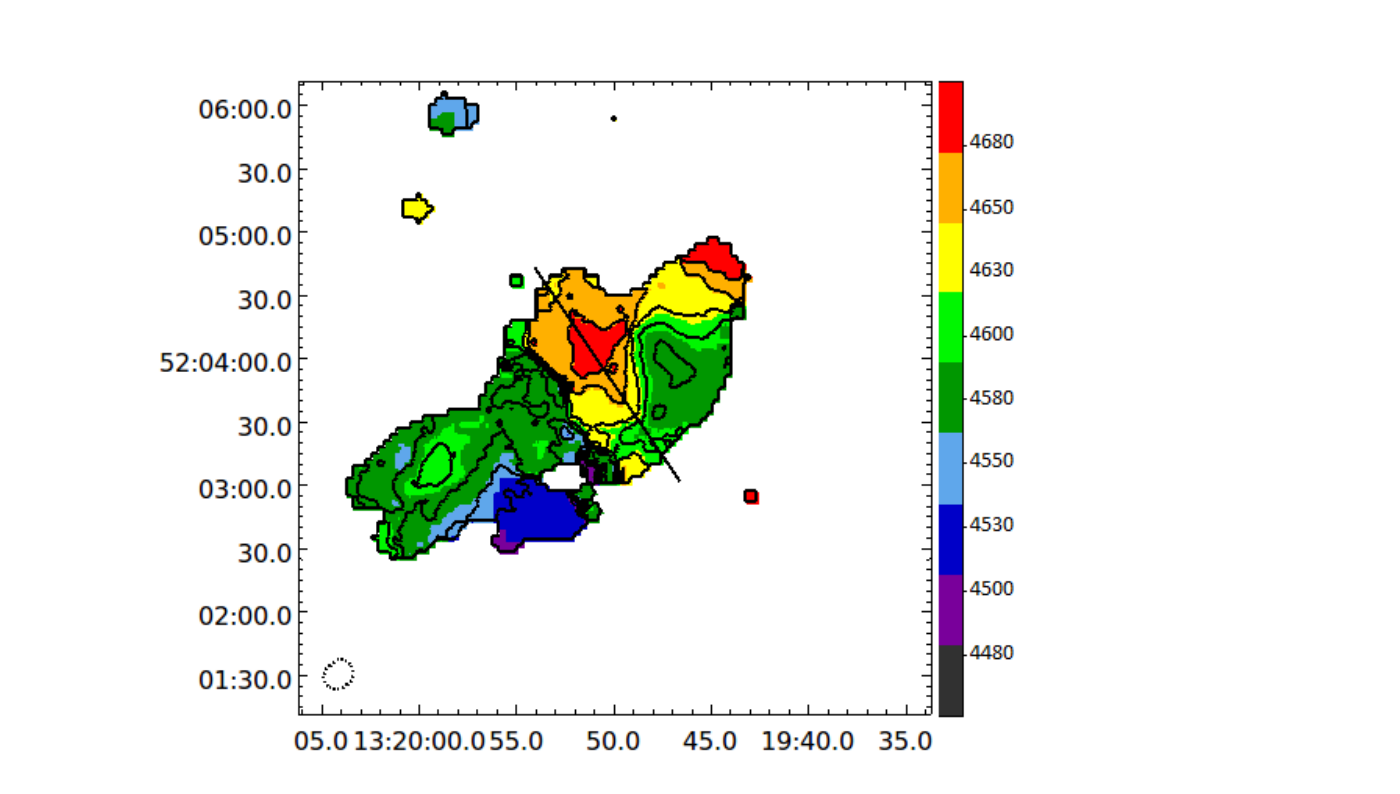}{0.57\textwidth}{(c)}
    \leftfig{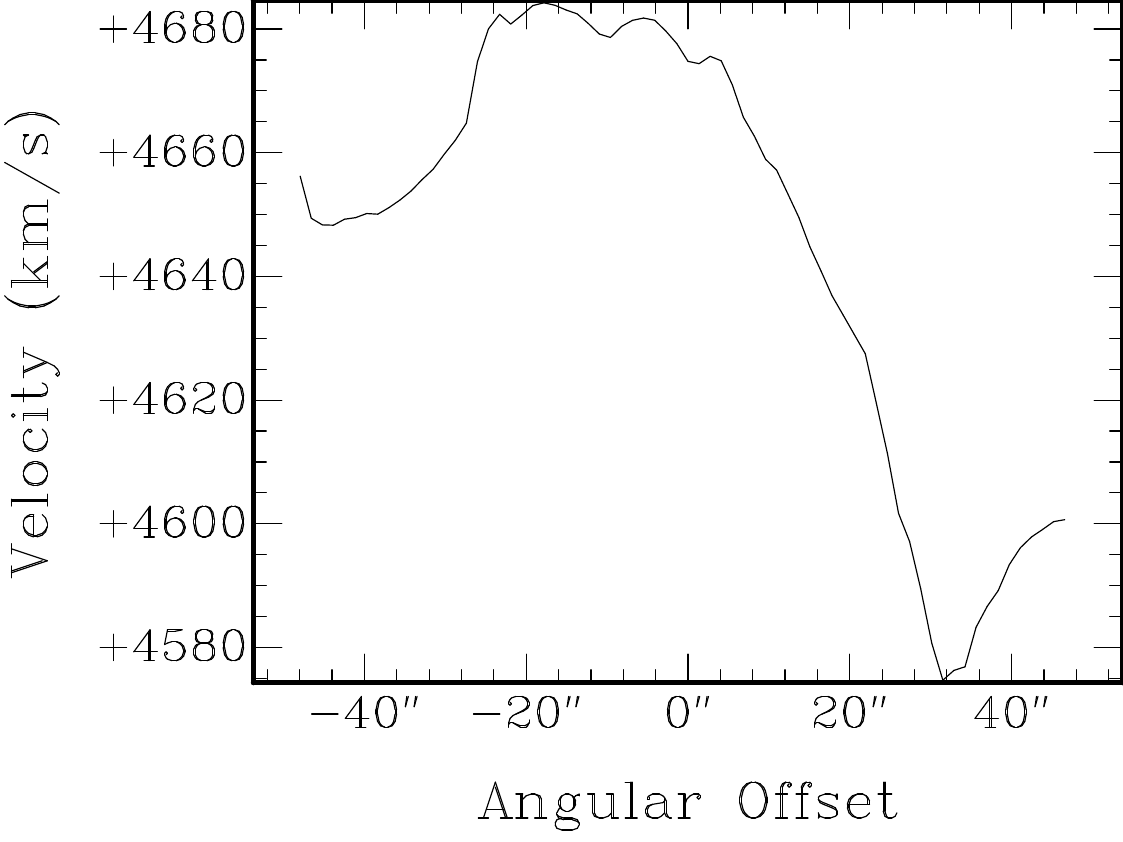}{0.37\textwidth}{(d)}}
  \gridline{\leftfig{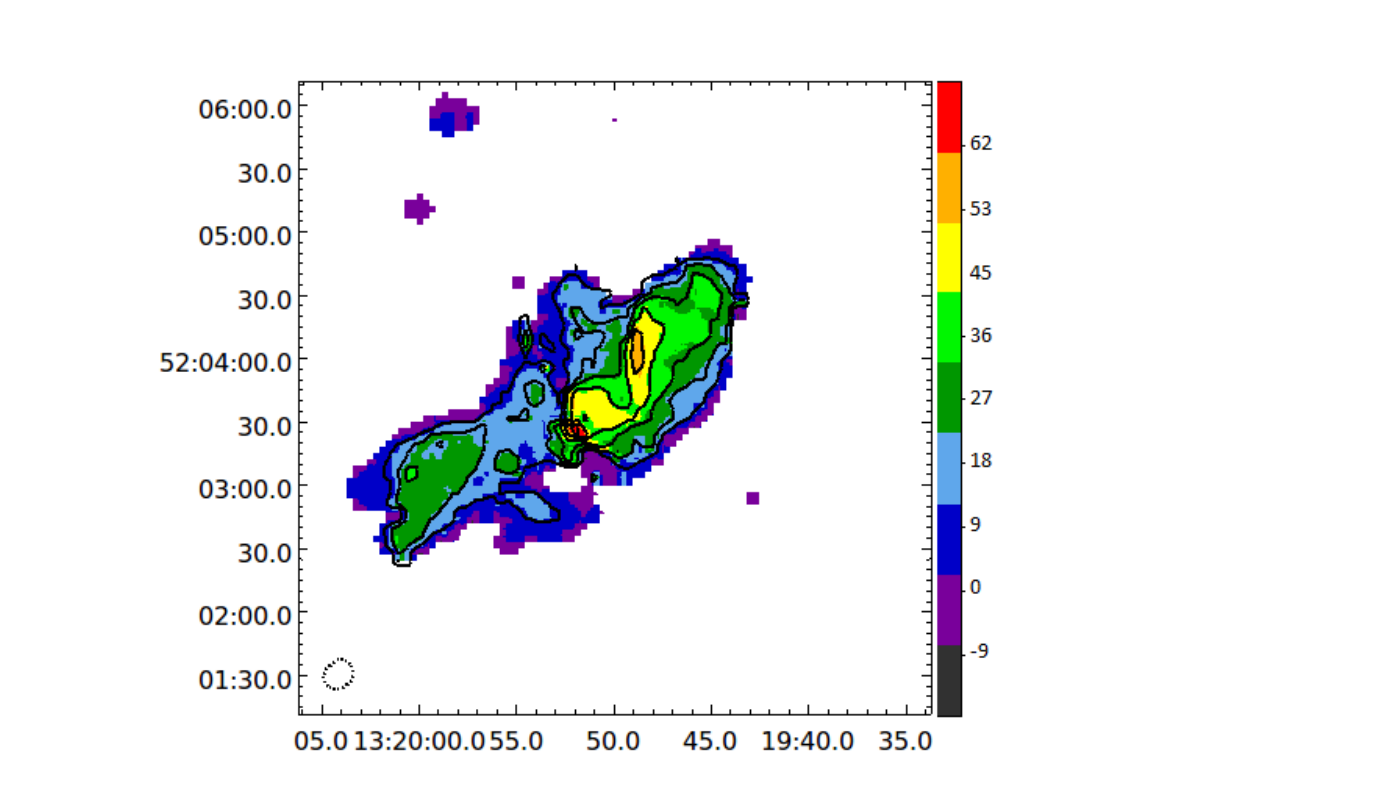}{0.57\textwidth}{(e)}
    \leftfig{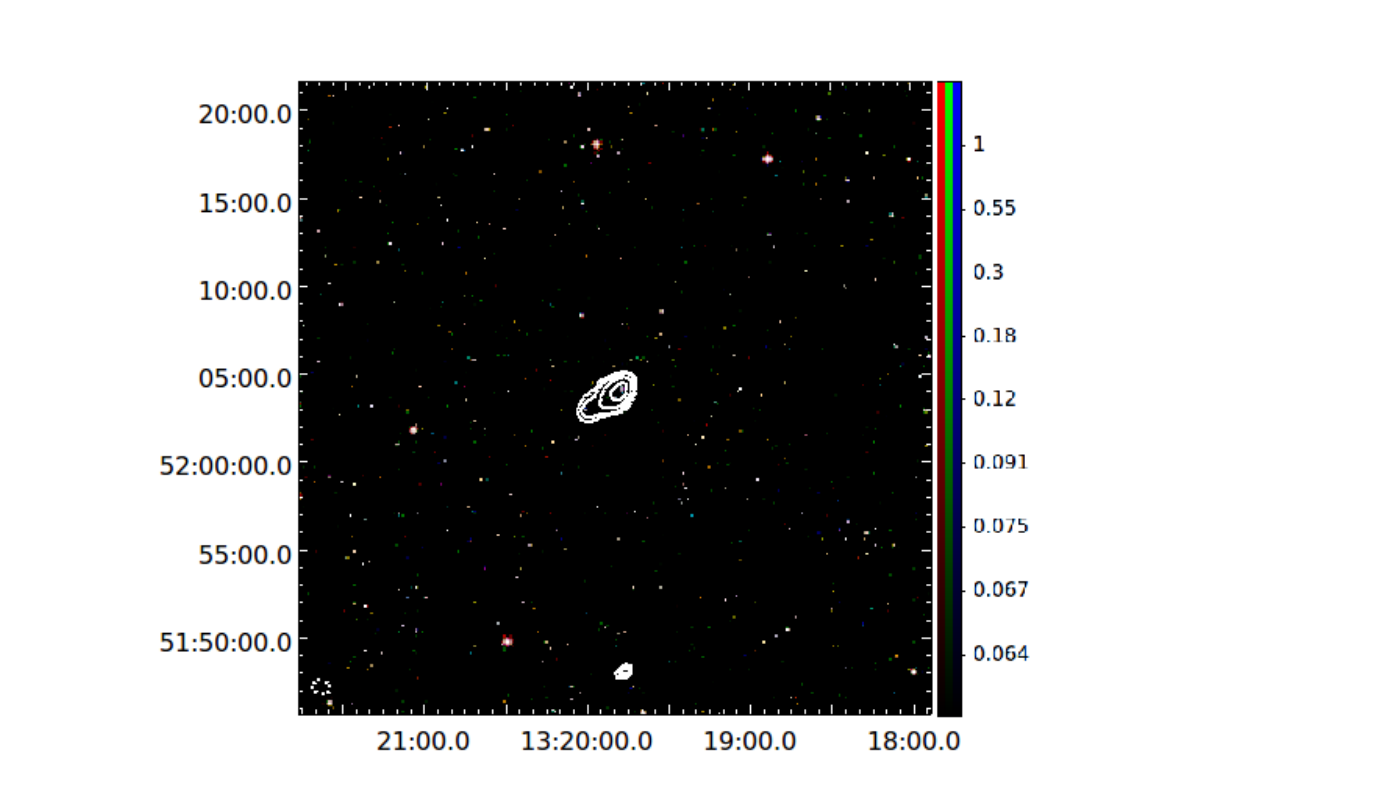}{0.57\textwidth}{(f)}}
  
\caption{\footnotesize SDSS1319+5203 (SBS 1317+523B): (a) Moment 0 contours
  made with a $15\arcsec \times 12\arcsec$ beam overlaid on a SDSS DR9
  \textit{gri} image. Contours represent H \textsc{I} intensities
  equivalent to column densities of $\rm 2^{n}\times10^{20}\ cm^{-2}$
  for n = 0, 1, 2, 3, 4, 5. Image units are analog to digital units
  (ADU). (b) Moment 0 map with the same contours as in (a). Map units
  are $\rm Jy\ Beam^{-1}\ m\ s^{-1}$. (c) Moment 1 map with a thick
  line showing the major axis. Contours are $25\ \rm km\ s^{-1}$. Map
  units are $\rm km\ s^{-1}$. (d) Velocities along the major axis
  slice shown in (c). (e) Moment 2 map with $10\ \rm km\ s^{-1}$
  contours. Map units are $\rm km\ s^{-1}$. (f) Low-resolution Moment
  0 contours made with a $63\arcsec \times 50\arcsec$ beam overlaid on
  SDSS DR9 \textit{gri} image. Contours represent H \textsc{I}
  intensities equivalent to column densities of $\rm
  2^{n}\times10^{20}\ cm^{-2}$ for n = 0, 1, 2, 3. Image units are the
  same as in (a). The horizontal and vertical axes of each map are
  right ascention and declination (J2000). Beam sizes are shown in the
  lower left corner of each map. \label{fig:1319}}

\end{figure*}
\begin{figure*}[htb!]

  \gridline{\leftfig{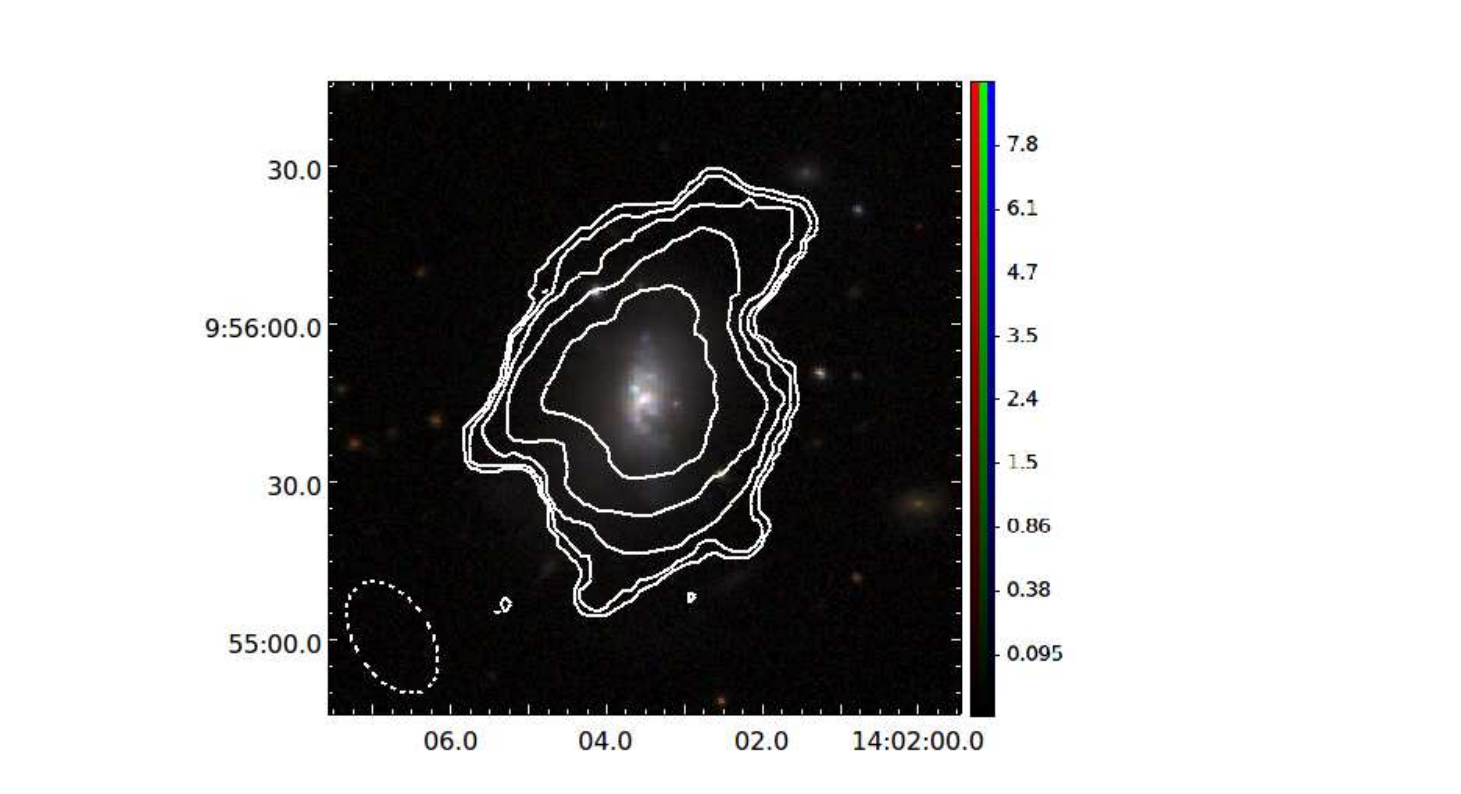}{0.57\textwidth}{(a)}
    \leftfig{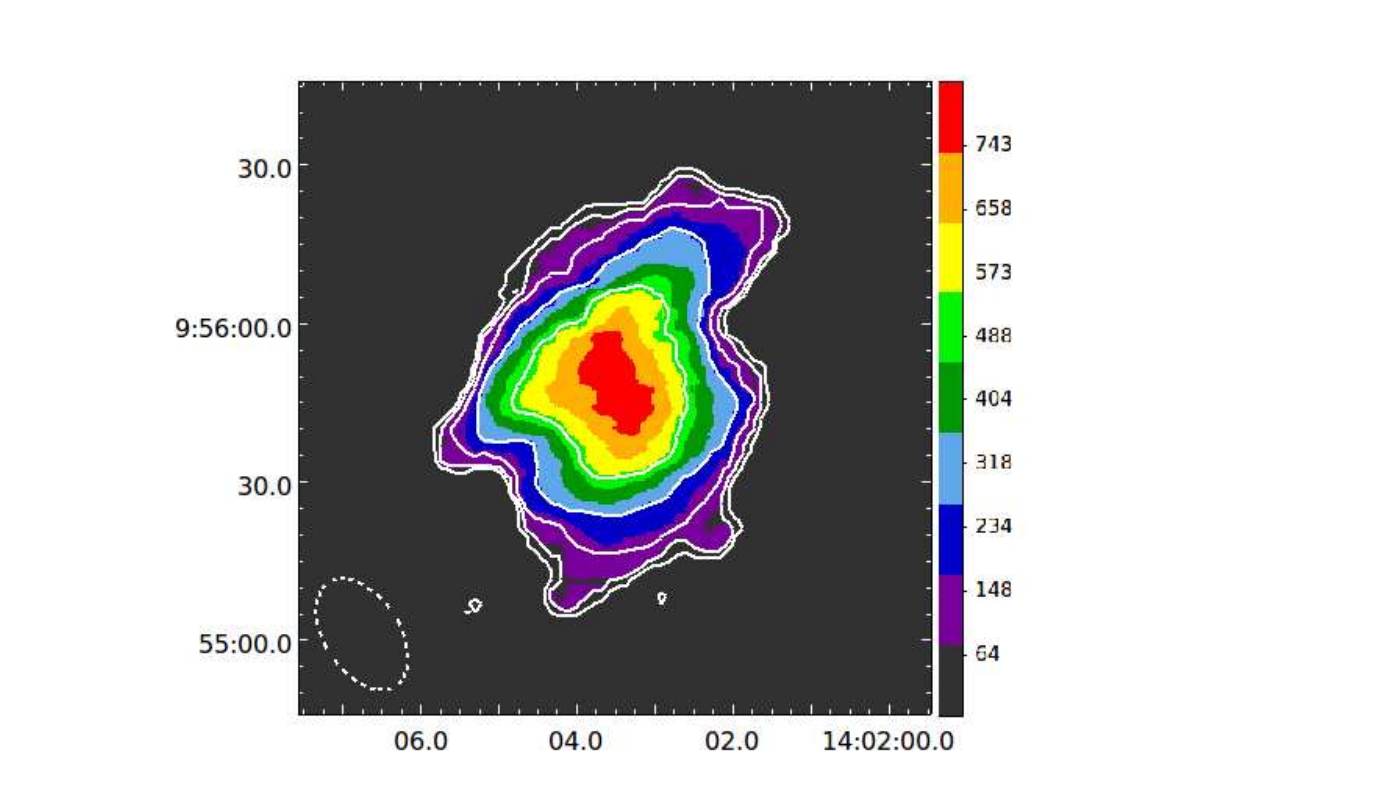}{0.57\textwidth}{(b)}}
  \gridline{\leftfig{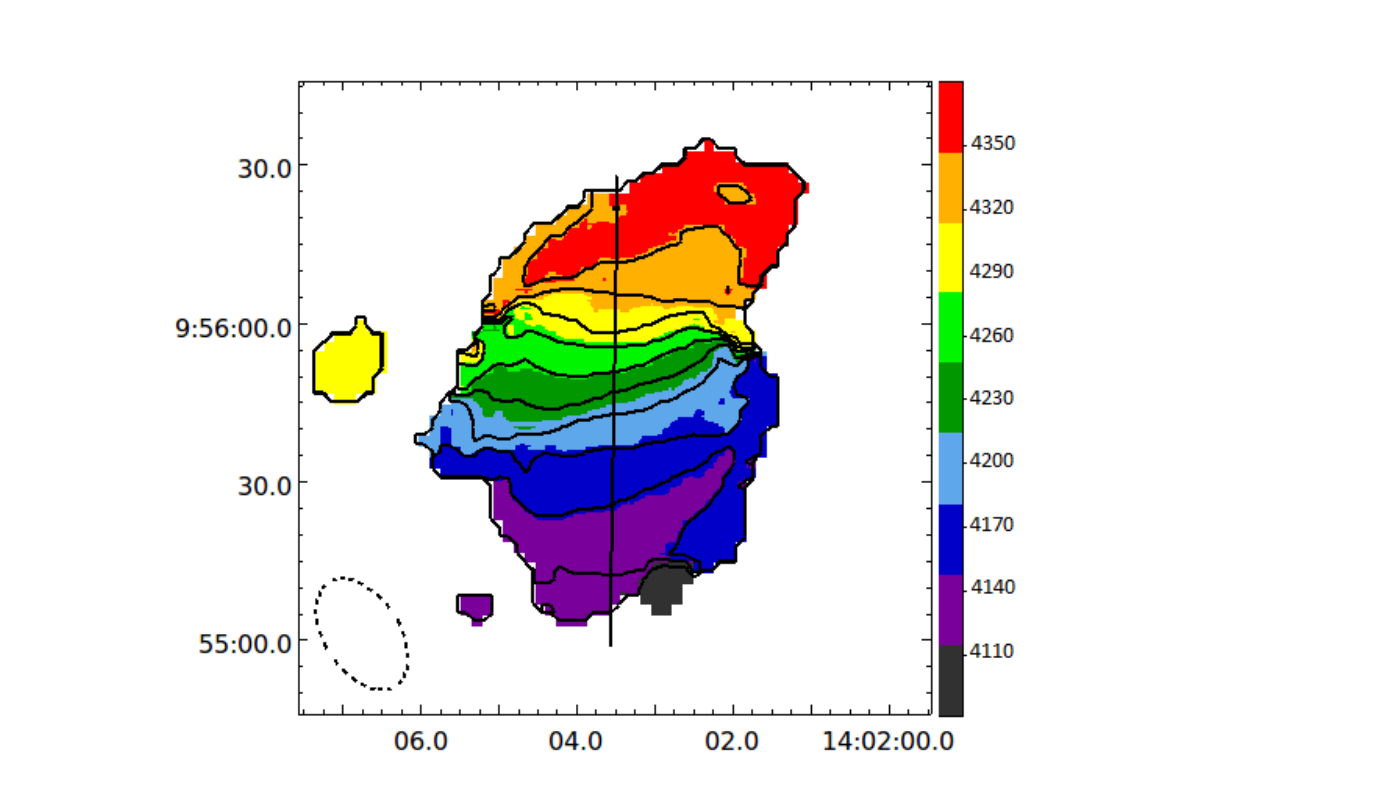}{0.57\textwidth}{(c)}
    \leftfig{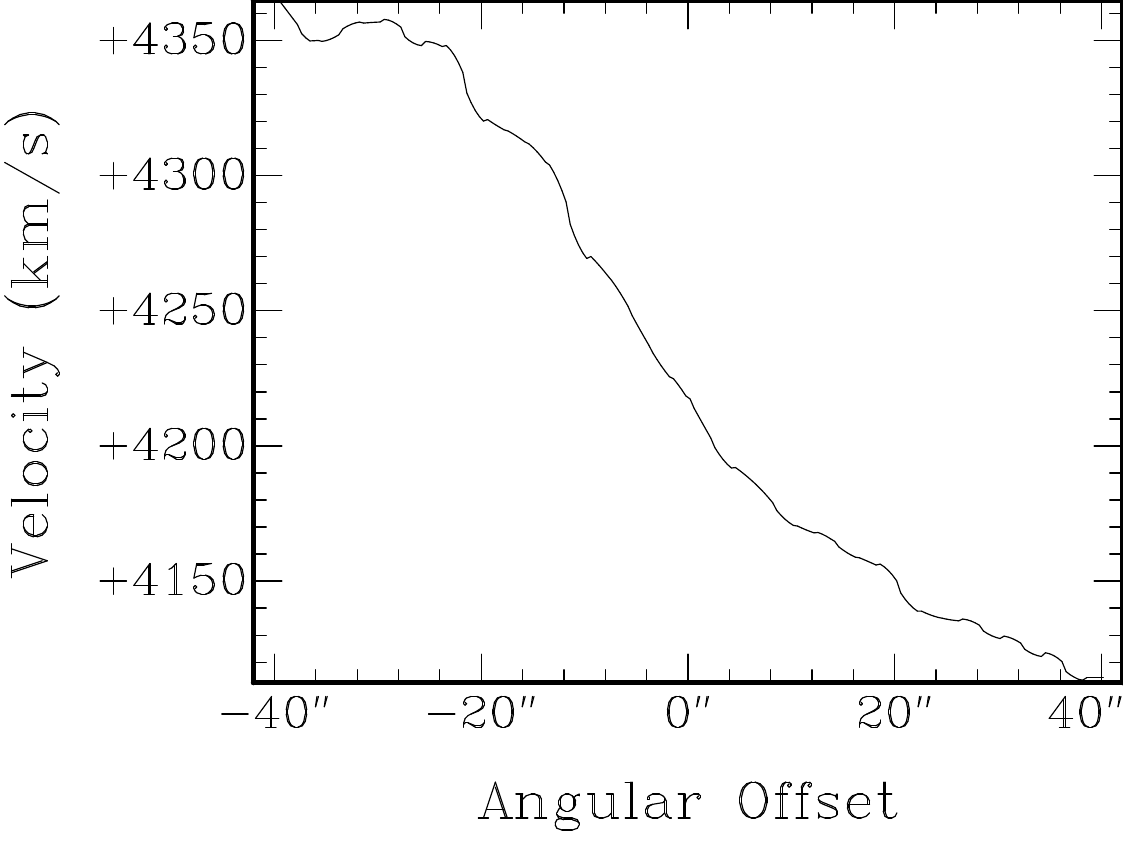}{0.37\textwidth}{(d)}}
  \gridline{\leftfig{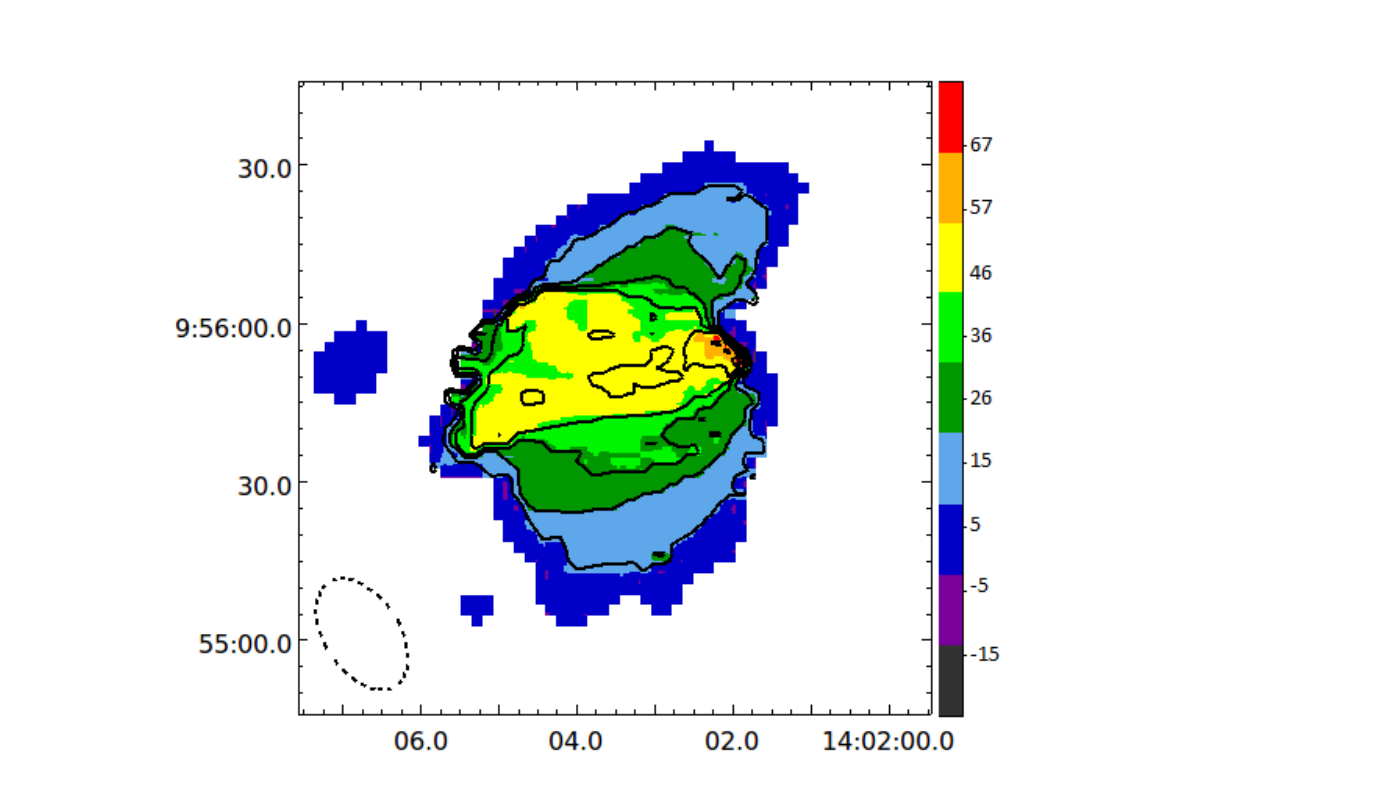}{0.57\textwidth}{(e)}
    \leftfig{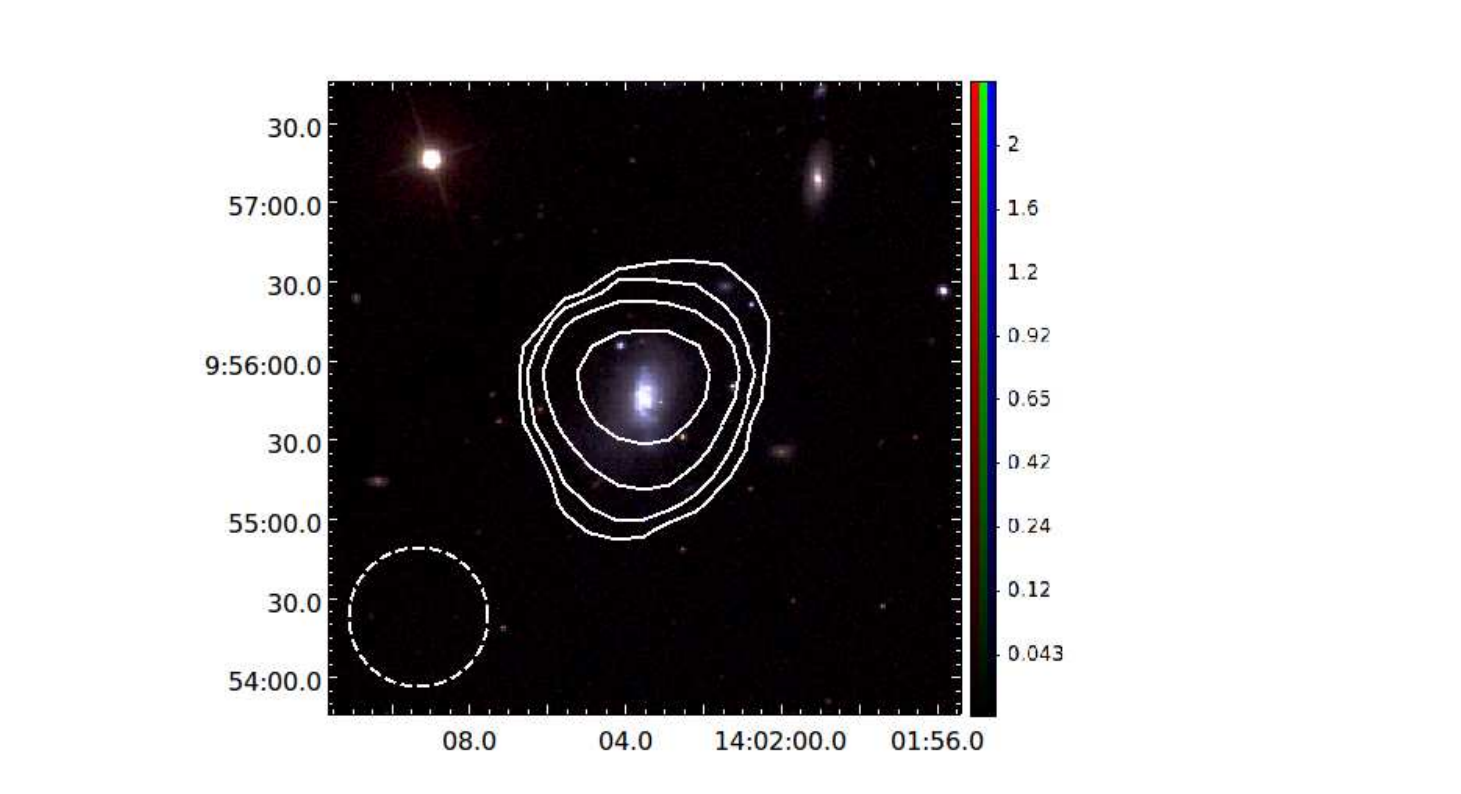}{0.57\textwidth}{(f)}}
  
\caption{\footnotesize SDSS1402+0955 (NGC 5414): (a) Moment 0 contours made
  with a $23\arcsec \times 14\arcsec$ beam overlaid on a SDSS DR9
  \textit{gri} image. Contours represent H \textsc{I} intensities
  equivalent to column densities of $\rm 2^{n}\times10^{20}\ cm^{-2}$
  for n = 0, 1, 2, 3, 4. Image units are analog to digital units
  (ADU). (b) Moment 0 map with the same contours as in (a). Map units
  are $\rm Jy\ Beam^{-1}\ m\ s^{-1}$. (c) Moment 1 map with a thick
  line showing the major axis. Contours are $25\ \rm km\ s^{-1}$. Map
  units are $\rm km\ s^{-1}$. (d) Velocities along the major axis
  slice shown in (c). (e) Moment 2 map with $10\ \rm km\ s^{-1}$
  contours. Map units are $\rm km\ s^{-1}$. (f) Low-resolution Moment
  0 contours made with a $53\arcsec \times 53\arcsec$ beam overlaid on
  SDSS DR9 \textit{gri} image. Contours represent H \textsc{I}
  intensities equivalent to column densities of $\rm
  2^{n}\times10^{20}\ cm^{-2}$ for n = 0, 1, 2, 3. Image units are the
  same as in (a). The horizontal and vertical axes of each map are
  right ascention and declination (J2000). Beam sizes are shown in the
  lower left corner of each map. \label{fig:1402}}

\end{figure*}
\begin{figure*}[htb!]

  \gridline{\leftfig{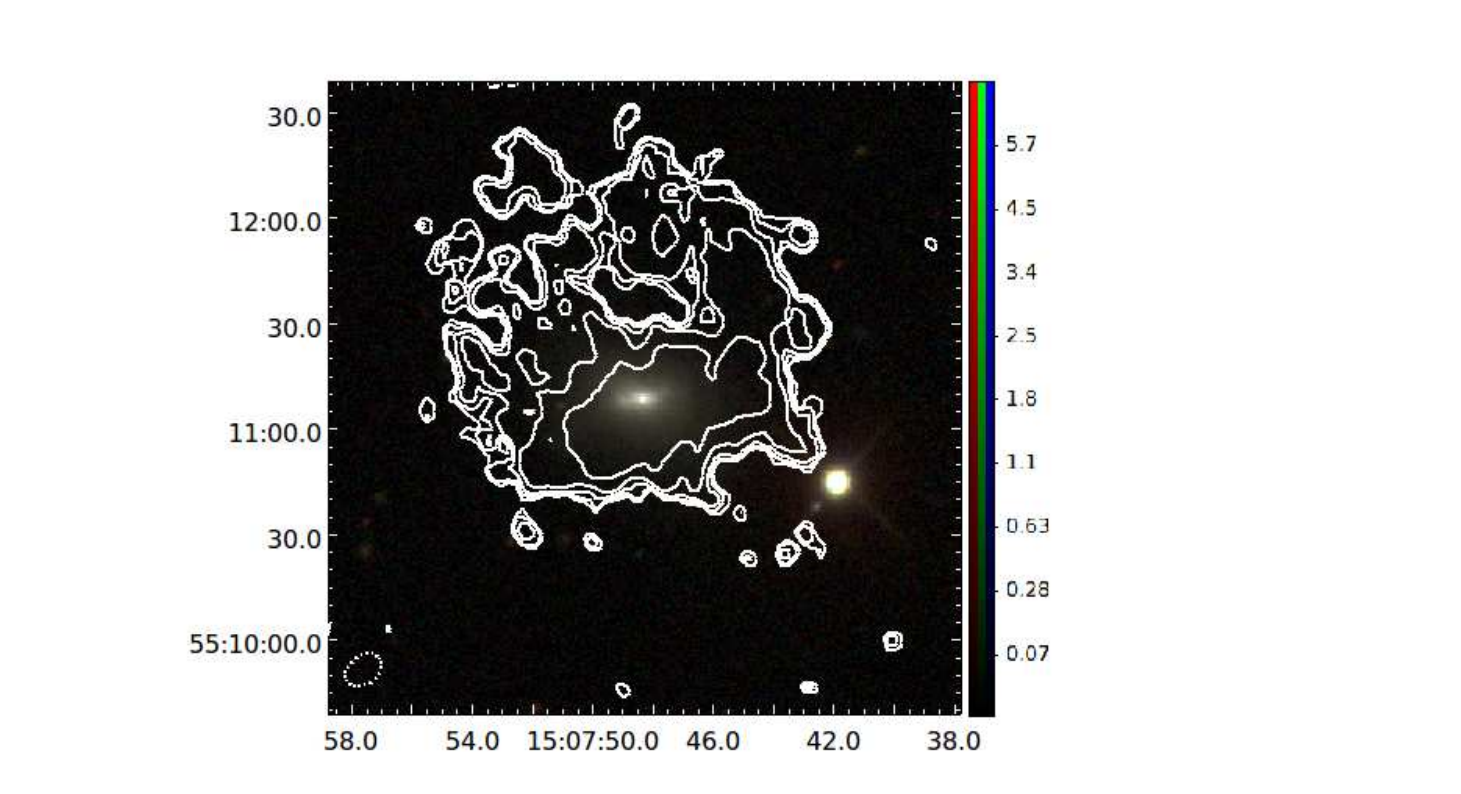}{0.57\textwidth}{(a)}
    \leftfig{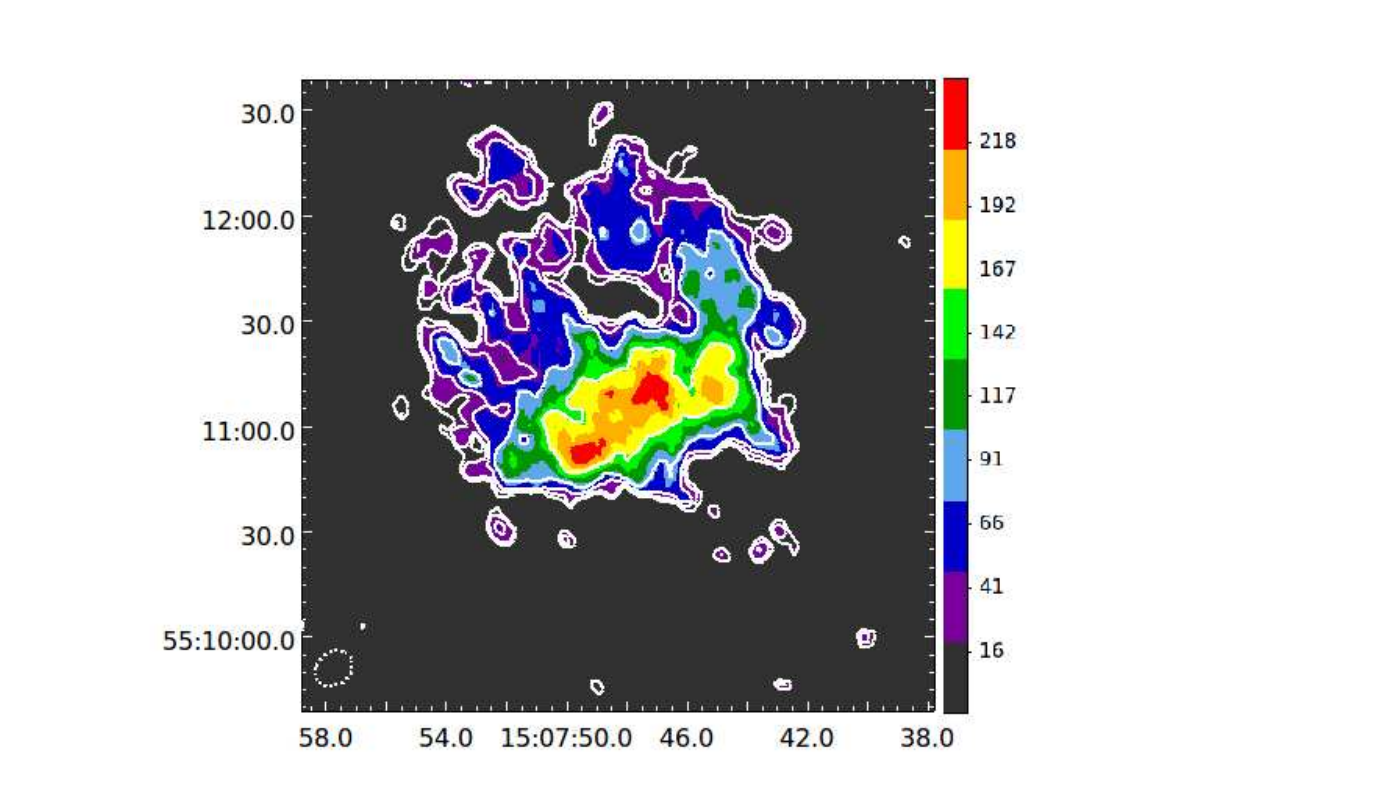}{0.57\textwidth}{(b)}}
  \gridline{\leftfig{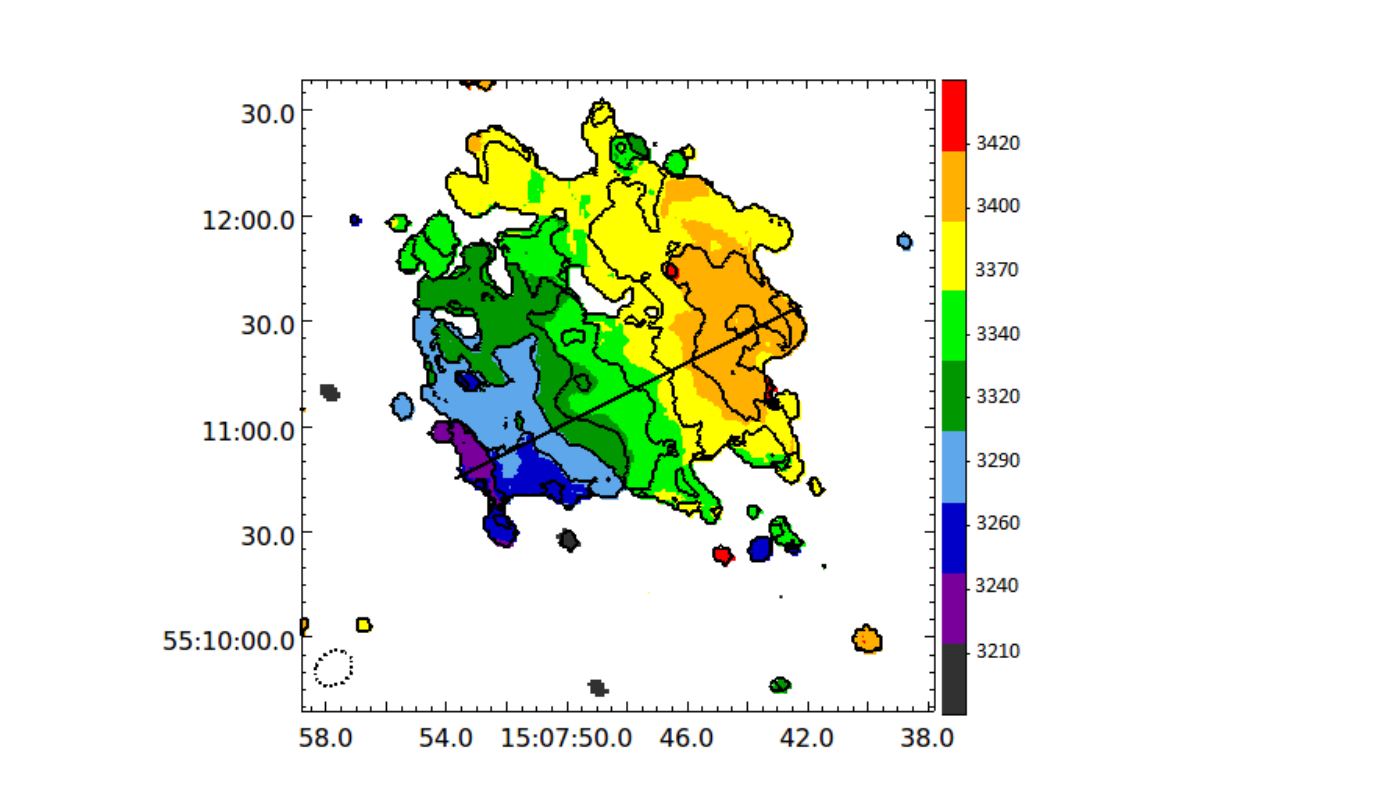}{0.57\textwidth}{(c)}
    \leftfig{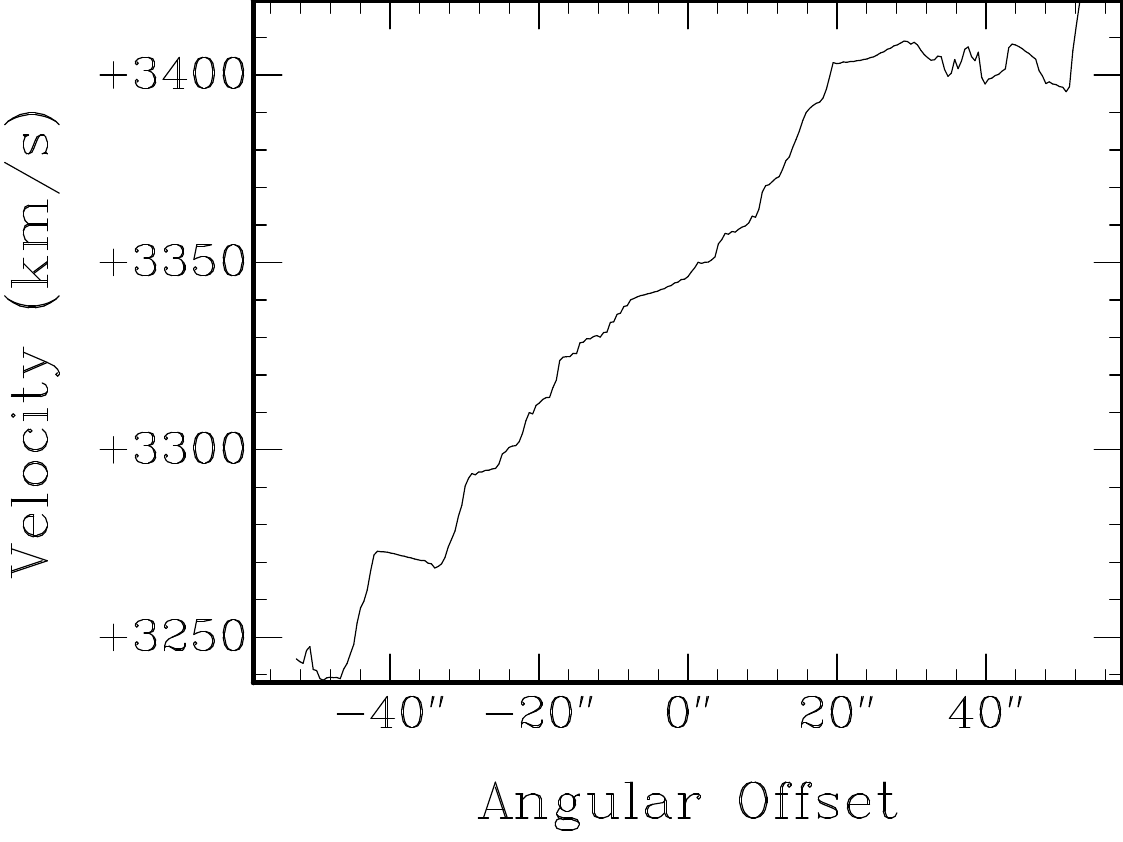}{0.37\textwidth}{(d)}}
  \gridline{\leftfig{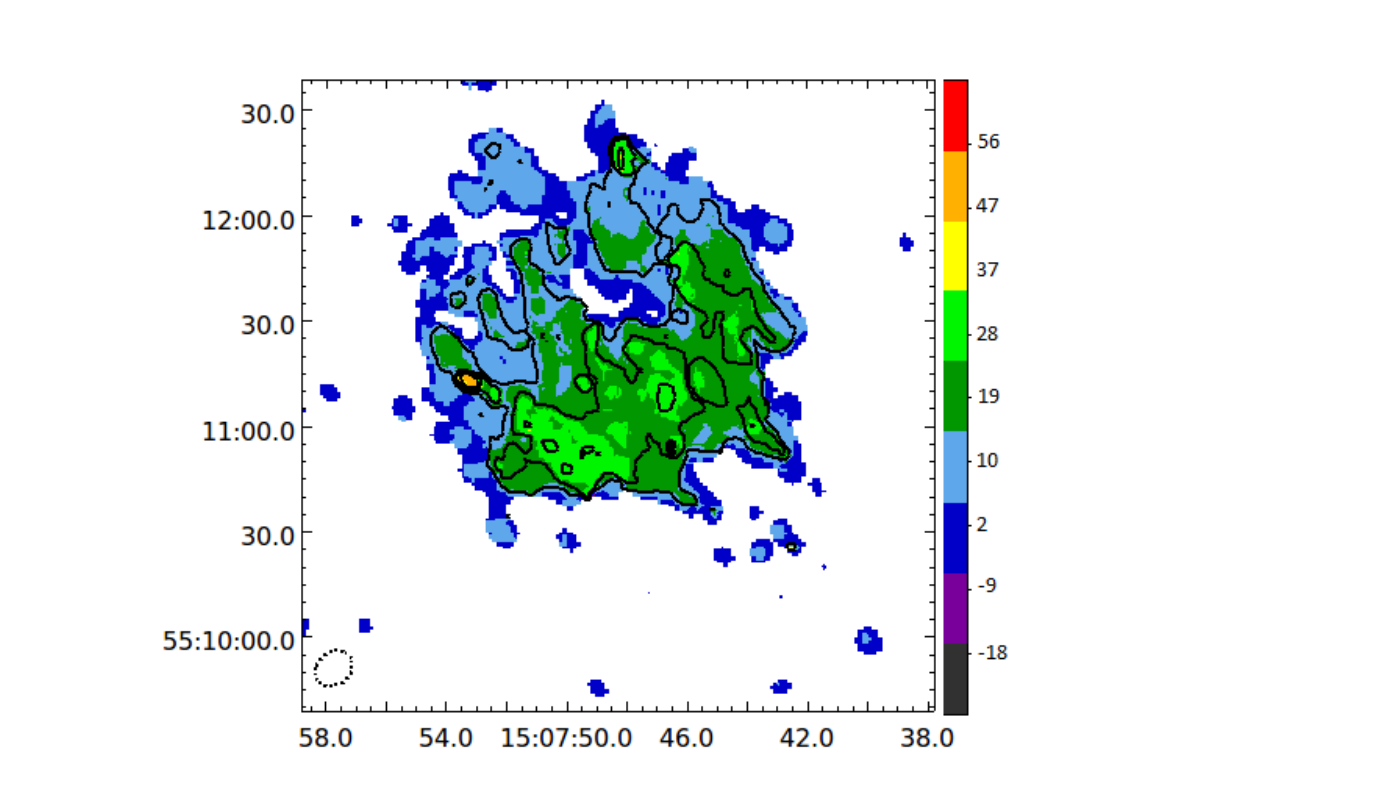}{0.57\textwidth}{(e)}
    \leftfig{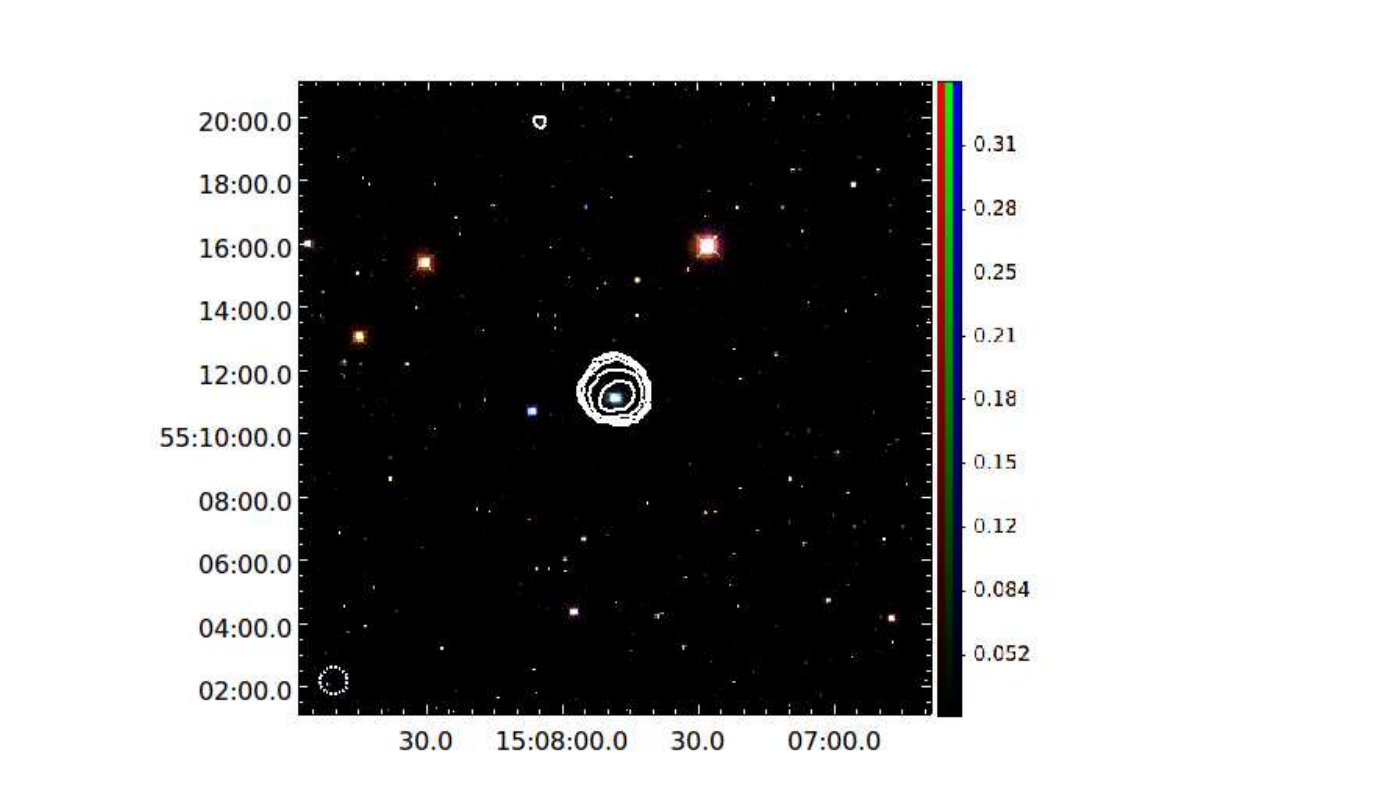}{0.57\textwidth}{(f)}}
  
\caption{\footnotesize SDSS1507+5511 (UGC 09737): (a) Moment 0
  contours made with a $11\arcsec \times 9\arcsec$ beam overlaid on a
  SDSS DR9 \textit{gri} image. Contours represent H \textsc{I}
  intensities equivalent to column densities of $\rm
  2^{n}\times10^{20}\ cm^{-2}$ for n = 0, 1, 2, 3, 4. Image units are
  analog to digital units (ADU). (b) Moment 0 map with the same
  contours as in (a). Map units are $\rm
  Jy\ Beam^{-1}\ m\ s^{-1}$. (c) Moment 1 map with a thick line
  showing the major axis. Contours are $25\ \rm km\ s^{-1}$. Map units
  are $\rm km\ s^{-1}$. (d) Velocities along the major axis slice
  shown in (c). (e) Moment 2 map with $10\ \rm km\ s^{-1}$
  contours. Map units are $\rm km\ s^{-1}$. (f) Low-resolution Moment
  0 contours made with a $52\arcsec \times 51\arcsec$ beam overlaid on
  SDSS DR9 \textit{gri} image. Contours represent H \textsc{I}
  intensities equivalent to column densities of $\rm
  2^{n}\times10^{20}\ cm^{-2}$ for n = -2, -1, 0, 1, 2. Image units
  are the same as in (a). The spatial scale of this figure was chosen
  to include a detected companion 9$\arcmin$ to the north. The
  horizontal and vertical axes of each map are right ascention and
  declination (J2000). Beam sizes are shown in the lower left corner
  of each map. \label{fig:1507}}

\end{figure*}

We used the high-resolution data cubes to make Moment 0 (total
intensity), Moment 1 (intensity-weighted velocity), and Moment 2
(velocity dispersion) maps, as well as low-resolution Moment 0 maps
that include detected companions, for each LCBG using the AIPS task
MOMNT. These moment maps are shown in Figures \ref{fig:0119} -
\ref{fig:1507}. We typically clipped the high-resolution moment maps
at the $2-3\sigma$ level, where we measured $\sigma$ from the RMS in
an emission-free channel. We chose this noise cut to maximize the
galaxy emission shown in the moment maps, while minimizing the noise
shown. We made an effort to include companions and preserve extended
structures with lower column densities when possible in order to more
fully show the morphology and environment of these galaxies.

\subsection{VLA observation and reduction of Mrk 325}

\begin{figure*}[htb!]

  \gridline{\leftfig{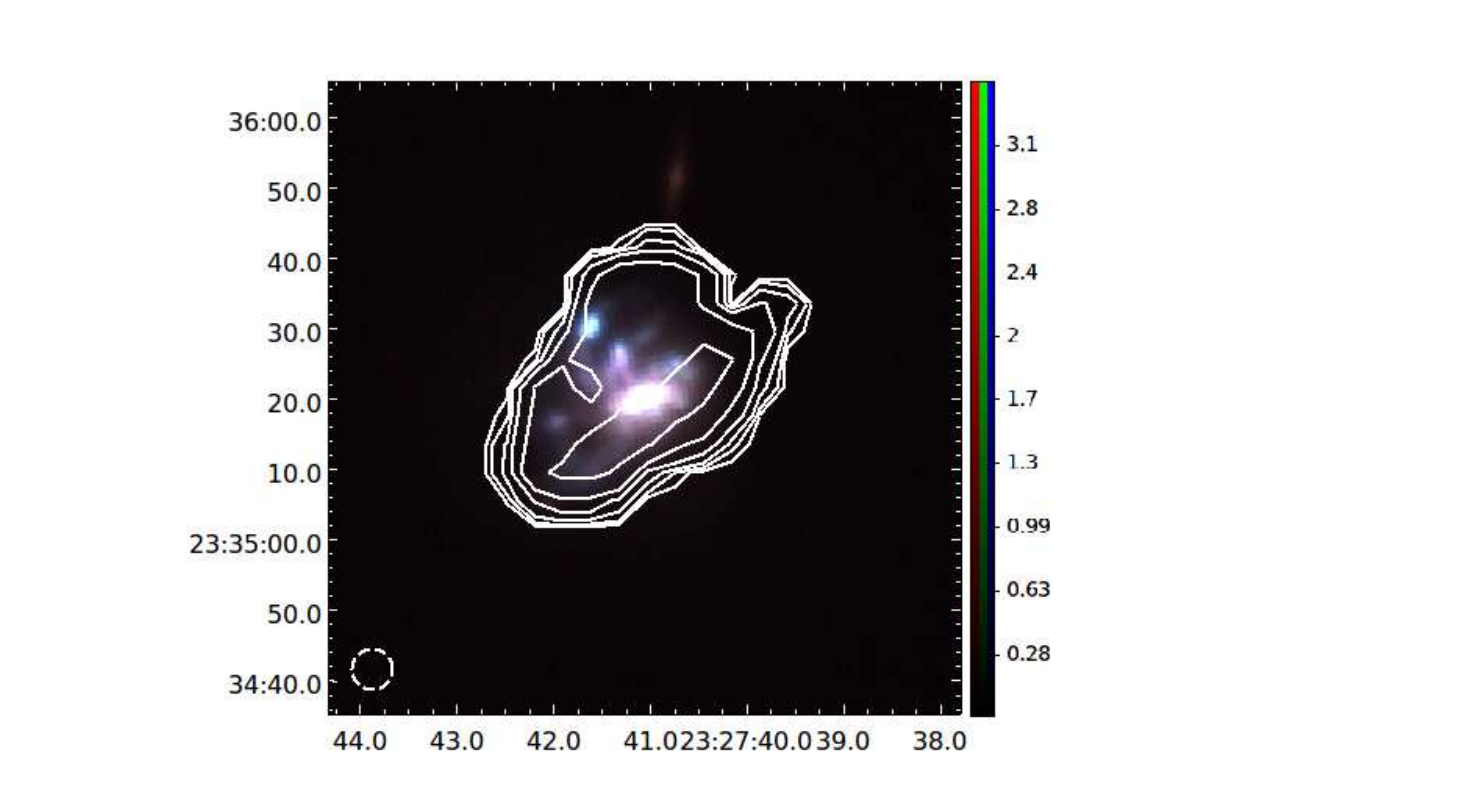}{0.57\textwidth}{(a)}
    \leftfig{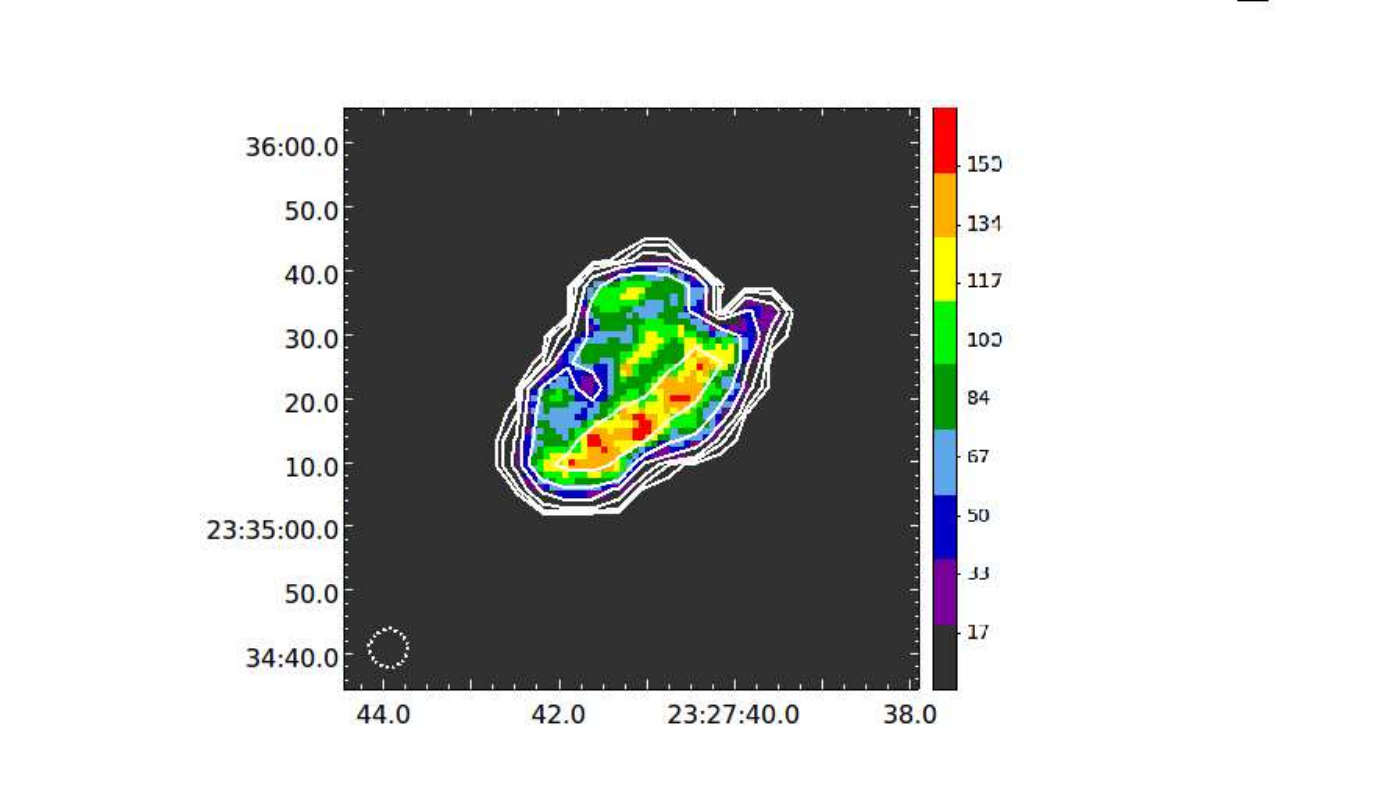}{0.57\textwidth}{(b)}}
  \gridline{\leftfig{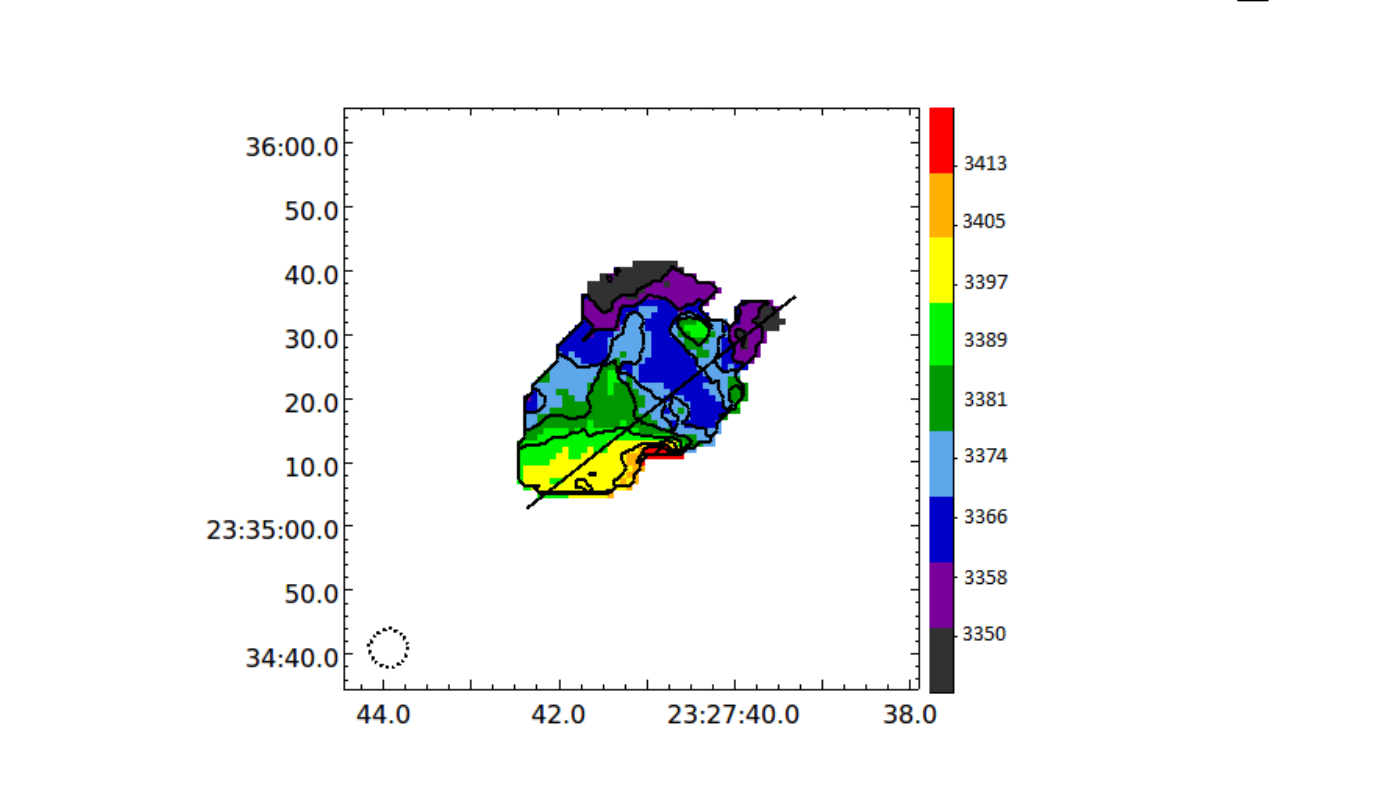}{0.57\textwidth}{(c)}
    \leftfig{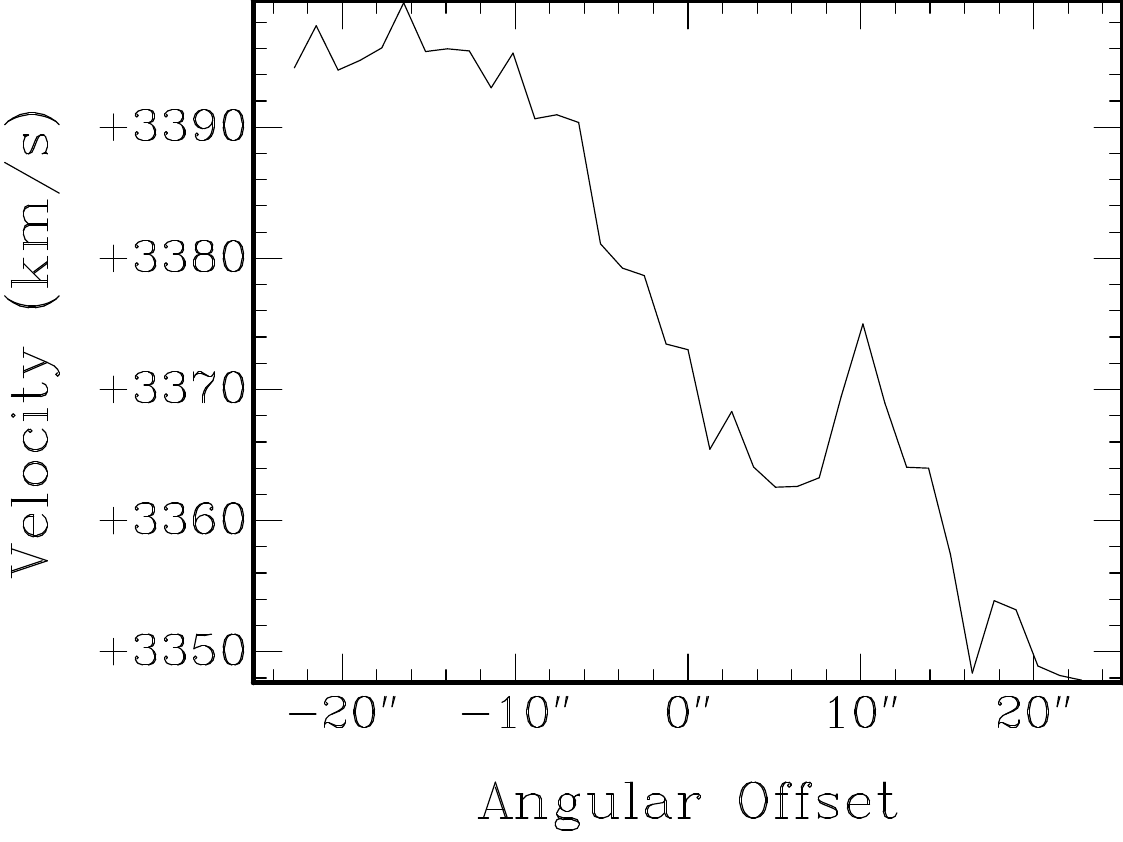}{0.37\textwidth}{(d)}}
  \gridline{\leftfig{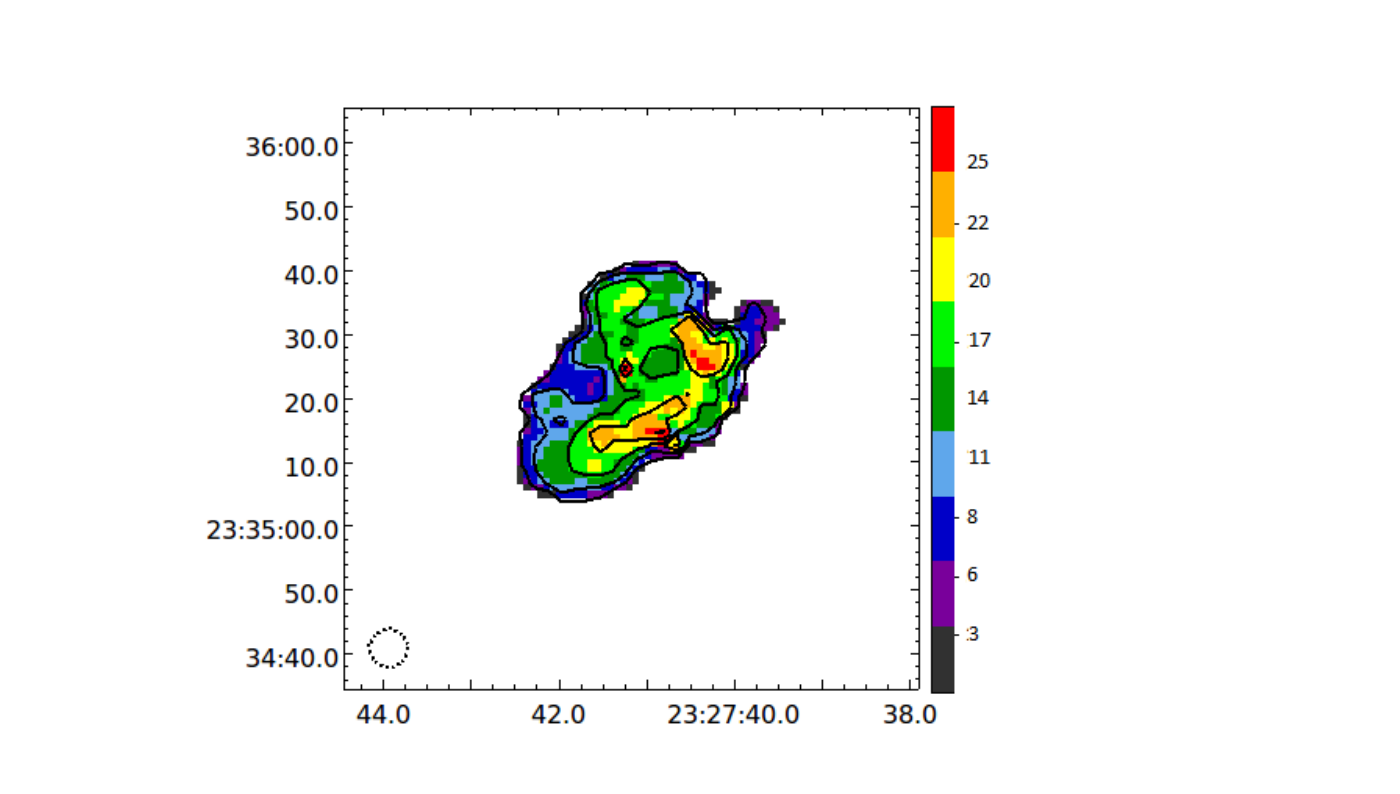}{0.57\textwidth}{(e)}
    \leftfig{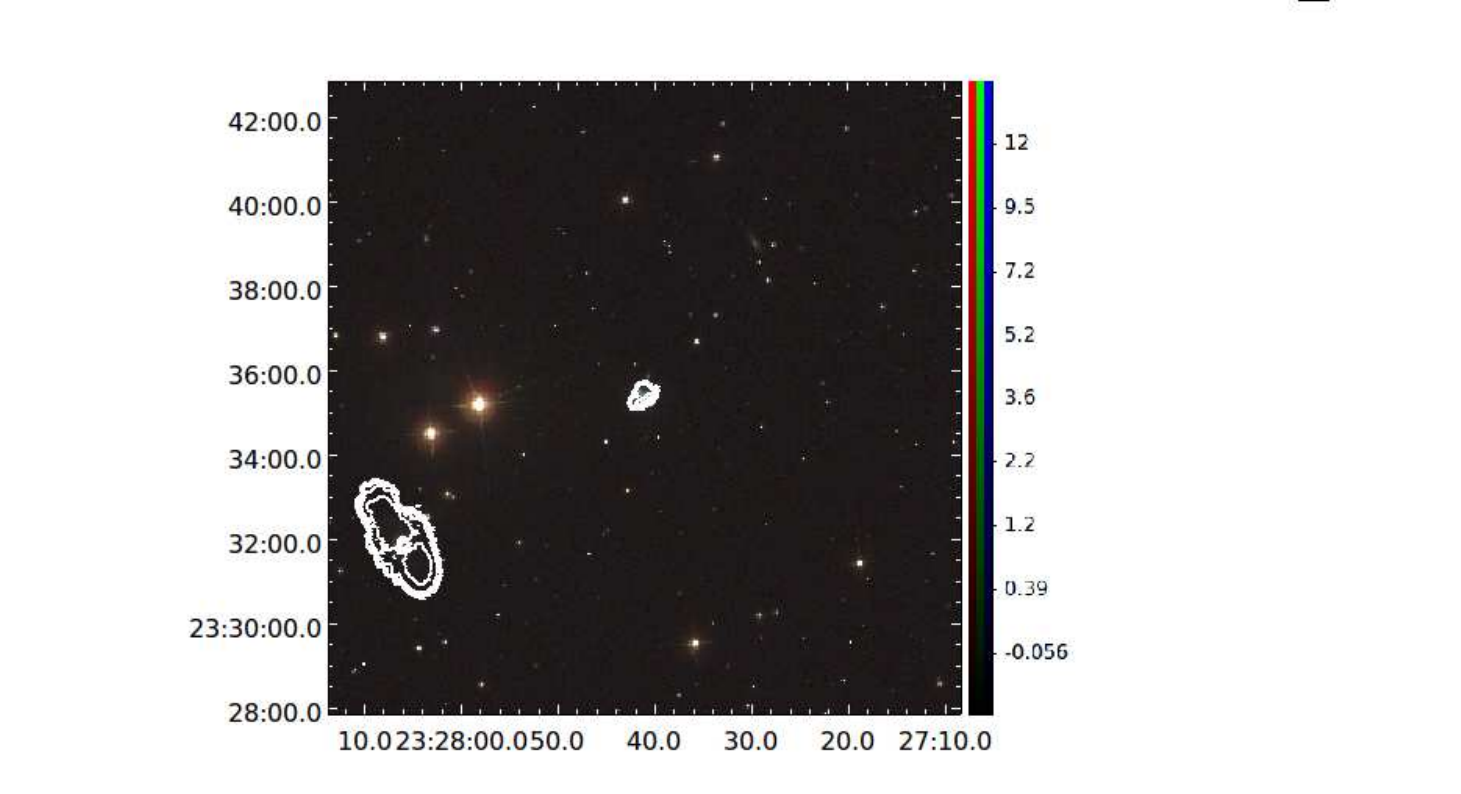}{0.57\textwidth}{(f)}}
  
\caption{\footnotesize Mrk 325: (a) Moment 0 contours made with a $6\arcsec
  \times 6\arcsec$ beam overlaid on a SDSS DR9 \textit{gri}
  image. Contours represent H \textsc{I} intensities equivalent to
  column densities of $\rm 2^{n}\times10^{20}\ cm^{-2}$ for n = 0, 1,
  2, 3, 4, 5. Image units are analog to digital units (ADU). (b)
  Moment 0 map with the same contours as in (a). Map units are $\rm
  Jy\ Beam^{-1}\ m\ s^{-1}$. (c) Moment 1 map with a thick line
  showing the major axis. Contours are $10\ \rm km\ s^{-1}$. Map units
  are $\rm km\ s^{-1}$. (d) Velocities along the major axis slice
  shown in (c). (e) Moment 2 map with $5\ \rm km\ s^{-1}$
  contours. Map units are $\rm km\ s^{-1}$. (f) Low-resolution Moment
  0 contours made with a $6\arcsec \times 6\arcsec$ beam overlaid on
  SDSS DR9 \textit{gri} image. Contours represent H \textsc{I}
  intensities equivalent to column densities of $\rm
  2^{n}\times10^{20}\ cm^{-2}$ for n = 0, 1, 2, 3, 4, 5. Image units
  are the same as in (a). For a $\theta \sim 1\arcmin$ moment 0 map of
  Mrk 325, see Figure 17 of Nordgren et al. (1997). The horizontal and
  vertical axes of each map are right ascention and declination
  (J2000). Beam sizes are shown in the lower left corner of each map. \label{fig:mrk325}}

\end{figure*}

We observed the final galaxy, Mrk~325, with the VLA in the B and C
configurations in December 2003 and November 2002 as part of projects
AP463 and AP438, respectively. We also used data from the VLA archive
taken as part of project AM361 in May 1992 and project AN62 in
November 1993. In all cases, we performed flux calibration via
observations of 3C48 or 3C286, and phase calibration through regular
observations of J2254+247 (B2251+244). We carried out the data
reduction for each configuration separately in the usual manner using
AIPS. Because the pointing center for the D configuration observations
was different than the B and C configuration data, we made the data
cubes by mosaicking the observations in
Miriad\footnote{http://www.atnf.csiro.au/computing/software/miriad/}
\citep{S95}. We made high-resolution and low-resolution moment maps
for Mrk 325 in the same way as described in Section 2.2, shown in
Figure \ref{fig:mrk325}.


\section{Results}

\subsection{H \textsc{I} content}

\begin{figure*}[htb!] 

\includegraphics[angle=0,width=1.0\textwidth]{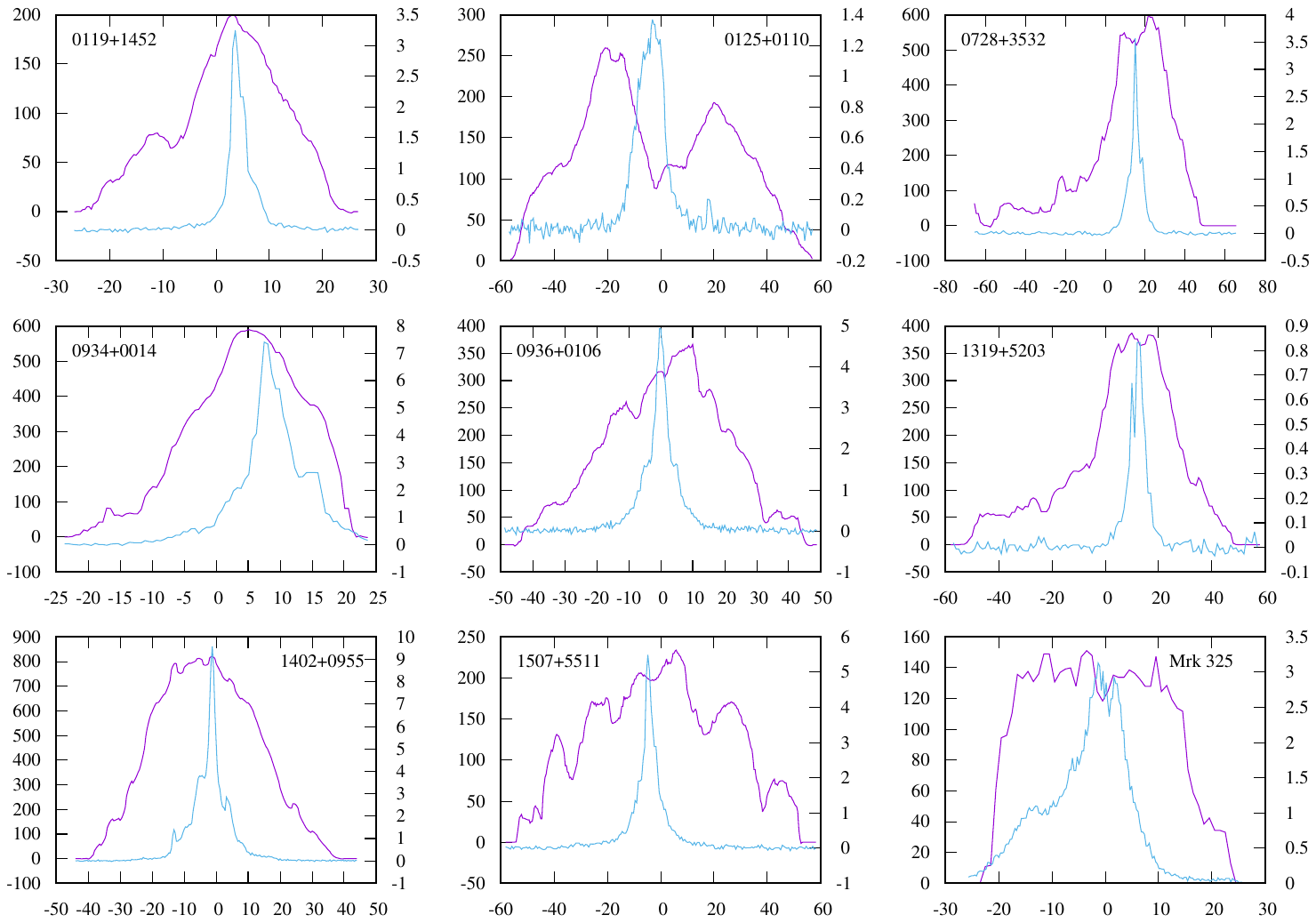}

\caption{H \textsc{I} (purple) and optical (blue) intensities measured
  along each galaxy's major axis (shown in panel c of Figure
  \ref{fig:0119} - \ref{fig:mrk325}). The H \textsc{I} intensities are
  measured from the Moment 0 maps in units of $\rm
  Jy\ beam^{-1}\ m\ s^{-1}$ (left vertical axes), and the optical
  intensities are measured from the SDSS g images in units of ADU
  (right vertical axes). The horizontal axes show angular distance
  along the major axis in arcseconds. The peak optical emission is
  within one beam major axis of the peak H \textsc{I} emission for all
  of the galaxies. \label{fig:radialprofile}}

\end{figure*}

The high-resolution Moment 0 maps in Figures 1-9 (panel a in each
figure) show heterogeneous H \textsc{I} morphologies. Despite the
variety of morphologies, all of the LCBGs in this sample have H
\textsc{I} emission whose maximum is within one beamwidth of with the
center of their optical emission, and extends beyond their stellar
radii (see Figure \ref{fig:radialprofile} for plots of the galaxies'
optical and H \textsc{I} emission along their major axes). Seven
(78$\%$) of the galaxies have companions that are detected in our H
\textsc{I} maps (see Table \ref{fig:companions}). In addition, seven
of the LCBGs (SDSS0119+1452, SDSS0728+3532, SDSS0934+0014,
SDSS1319+5203, SDSS1402+0955, SDSS1507+5511, and Mrk 325) have H
\textsc{I} intensity contours that appear asymmetric.

We measured H \textsc{I} profiles for each LCBG using the AIPS task
ISPEC. We used the low-resolution data cubes in order to avoid missing
short spacings. We chose spatial boundaries for ISPEC using the
$2-3\sigma$ extent of each LCBG's H \textsc{I} emission in
low-resolution Moment 0 maps. In the case of multiple galaxies in a
common H \textsc{I} envelope, we measured the H \textsc{I} profile of
the entire envelope because identifying boundaries for each galaxy
while excluding H \textsc{I} emission associated with other galaxies
or tidal features in the envelope introduced large uncertainties. We
note that this means that measurements of quantities such as $\rm
M_{HI}$ using the H \textsc{I} profiles of LCBGs in larger envelopes
(SDSS0728+3532 and SDSS1319+5203) also encompass the entire envelope,
including multiple galaxies.

We calculated an integrated flux for each galaxy by summing the flux
in each channel within the first crossing at 0 mJy on each side of the
peak and multiplying the sum by one channel width. We found
uncertainties on the integrated fluxes following the method of
\citet{Chandra04}. To do this, we measured the RMS of the emission
within an aperture that did not spatially coincide with a galaxy's
position. We made this measurement in a velocity channel of each
galaxy's low-resolution cube that did not contain any H \textsc{I}
emission from the target galaxy or any of its companions. We then
added this value in quadrature with 10$\%$ of the galaxy's peak flux
(a conservative estimate of the GMRT's flux calibration uncertainty)
so that the uncertainty in each galaxy's integrated flux is $\rm
\delta \int{S\ dv} = \sqrt{(S_{RMS})^{2} +
  (0.1\ S_{peak})^{2}}$. After finding their integrated H \textsc{I}
fluxes and uncertainties, we then calculated $\rm M_{HI}$ for each
galaxy (or group of galaxies, in the case of systems with multiple
galaxies sharing a common H \textsc{I} envelope) using the equation

\begin{equation}
\rm{\left(\frac{M_{HI}}{\rm M_{\sun}}\right) = 2.356 \times 10^{5} \left(\frac{\rm{D_{HI}}}{{Mpc}}\right)^{2} \frac{\int {\rm{S \ dv}}}{{Jy \  km \  s^{-1}}}}
\end{equation}

\noindent where $\rm \int {S \ dv}$ is the integrated H \textsc{I}
flux within the spectrum's crossing of 0 mJy and $\rm D_{HI}$ is the
galaxy's distance derived from dividing the recession velocity by the
Hubble constant. This equation assumes that a galaxy's H \textsc{I} is
optically thin, and that it is at low redshift, which are reasonable
assumptions for this sample. The H \textsc{I} profile properties for
each galaxy are listed in Table \ref{fig:spectrum}.

We compare the $\rm M_{HI}$ we measure for the LCBGs in our sample to
those measured from the integrated line profiles observed by
\citet{CG04} for the same LCBGs using single-dish observations in
Table \ref{fig:spectrum}. Many of the LCBGs that \citet{CG04} detected
had optical companions within the GBT beam. If these companions
contain H \textsc{I}, they will add emission to the observed H
\textsc{I} spectrum, and thus increase the measured $\rm M_{HI}$ above
what would be measured if the target LCBG was able to be observed by
itself. We flag the galaxies that \citet{CG04} identified as having
companions within the GBT beam in Table \ref{fig:spectrum}.

In contrast to previous single-dish observations, our observations can
spatially resolve the LCBGs from their companions. Thus, the $\rm
M_{HI}$ that we measure by integrating over the intensities measured
in each velocity channel within the spatial boundaries of each
galaxy's H \textsc{I} map are more likely to reflect the true $\rm
M_{HI}$ of the target galaxies than those measured from integrating
over the H \textsc{I} spectrum observed in an unresolved single
pointing. In addition, having unresolved companions or tidal features
in the beam can act to broaden a galaxy's observed linewidth, and thus
increase its inferred rotation velocity. Since LCBGs' possible
evolutionary scenarios depend on whether they are rotation-dominated,
dispersion-dominated, or show signatures of interactions, it is
important to determine whether LCBGs' linewidths can be interpreted as
the result of rotation. We discuss this further in Section 4.2.

\subsection{Velocity measurements}

As is shown in the high-resolution Moment 1 maps (panel c in Figures
\ref{fig:0119} - \ref{fig:mrk325}), all of the LCBGs in our sample
show a velocity gradient in their Moment 1 maps, which is evidence of
rotation. We measured systemic and rotation velocities ($\rm V_{rot}$)
for the LCBGs in our sample using the high-resolution Moment 1 maps
for each LCBG by measuring a slice of velocities along the galaxies'
major axes (see panel d of Figures \ref{fig:0119} - \ref{fig:mrk325}).

We determined the major axis of each galaxy using a visual inspection
of their Moment 1 maps to identify features of rotation. We then used
the program KPVSLICE from the
Karma\footnote{http://www.atnf.csiro.au/computing/software/karma/}
package of reduction tools to produce position-velocity diagrams from
the data cube along this major axis. This method produced measured,
rather than fit, rotation curves from which we measured $\rm V_{rot}$
at the half-light radius ($\rm R_{eff}$), the extent of ongoing star
formation ($\rm{R_{25}(B)}$, the radius at which the galaxy has $\rm
SBe(B) = 25\ mag \ arcsec^{-2}$) and the extent of neutral hydrogen
($\rm{R_{HI}}$, calculated as half of the galaxy's diameter across its
major axis between locations with a column density of $\rm 1\
M_{\odot}\ pc^{-2}$, or $\rm N_{HI} = 1.26 \times 10^{20}
\ cm^{-2}$). We note that while this method produced easily measurable
and reproducible values of $\rm V_{rot}$ that do not depend on the
velocities projected onto the Moment 1 map because they are measured
directly from the data cube, these velocities are only valid along the
H \textsc{I} major axis. We report the uncertainty on $\rm V_{rot}$ as
one channel width corrected for optical inclination. We also
calculated recession velocities as the velocity halfway between the
velocities at each $\rm{R_{HI}}$ edge along the major axis, and $\rm
M_{dyn}$ using $\rm V_{rot}$ at $\rm{R_{25}(B)}$ and $\rm{R_{HI}}$. We
report these values in Table \ref{fig:pvcuts}.


We attempted to fit rotation curves using the AIPS task GAL and the
tilted-ring fitting code $\rm
^{3D}Barolo$\footnote{http://editeodoro.github.io/Bbarolo/} \citep{DF15}
to each of these LCBGs to determine their H \textsc{I} centers, $\rm
V_{rot}$, recession velocities, and inclinations. While we were able
to fit rotation curves to the galaxies if we assumed a rotation curve
shape and held some parameters fixed, it was not possible to fit
well-constrained rotation curves for the galaxies that allowed the
centers, extents, position angles, and inclinations of the galaxies to
be free parameters and did not assume a rotation curve shape, even
when using the Moment 0 maps as weights.

The $\rm V_{rot}$ values derived from measuring along the major axis
of each galaxy are less dependent on models that have systematic
uncertainties than rotation curve fits, and are more easily
reproduced. Thus, we use the velocities along the major axis shown in
Table \ref{fig:pvcuts} when discussing $\rm V_{rot}$ in the remainder
of the paper.

We then calculated $\rm M_{dyn}$ for each LCBG using the equation
\begin{equation}
\rm M_{dyn}=\frac{V_{rot}^2 \times R}{G}
\end{equation}
where R is the radius at which $\rm V_{rot}$ is measured (and within
which $\rm M_{dyn}$ applies). $\rm V_{rot }$ is corrected for
inclination by
\begin{equation}
\rm V_{rot}=\frac{V_{measured}}{\sin{i}}
\end{equation}
where $i$ is the optical inclination calculated using $i = \rm
cos^{-1}(expAB_{i})$, where $\rm expAB_{i}$ is the ratio of each
galaxy's minor and major axis lengths in the SDSS \textit{i} band
using an exponential galaxy profile as reported in each galaxy's SDSS
DR9 photometry table.

We note that the $\rm V_{rot}$, and thus $\rm M_{dyn}$, that we
calculate depend on the galaxies' inclinations. Since we do not have
the spatial or velocity resolution to reliably fit rotation curves and
inclination models to the LCBGs in our sample, we have not been able
to accurately measure the inclinations of the galaxies' gas. We are
thus restricted to the same assumption that \citet{CG04} made: the gas
in these galaxies is inclined at the same angle with respect to our
line of sight as their optical components. Since the galaxies appear
to have disklike rotation, we do not have evidence that this is an
unreasonable assumption. \citet{dB08} compared the optical
inclinations of nearby galaxies for which they measured
high-resolution H \textsc{I} maps to inclinations measured by fitting
ellipses to the galaxies' H \textsc{I} contours and inclinations
derived from rotation curve fitting. For these galaxies, the average
difference between the optical inclinations and the inclinations from
the H \textsc{I} disk contours is $-0.6^{\circ}$ $\pm$
$14^{\circ}$. The average difference between the optical inclinations
and the inclinations derived from rotation curve fitting was
$-2.3^{\circ}$ $\pm$ $17^{\circ}$. They determined that for finding
the average inclination of a disk, using H \textsc{I} contours is just
as reliable as using the optical inclination. We use optical
inclinations to correct $\rm V_{rot}$ in this paper, and acknowledge
that this introduces a systematic uncertainty into the $\rm V_{rot}$
values we report.


\subsection{Velocity dispersions}

We calculated the average intensity-weighted velocity dispersions,
$\sigma$, of the LCBGs in our sample by taking the average pixel
values of the Moment 2 maps at four locations: (1) within a circle
bordered by the half-light radius, $\rm R_{eff}(B)$, (2) within a
circle bordered by $\rm R_{25}$, (3) for the whole disk within $\rm
R_{HI}$, and (4) outside of region within $\rm R_{25}$. We chose the
$\rm R_{25}$ radius limit because it has been previously found that
$\rm R_{25}$ generally signifies the outer limit of active star
formation in dwarf and spiral galaxies \citep{T09}. Thus, $\sigma$
within $\rm R_{25}$ is a measure of the gas properties that affect and
are affected by galaxies' star formation (primarily by processes such
as supernovae that trace recent star formation), while $\sigma$
outside of $\rm R_{25}$ probes the kinematics of the galaxies beyond
the region where they actively form stars \citep{T09}. These values of
$\sigma$ are tabulated in Table \ref{fig:sigmatable}. 

Since the LCBGs' $\rm R_{eff}$ are not well-resolved by the beams in
the high-resolution maps ($\rm R_{eff}/R_{beam} = 1.0$ on average,
with a range of 0.45 to 1.9), beam smearing is likely to affect the
velocity dispersions within $\rm R_{eff}$. To correct for the effects
of beam smearing in the centers of the LCBGs that we observed, we
applied a correction to $\rm \sigma_{R_{eff}}$ following \citet{DF15}
and \citet{Stott16}. To find the intrinsic velocity dispersion of the
H \textsc{I}, we used the equation $\rm \sigma_{HI} =
\sqrt{\sigma_{obs}^{2} - \sigma_{inst}^{2}}$. In this equation, $\rm
\sigma_{obs}$ is the average velocity dispersion within $\rm R_{eff}$
measured from the Moment 2 map and corrected for beam smearing. This
beam smearing correction is done by subtracting the velocity gradient
within $\rm \sigma_{R_{eff}}$ ($\rm \sigma_{obs} = \sigma_{meas} -
\Delta V / \Delta R$ where $\Delta R$ is the number of pixels along
$\rm R_{eff}$ and $\Delta V$ is half of the velocity gradient between
opposite points at $\rm R_{eff}$ along the major axis, uncorrected for
inclination). $\rm \sigma_{inst}$ is the estimated contribution to the
observed velocity dispersion from instrumental effects due to the
finite velocity resolution of the telescopes ($\rm \sigma_{inst} =
W_{chan}/\sqrt{2 ln 2}$). This correction decreased $\rm
\sigma_{R_{eff}}$ by an average of $27\%$ with respect to the
uncorrected average velocity dispersion within $\rm R_{eff}$, with a
standard deviation of $10\%$. For SDSS 0125+0110, we only report a
$\rm 3\sigma$ upper limit for the velocity dispersion within $\rm
R_{eff}$ because subtracting the correction from the measured average
velocity dispersion within $\rm R_{eff}$ resulted in a negative
number. We do not expect beam smearing to have a significant
contribution to the velocity dispersion at larger radii ($\rm R_{25}$
is on average 4.4 times larger, and $\rm R_{HI}$ is on average 6.8
times larger, than the beam radius along the major axis for the LCBGs
in our sample), so we only apply the correction to $\rm
\sigma_{R_{eff}}$.

We find that the areas of highest $\sigma$ tend to coincide with the
optical centers of the LCBGs in our sample, similar to what
\citet{T09} measured for spiral galaxies. This is true not only for
the relatively isolated LCBGs, but also for LCBGs with companions
(even those with companions in a common H \textsc{I} envelope, with
obvious evidence of gas interactions and disturbed morphology).

We also calculated the ratio of each LCBG's $\rm V_{rot}$ (corrected
for inclination) at a given radius to its average $\sigma$ inside of
that radius, $\rm V_{rot}/\sigma$, to determine the relative
contributions of ordered rotation and disordered motion of each
galaxy's H \textsc{I} emission. A galaxy's $\rm V_{rot}/\sigma$ values
are indicative of whether it has bulge-like or disk-like behavior,
with values of $\rm V_{rot}/\sigma < 1$ signifying that random motions
of the individual gas clouds dominate over rotation. Such values of
$\rm V_{rot}/\sigma$ are typically present in a galaxy's bulge, if it
has one, while values of $\rm V_{rot}/\sigma \sim 1$ can be found in
``pseudobulges'', which are built up by internal processes and
maintain some rotation \citep[for a review, see][]{KK04}. We also
wanted to investigate possible gravitational instabilities in the
disks as a potential trigger for star formation, which can be traced
by comparing $\rm V_{rot}/\sigma$ to the amount of gas available in
the disk. In Table \ref{fig:sigmatable}, we show values of $\rm
V_{rot}/\sigma$ within several radii. We find that $\rm
V_{rot}/\sigma$ increases at larger radii, with the highest values
occuring when $\rm \sigma$ is measured outside of $\rm R_{25}$, and
the lowest values occurring within $\rm R_{eff}$. We discuss the
implications of LCBGs' $\rm V_{rot}/\sigma$ values in Section 4.3.

\section{Discussion}

\subsection{Companions, mergers, and interactions}

It has been hypothesized that LCBGs' bright, blue, strongly
star-forming appearances are due to star formation triggered by major
and minor mergers \citep{AO01, O01}. These authors point out that
LCBGs tend to have asymmetrical stellar distributions and non-uniform
rotation curves, which are suggestive of mergers and interactions. In
contrast, \citet{W04} find that the majority of their sample of local
LCBGs have symmetric morphologies. It is known that star formation can
be triggered by mergers and interactions, so it would not be
surprising if LCBGs were merger-driven. However, \citet{CGletter}
found using optical data that only 20$\%$ of the galaxies in their
sample of local (D $<$ 76 Mpc) LCBGs are in merging systems. In our
sample, two of the nine galaxies have H \textsc{I} gas that overlaps
with the gas of another galaxy, which is consistent with the merger
rate of the \citet{CGletter} sample. The relatively low percentage of
LCBGs in merging systems suggests that the LCBGs in our sample do not
seem to require ongoing mergers to trigger their star formation.

Even though the LCBGs in our sample and the sample observed by
\citet{CGletter} are not preferentially undergoing current major
mergers, LCBGs have been observed to be commonly found in denser
environments where close encounters with other galaxies that disturb
their gas are more likely. \citet{CGletter} found that in their
sample, 40$\%$ of LCBGs are found in clusters. \citet{C11} found that
from $\rm 0.5<z<1.0$, LCBGs are more likely to reside in denser
environments at lower redshifts than at higher redshifts, and
\citet{C14} found that LCBGs tend to reside in the outer regions of
clusters. Those authors hypothesize that in intermediate-redshift
clusters, LCBGs are gas-rich blue galaxies whose star formation is
triggered during their first infall into the cluster \citep{C11,
  C14}. The LCBGs in our sample tend to have other galaxies nearby. In
our sample, seven out of nine LCBGs have companions within one GBT
beamwidth at 1.4 GHz, which is 167 kpc across at the average distance
of the LCBGs in our sample, and three of the nine LCBGs have
companions within one Arecibo beamwidth at 1.4 GHz, which is 56 kpc
across at the average distance of the LCBGs in our sample. Six out of
those seven LCBGs with companions have companions that we detect in
our H \textsc{I} maps, and five of those seven have detected
companions within $\sim$10 times the LCBGs' H \textsc{I} radii ($\rm
R_{HI}$) and within 100 $\rm km\ s^{-1}$ of the LCBGs' systemic
velocities. Seven of the nine LCBGs have disturbed gas properties that
may be the result of an interaction with a companion, such as
irregular morphologies, H \textsc{I} major axes that are offset in
position angle from optical major axes, and disturbed velocity
fields. Because this is not the case for every LCBG in our sample, we
do not have strong evidence from this study that star formation in
LCBGs \emph{must} be triggered solely by interactions, though
interactions may contribute to the star formation properties of some
LCBGs. We discuss an additional potential cause of star formation in
LCBGs in Section 4.4.1.

\subsection{Comparison with single-dish results}

One of the primary goals of this study was to investigate how results
from single-dish observations of nearby LCBGs compare to those derived
from resolved maps. Since LCBGs are unresolved at the distances at
which they are common, it is important to determine whether unresolved
observations of these galaxies are sufficient to describe their global
properties and predict their evolutionary paths. Eight of the LCBGs in
our sample were observed with the GBT by \citet{CG04} at a resolution
of $\sim 9\arcmin$, which is large with respect to their $\rm
R_{HI}$. We compare the $\rm M_{HI}$ of these galaxies derived from
our resolved observations and the unresolved observations of
\citet{CG04} in the last column of Table \ref{fig:spectrum}
\citep[SDSS0125+0110 was not observed by][and so we exclude it from
  these comparisons]{CG04}. We find that for six of the eight LCBGs
common to both samples, the single dish observations generate an equal
or greater $\rm M_{HI}$ than what we calculate from resolved
observations. We recover more H \textsc{I} emission for two of the
LCBGs in our sample (SDSS0728+3532 and SDSS1319+5203) than was
measured by \citet{CG04}, which is likely due to those galaxies
residing in H \textsc{I} envelopes that include other galaxies (the H
\textsc{I} masses we report for those two LCBGs are for the entire
envelope), though their H \textsc{I} envelopes are unresolved with the
GBT beam. On average, the $\rm M_{HI}$ that we measure is 76$\%$ of
what \citet{CG04} measured in their single dish observations, although
there is a large dispersion (40$\%$) between the values obtained in
both measurements.

We note that we used slightly different distances to calculate $\rm
M_{HI}$ than \citet{CG04} did. The recession velocities we measured,
and thus the distances we calculated, were on average $\rm
5\ km\ s^{-1}$ lower than those measured by \citet{CG04}. This
translates to a difference of $\rm 0.1\ Mpc$, and no galaxy had a
difference of more than $\rm 1\ Mpc$. If we calculate $\rm M_{HI}$
using distances derived from the same $\rm V_{sys}$ that \citet{CG04}
used, we measure $78 \% \pm 42 \%$ of the H \textsc{I} that
\citet{CG04} did. We will use the distances employed by \citet{CG04}
for the remainder of the comparisons in this section.

In comparison, in resolved observations of five LCBGs with the VLA,
\citet{CG07} measured values of $\rm M_{HI}$ that were on average
61$\%$ of the measured single dish values from \citet{CG04}, with a
similarly large dispersion. If we remove the two galaxies that reside
in larger H \textsc{I} envelopes from consideration, we recover on
average 59$\%$ of the $\rm M_{HI}$ that \citet{CG04} measured for the
remaining six galaxies, with a dispersion of 26$\%$, which is
consistent with the \citet{CG07} result.

We also compare the inclination-corrected $\rm V_{rot}$ and $\rm
M_{dyn}$ values we derive from our data cubes to those calculated from
$\rm W_{20}$ corrected for inclination in the \citet{CG04} sample in
Table \ref{fig:compare}. With the exception of SDSS1507+5511, the
single-dish $\rm V_{rot}$ values calculated by \citet{CG04} using half
of $\rm W_{20}$ corrected for inclination are larger than the
inclination-corrected $\rm V_{rot}$ values that we measure using a cut
along the major axis ($\rm \langle 0.5 \times W_{20}/V_{rot}\rangle =
$ 2.2 $\pm$ 1.3). Since the observations of \citet{CG04} were made
with beam sizes large enough to include contributions from companion
galaxies in the case of SDSS0119+1452, SDSS0934+0014, SDSS0936+0106,
SDSS1319+5203, and SDSS1402+0955, and tidal features in the case of
SDSS0728+3532 and SDSS1319+5203, their measurements of $\rm W_{20}$
are not spatially resolved enough to distinguish the velocity
contributions of the LCBGs from the contributions of their nearby
companions.

When calculating $\rm M_{dyn}$, \citet{CG04} estimated that $\rm
R_{HI} = 2\times R_{25}$, following \citet{B94}, since they did not
have measured values of $\rm R_{HI}$. We compare the estimated and
measured values of $\rm R_{HI}$ in Table \ref{fig:compare}. The $\rm
R_{HI}$ values that we measure are on average 81$\%$ of those used in
\citet{CG04}, though the scatter is relatively large ($\rm \langle
R_{HI}^{GMRT} / R_{HI}^{est.} \rangle = 0.81 \pm
0.42$). SDSS0728+3532, SDSS0936+0106, and SDSS1319+5203 have measured
$\rm R_{HI}$ values that are larger than those that \citet{CG04}
estimated. We use our measured $\rm R_{HI}$ to calculate $\rm M_{dyn}$
here.

With the exception of SDSS0936+0106 and SDSS 1507+5511, the $\rm
M_{dyn}$ within the estimated $\rm R_{HI}$ calculated by \citet{CG04}
are larger than those that we calculate here, owing to the larger
$ \rm V_{rot}$ and $\rm R_{HI}$ values estimated using single dish
observations. On average, the single-dish $\rm M_{dyn}$ values are
10.5 times larger than $\rm M_{dyn}$ measured along the galaxies'
major axes, with a large scatter ($\rm
\sigma_{(M_{dyn}^{GBT}/M_{dyn}^{GMRT})} = 10.3$).

The H \textsc{I} mass fractions, $\rm f_{HI} = M_{HI}/M_{dyn}$, that
we calculate using our resolved observations are on average nine times
larger than those calculated from single-dish measurements, though
with an equivalent standard deviation ($\rm \langle
f_{HI}^{GMRT}/f_{HI}^{GBT}\rangle =$ 6.1 $\pm$ 6.1. Only SDSS0936+0106
and SDSS1507+5511 have smaller $\rm f_{HI}$ when using resolved data
than the $\rm f_{HI}$ values derived from single-dish
observations. For two of the LCBGs, SDSS0728+3532 and SDSS1319+5203,
the $\rm M_{HI}$ values that we calculate encompass the entire,
multi-galaxy H \textsc{I} envelopes in which these galaxies reside,
while $\rm M_{dyn}$ only encompasses the LCBGs. As a result, the $\rm
f_{HI}$ that we calculate are likely much higher than the true values
(for example the $\rm f_{HI}$ values of SDSS1319+5203 is 3.5, which is
unphysically high).

The major advantages that our current study has over those undertaken
with single dishes are that (1) our improved spatial resolution
enables us to distinguish individual galaxies from their nearby
companions, (2) mapping the galaxies allows for their rotation axes to
be identified and their $\rm V_{rot}$ to be measured rather than
estimated from linewidths, and (3) mapping the galaxies also makes
measuring their $\rm R_{HI}$ possible, enabling calculations of their
$\rm M_{dyn}$ to be made with fewer assumptions. We generally
calculate lower $\rm M_{dyn}$ and higher $\rm f_{HI}$ than what was
calculated from single-dish measurements by \citet{CG04}. This result
strengthens their assertion that LCBGs are gas-rich galaxies with
smaller $\rm M_{dyn}$ than elliptical galaxies.

\begin{figure*}[htb!]

\begin{center}
\includegraphics[angle=0, width=0.8\textwidth]{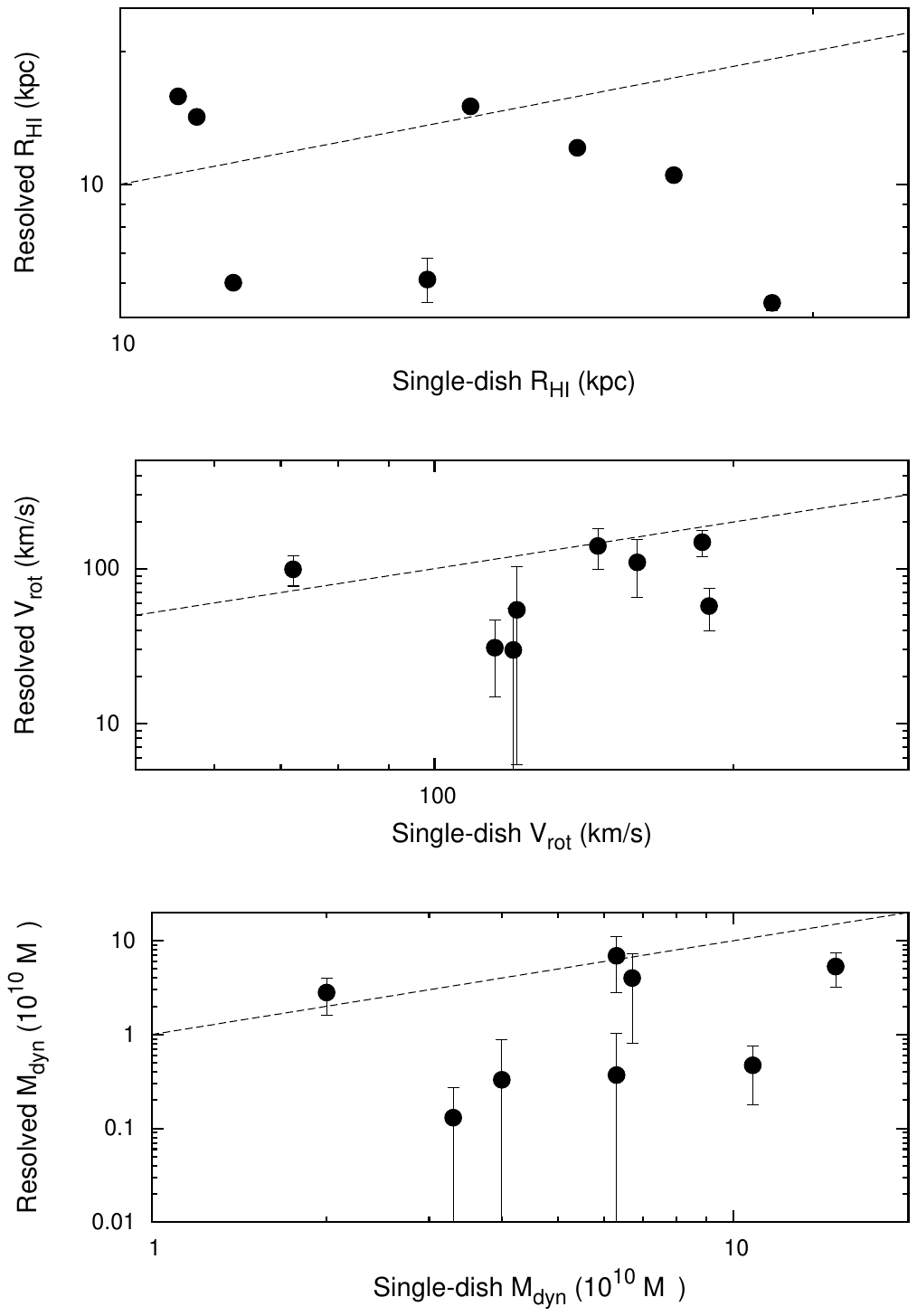}
\end{center}

\caption{$\rm R_{HI}$ (top), $\rm V_{rot}$ (middle), and $\rm M_{dyn}$
  (bottom) using data from \citet{CG04} and our measurements for the
  LCBGs common to both samples. \citet{CG04} estimated $\rm R_{HI}$ to
  be $\rm R_{HI} = 2 \times R_{25}$, and used half of the width of
  each galaxy's single-dish H \textsc{I} spectrum corrected for
  inclination and random motions as $\rm V_{rot}$. The dashed black
  lines show a 1:1 relationship between the two data sets. In some
  cases, error bars are smaller than point sizes. \label{fig:compareHIprops}}

\end{figure*}

From comparing the H \textsc{I} properties of the LCBGs in our sample
to those measured with a single dish, we find that the $\rm V_{rot}$,
$\rm R_{HI}$, and $\rm M_{dyn}$ that we measure are not related by a
simple scale factor to those estimated using single dish linewidths
and $\rm R_{25}$. See Figure \ref{fig:compareHIprops} for a visual
representation of the scatter in the H \textsc{I} properties that we
measure when compared to those reported by \citet{CG04}. We note that
our sample size is small, so we cannot rule out a characteristic
relationship between $\rm R_{25}$ and $\rm R_{HI}$ or between
single-dish linewidths and $\rm V_{rot}$ in LCBGs, though we do not
find such a relationship here.

\subsubsection{Comparison with stellar masses}
As a constraint on the $\rm M_{dyn}$ that we have calculated, we have
also calculated stellar masses, $M_{*}$, for each LCBG using the
equation $\rm log(M_{*}/L) = a_{\lambda} + b_{\lambda} \times Color$
given in \citet{B01}, where $\rm a_{\lambda}$ and $\rm b_{\lambda}$
are constants dependent on the wavelength of measured luminosity and
are tabulated in Table 1 of \citet{B01}. We used the $\rm B-V$ colors
that we listed in Table \ref{fig:obssummary}, and K-band magnitudes
from the Two Micron All Sky Survey (2MASS) catalog \citep{S06}. We
note that three LCBGs, SDSS0119+1452, SDSS0934+0014, and Mrk 325, had
$\rm M_{dyn}$ values lower than their stellar masses. \citet{B01}
state that the scatter on their color - $\rm M_{*}/L$ relation is
$\sim 10 \%$, which is smaller than the difference between the $\rm
M_{*}$ and $\rm M_{dyn}$ values for these galaxies, so it is not
likely that uncertainties on the color - $\rm M_{*}/L$ relation are
responsible for this unphysical result. There are two possible reasons
for these galaxies having larger $\rm M_{*}$ than $ \rm
M_{dyn}$. First, if a galaxy's H \textsc{I} is more face-on than its
optical emission, then we have likely underestimated its $\rm M_{dyn}$
due to under-correcting its rotation velocity for inclination. Since
these three galaxies were the most difficult to identify axes of
rotation for, it is likely that the uncertainty in their rotation
velocities is higher than for the other LCBGs. Second, the equation
used to calculate $\rm M_{*}$ is a relationship between $\rm M_{*}/L$
and galaxy colors determined by a model for several combinations of
colors and optical and near-infrared absolute magnitudes. As was shown
in \citet{CG04}, LCBGs are more likely to have lower $\rm M/L$ than
the average for local galaxies, and \citet{B01} find that bluer colors
correlate with lower $\rm M/L$. If LCBGs significantly deviate from
the color-$\rm M/L$ relationship that \citet{B01} have derived (for
example, if the $\rm B-V$ that we use in our calculations is redder
than the average $\rm B-V$ for a galaxy's disk), we may be
overestimating their stellar masses.

\subsection{Tully-Fisher relation}
The Tully-Fisher (T-F) relation \citep{TF77} posits that for rotating
galaxies, intrinsic luminosity is proportional to the galaxy's $\rm
V_{rot}$ raised to the fourth power. \citet{CG04} showed that not all
of the LCBGs in their sample follow the T-F relation when they
estimated $\rm V_{rot}$ using single-dish linewidths. It would be
expected that a galaxy with active star formation could have a
temporarily elevated B-band intrinsic brightness relative to the
expected brightness from the T-F relation given its $\rm
V_{rot}$. \citet{CG04} did see some evidence of that effect in their
sample. In addition, they also found that some LCBGs are less
intrinsically bright in the B band than their $\rm V_{rot}$ would
suggest, which would not be expected for star-forming galaxies. Since
a galaxy undergoing active star formation becomes fainter in the B
band once its star-forming episode ends, an LCBG that is fainter than
would be expected for a galaxy on the T-F relation would never become
bright enough to evolve onto the T-F relation. However, if a galaxy's
$\rm V_{rot}$ is overestimated by its single-dish linewidth, the
galaxy could appear to be too faint to follow the T-F relation given
its (overestimated) $\rm V_{rot}$. This scenario could happen if, for
example, an unresolved nearby companion or tidal feature exists whose
recession velocity overlaps with the rotation velocity range of the
target galaxy. \citet{CG04} found that six of the ten LCBGs in their
sample that are too faint to follow the T-F relation have
companions. Since our resolved observations enable us to measure the
$\rm V_{rot}$ values of the LCBGs in our sample, we revisit whether
LCBGs follow the T-F relation using our velocity measurements.

\begin{figure*}[htb!]
\begin{center}
\includegraphics[angle=0, width=0.75\textwidth]{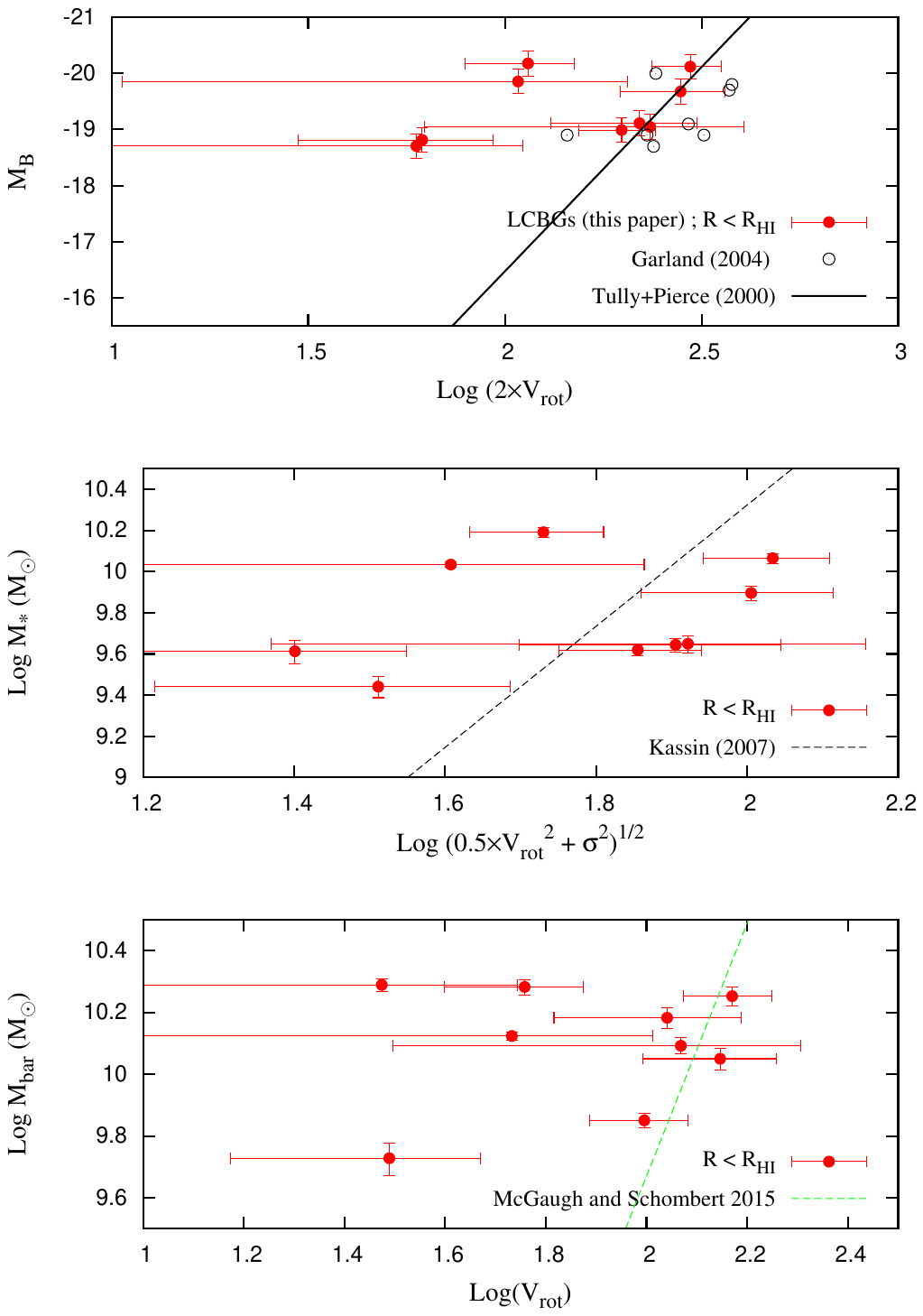}
\end{center}

\caption{\footnotesize \textit{Top}: A version of the Tully-Fisher
  relation, described in \citet{TP00}. $\rm M_{B}$ for the LCBGs in
  our sample (filled circles) are calculated as described in
  \citet{CG04} using SDSS g and r magnitudes and distances taken from
  Table \ref{fig:obssummary}. Their $\rm V_{rot}$ values are taken
  from cuts across the galaxies' major axes at $\rm R_{HI}$ and
  corrected for inclination. The same LCBGs (with the exception of
  SDSS0125+0110) are plotted with $\rm M_{B}$ and linewidths corrected
  for random motions and inclinations taken from Table 1 and Table 3
  of \citet{CG04} (open circles). The Tully-Fisher relation as
  described in \citet{TP00} is plotted with the black line. Four of
  the LCBGs in our sample are much brighter in the B band than their
  $\rm V_{rot}$ values would suggest, while six of the LCBGs in the
  \citet{CG04} sample are faint in the B band with respect to their
  linewidths. We find that the LCBGs in our sample either follow the
  Tully-Fisher relation or have the ability to evolve onto it once
  their star formation is quenched and their $\rm M_{B}$ subsequently
  fade. We interpret the galaxies lying to the right of the
  Tully-Fisher relation from the \citet{CG04} sample as having
  overestimated rotation velocities due to the inclusion of
  non-rotation H \textsc{I} features or companion galaxies in the
  beam. \textit{Middle}: Stellar mass Tully-Fisher relation as
  described in \citet{K07}. $\rm M_{*}$ are calculated using K-band
  magnitudes and B-V colors. $\rm V_{rot}$ values are calculated the
  same way as in the top figure. $\sigma$ values are the average
  $\sigma$ within $\rm R_{HI}$. The same LCBGs that lie to the left of
  the T-F relation on the top plot also lie to the left of the stellar
  mass T-F relation. \textit{Bottom:} Baryonic Tully-Fisher relation
  as described in \citet{MS15}. $\rm M_{bar}$ is the sum of the
  stellar mass and gas mass for each galaxy. When calculating the gas
  mass, we assumed a 10$\%$ contribution from $\rm H_{2}$. $\rm
  V_{rot}$ values are calculated the same way as in the top and middle
  figures. \label{fig:TF}}

\end{figure*}

We have plotted the LCBGs in our sample in Figure \ref{fig:TF} along a
version of the T-F relation described in \citet{TP00}. In this plot,
we use $\rm V_{rot}$ as measured along the galaxies' major axes and
corrected for optical inclination, as well as their $\rm M_{B}$ listed
in Table \ref{fig:obssummary}. We also plotted the corresponding
linewidths and $\rm M_{B}$ calculated for those LCBGs in
\citet{CG04}. Five of the nine LCBGs in our sample appear to follow
the T-F relation (within error bars), while four LCBGs are brighter
than anticipated given their $\rm V_{rot}$. Three of these four LCBGs
have $\rm M_{*} > M_{dyn}$ as discussed in Section 4.2.1. None of the
LCBGs in our sample have lower than expected luminosities given their
$\rm V_{rot}$ values, while six of the LCBGs have low luminosities
with respect to rotation velocities inferred from their linewidths as
measured in \citet{CG04}. Since the average $\rm V_{rot}$ derived from
half of the galaxies' single-dish linewidths is nearly three times the
$\rm V_{rot}$ values that we measure, we can infer that the cause of
some LCBGs lying to the right of the T-F relation in single-dish
measurements is likely due to uncertainties in estimating $\rm
V_{rot}$ from single-dish linewidths.

In addition to a temporarily elevated luminosity due to ongoing star
formation, one potential cause of some LCBGs being positioned to the
left of the T-F relation could be disturbed H \textsc{I} velocity
fields due to mergers or interactions. Eight of the nine LCBGs in our
sample have nearby companions or show signs of disturbed gas
morphology, though their optical morphologies remain disk-like. If a
galaxy's $\rm V_{rot}$ is not accurately measured by taking a cut
along its major axis and correcting for inclination (but is instead
underestimated), the galaxy could appear to be too luminous for its
measured $\rm V_{rot}$. We also note that, as we discussed in Section
3.2, our measured $\rm V_{rot}$ values include a correction for
optical inclination that could introduce a systematic error into the
reported $\rm V_{rot}$ values. Since the four LCBGs that lie to the
left of the T-F relation also have $\rm M_{*} > M_{dyn}$, it is likely
that we are underestimating their dynamical masses, and thus also
underestimating their $\rm V_{rot}$.

Another possible cause of deviation from the T-F relation, which does
not exclude a merger scenario, could be due to the formation of a
bulge or pseudobulge \citep{T14}. If LCBGs are undergoing their final
major burst of star formation while they build a bulge and transition
to more quiescent S0 or dE-type galaxies, we may be able to see
evidence of this transformation in their H \textsc{I}
properties. Earlier-type spiral galaxies have higher mass-to-light
ratios than later-type spirals, so their T-F relations tend to be
flatter than the average T-F relation for spiral galaxies
\citep{T14}. None of the LCBGs in our sample appear to have higher
mass to light ratios than the T-F relation would suggest (see Figure
\ref{fig:TF}), so we do not see evidence that the LCBGs in our sample
have prominent bulges like Sa-type galaxies.

As the LCBGs in our sample either follow the T-F relation or have the
potential to evolve onto it once their blue luminosities fade due to
decreased star formation activity, we can infer that the LCBGs in our
sample are likely to be rotation-supported. An additional
consideration to include in our analysis is the effect of velocity
dispersion on the galaxies' rotation velocities. Since we measured the
LCBGs' rotation velocities at the edges of the extent of the galaxies'
H \textsc{I}, where their velocity dispersions are relatively low (see
the Moment 2 maps in Figures \ref{fig:0119} - \ref{fig:mrk325}), the
effects of velocity dispersions on the galaxies' rotation velocities
that we have measured are not likely to be significant.  It is
possible that the rotation velocities measured by \citet{CG04} from
single-dish linewidths could be affected by the inclusion of velocity
dispersion. \citet{CG04} addressed this effect by incorporating a
correction for velocity dispersion in their reported linewidths, with
the assumption that the random motions of the gas contributed $\rm
\sim 38\ km\ s^{-1}$ to the measured single-dish linewidth. If this
correction was insufficient, the additional contribution from velocity
dispersion would increase the measured linewidth relative to what
would be measured due to pure rotation. This increase could contribute
to the data points in Figure \ref{fig:TF} from \citet{CG04} that lie
to the right of the T-F relation, where it is impossible to evolve
onto the T-F relation solely due to quenching of star formation.

To investigate whether the four LCBGs that lie to the left of the T-F
relation have kinematics that are not dominated by rotation, we
plotted the LCBGs in our sample along the stellar mass T-F relation
described by \citet{K07}. This relationship correlates $\rm M_{*}$
with the kinematic property $S_{0.5}$, which is defined as $S_{0.5} =
(0.5V_{rot}^2 + \sigma^{2})^{1/2}$ \citep{W06,K07}. Since we
calculated $\rm M_{*}$ using near-infrared magnitudes, which are not
as sensitive to recent star formation as B-band magnitudes, the LCBGs
that lie to the left of the T-F relation may lie closer to the stellar
mass T-F relation if recent star formation is significantly elevating
the B-band magnitudes in these galaxies. In Figure \ref{fig:TF}, we
find that the same four LCBGs that lie to the left of the T-F relation
also lie to the left of the stellar mass T-F relation when we measure
$\rm V_{rot}$ and $\sigma$ at $\rm R_{HI}$, and three of those four
LCBGs (SDSS1319+5203 is the exception) lie to the left of the stellar
mass T-F relation at $\rm R_{25}$. In addition, the five LCBGs that
follow the T-F relation lie to the right of the stellar mass T-F
relation.

Finally, we also plotted the baryonic T-F relation as described in
\citet{MS15} in the bottom panel of Figure \ref{fig:TF} to investigate
whether combining the LCBGs' gas masses and stellar masses produced a
different result when plotted against their $\rm V_{rot}$. The result
was the same as for the other two versions of the T-F relation: the
same four galaxies lie to the left of the baryonic T-F relation.

Since including both ordered rotation and disordered motions in
$S_{0.5}$, and both gas and stellar masses in the baryonic T-F
relation, did not move the four LCBGs to the left of the T-F relation
onto the stellar mass or baryonic mass T-F relations, it is not likely
to be the case that these four LCBGs are dominated by disordered
motions at large radii or have unusually large stellar masses. It is
instead likely that we are underestimating their dynamical masses. It
is also possible that these galaxies have recently undergone a merger
or interaction that has disturbed their kinematics, making their
rotation more difficult to measure and causing lower measured $\rm
V_{rot}$ and $\rm M_{dyn}$ than would be measured for more settled
disks. We discuss this possibility further in Section 4.5.

\subsection{Disk and bulge kinematics}

\begin{figure*}[htb!]

\begin{center}
\includegraphics[angle=0, width=0.9\textwidth]{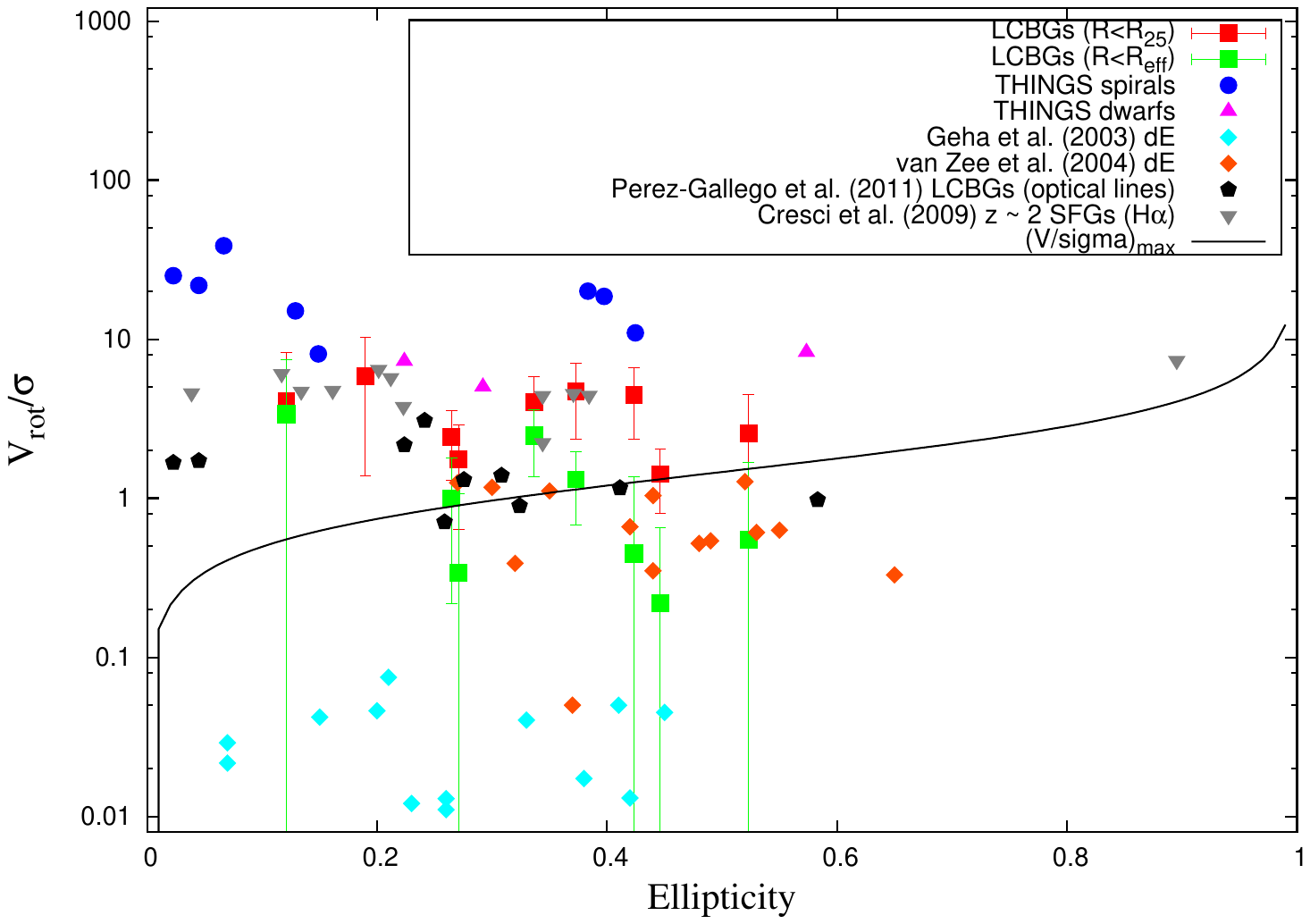}
\end{center}

\caption{\footnotesize $\rm V_{rot}/\sigma$ plotted against
  ellipticity ($\epsilon = 1-b/a$) for a variety of galaxy
  samples. The solid black curve is the maximum value of $\rm
  V_{rot}/\sigma$ allowed for elliptical galaxies. Galaxies above this
  curve are too rotation-supported to be classified as elliptical
  galaxies. The square points represent the LCBGs in our sample
  measured within $\rm R_{25}$ (red squares) and $\rm R_{eff}$ (green
  filled squares for galaxies with $\rm R_{eff}$ smaller than the beam
  size, and green open squares for galaxies with $\rm R_{eff}$ larger
  than the beam size). The velocity dispersions within $\rm R_{eff}$
  have been corrected for beam smearing. Archival data, including
  spiral galaxies from the THINGS sample (filled circles), dwarf
  galaxies from the THINGS sample (filled triangles), dwarf elliptical
  galaxies with a measured rotation component from \citet{Geha03}
  (filled diamonds) and \citet{vZ04} (open diamonds), compact
  star-forming galaxies with disk components at $\rm z \sim 2$
  \citet{Cresci09}, and for LCBGs in the \citet{PG11} sample (filled
  stars) are also plotted. $\epsilon$ values for the nine LCBGs in our
  sample are calculated using the same SDSS i-band axis ratios that we
  used to calculate the galaxies' inclinations. The error bars on
  $\epsilon$ calculated this way are smaller than the point
  sizes. $\epsilon$ values for the previously published samples are
  taken from Hyperleda except for the \citet{Geha03} sample, where
  $\epsilon$ is taken from Table 3 of that paper, the \citet{vZ04}
  sample, where $\epsilon$ is taken from Table 1 of that paper, and
  the \citet{Cresci09} sample, where $\epsilon$ is calculated from the
  inclinations listed in Table 2 of that paper. LCBG $\rm V_{rot}$
  values are measured using a cut along the galaxies' major
  axes. THINGS \citep{T09} $\sigma$ values are the average H
  \textsc{I} $\sigma$ values measured within $\rm R_{HI}$. THINGS $\rm
  V_{rot}$ values are taken using half of $W_{20}$ corrected for
  inclination from \citet{W08}. \citet{Geha03} dwarf elliptical $\rm
  V_{rot}$ and $\sigma$ values are measured from optical absorption
  lines within 0.5-1 $\rm R_{eff}$. \citet{vZ04} dwarf elliptical $\rm
  V_{rot}$ and $\sigma$ values are measured from optical absorption
  lines within the last point where a rotation curve could be
  fit. \citet{Cresci09} $\rm V_{rot}$ and $\sigma$ values are measured
  from $\rm H\alpha$ emission lines within a ``scale radius'' fit from
  a Gaussian profile along the major axis of the galaxies' intensity
  maps. \citet{PG11} $\rm V_{rot}$ values are measured from rotation
  curves fit to H$\alpha$ velocity maps, and $\sigma$ values are
  measured from [O\textsc{III}]$\lambda$5007 maps. \label{fig:vsigmaell}}

\end{figure*}

One way we can further infer whether LCBGs contain significant bulges
is to compare their ratios of ordered to disordered motion to their
ellipticity, $\epsilon$. Using virial theorem arguments, $\rm V_{rot}/
\sigma$ can be related to $\epsilon$ by

\begin{equation}
\rm \frac{V_{max}}{\sigma} = \frac{\pi}{4} \sqrt{2[(1-\epsilon)^{-0.9}-1]}
\end{equation}

\noindent where $\rm \epsilon = 1 - b/a$ \citep{SG}. We plot this
relation, which indicates the maximum ratio of ordered motions to
random motions allowable for a given flatness of elliptical galaxies,
in Figure \ref{fig:vsigmaell}. When measured within $\rm R_{25}$, all
of the LCBGs in our sample lie above this relation, along with spiral
galaxies and late-type dwarf galaxies from the THINGS sample
\citep{W08} and disklike star-forming galaxies at higher redshifts
\citep{Cresci09}, which shows that they rotate faster (or have smaller
values of $\sigma$) than is permitted for elliptical galaxies. By
contrast, all of the dwarf elliptical galaxies with a significant
rotation component studied by \citet{Geha03}, and most of the dwarf
ellipticals studied by \citet{vZ04}, lie below this relation. When we
measured $\rm V_{rot} $ at $\rm R_{eff}$ and the average $\sigma$
within $\rm R_{eff}$, the $\rm V_{rot}/\sigma$ values of the galaxies
in our sample lie near the relation, which implies that the gas in the
central portion of LCBGs has approximately the maximum $\rm V_{rot}$
possible for elliptical galaxies. While the stellar kinematics of
LCBGs may not be identical to the gas kinematics at this radius, we
expect both the stars and the gas to trace a similar potential so
close to the galaxies' centers since their optical emission is
centrally concentrated, and the maximum intensities of their H
\textsc{I} emission are within a beamwidth of their peak optical
emission \citep[e.g.][]{N07, S16}.  \citet{B04} show that a sample of
very blue ($\rm B-V \sim 0.25$), very compact (SBe(B) $\sim$ 19 $\rm
mag \ arcsec^{-2}$) intermediate-redshift LCBGs lie below the relation
(see their Figure 2), which suggested to them that LCBGs may evolve
into dwarf elliptical galaxies once their star formation has been
quenched. That study differs from ours in that it surveyed an extreme
subset of intermediate-redshift LCBGs and measured ionized gas rather
than H \textsc{I}. \citet{PG11} measured $\rm V_{rot}/\sigma$ for
ionized gas using optical emission lines for a sample of local LCBGs
that has two galaxies in common with our sample (SDSS1507+5511 and Mrk
325). When compared with the ellipticities of those galaxies, the
LCBGs in their sample behave in a way similar to the LCBGs in our
sample measured at $\rm R_{eff}$. The $\rm V_{rot}/\sigma$ values that
we have measured make the presence of large-scale classical bulges
that contain gas unlikely at present in LCBGs, though the gas in the
innermost regions of LCBGs may display bulge-like behavior. This
suggests that if the local LCBGs in our sample are representative of
the population of LCBGs that is common at $\rm z\sim 1$, those
higher-redshift LCBGs must also be dominated by ordered rotation. If
this was the case, then LCBGs at higher redshifts are likely disk
galaxies with extensive star formation in their disks, rather than
irregular or spheroidal galaxies.

\subsubsection{Disk instabilities and central bulges}

Some studies have hypothesized that LCBGs are bright, star-forming
bulges (or bulge progenitors) of disk galaxies \citep{BvZ01,H01}. This
scenario is consistent with the centrally-peaked $\sigma$ values and
relatively low central $\rm V_{rot}/\sigma$ values that we find for
many of the LCBGs in our sample. The motion of the gas at large radii
in the galaxies in our sample is dominated by rotation ($\rm
V_{rot}/\sigma > 1$) even within $\rm R_{25}$, and even for the LCBGs
with disturbed velocity fields, so we rule out the existence of
gas-rich ``classical'' bulges in our sample. Even so, it is possible
that the LCBGs in our sample have or are developing central bars or
``pseudobulges'' \citep{KK04}, which are less supported by random
motions than they are by rotation.

One mechanism for developing a central bulge in star-forming galaxies
is the ``clump-origin bulge'' \citep[e.g.][]{N98, N99, N00, N01, D09,
  E09, I12}. In this scenario, gas infalling onto a galaxy's disk
develops overdensities within the disk that contract and become
star-forming clumps. As the clumps orbit along with the rest of the
disk, they move toward the center of the galaxy due to dynamical
friction and can merge with other clumps. When these clumps merge with
each other, star formation rates in the clumps increase briefly, which
gives the clumps a bright, blue appearance and drives up the galaxies'
global star formation rates. Finally, the few large clumps that remain
merge in the center of the galaxy, causing either a small clump-origin
nuclear bulge or bar that maintains some of the the angular momentum
that the clumps had in the disk. At this point, the star formation
rate of the clumps rapidly declines \citep[for a visual illustration
  of this process, see Figure 1 of ][]{I12}. The lifetimes of the
clumps are governed by their size (more massive clumps are less likely
to disperse due to outward pressure from their star formation), as
well as their distance from the center of the galaxy (clumps that have
less distance to travel as they move toward the center are more likely
to reach the center of the galaxy intact). Clumpy galaxies have been
observed at a range of redshifts, including galaxies that resemble
LCBGs. For example, \citet{O09} found that star-forming clumps,
including large, bright central clumps, are common in a sample of
Lyman Break Analog galaxies at z$\sim 0.1-0.3$ that have similar
effective radii and dynamical masses to the LCBGs in our
sample. \citet{CGletter} found that $40\%$ of local LCBGs are clumpy,
likely due to the buildup of accreted gas from interactions with
companions or material in galaxy clusters.

To determine whether conditions in the LCBGs' disks are conducive to
the formation of clumps, we can use the gas properties we measure to
calculate the galaxies' Toomre parameters \citep{T64} for the
stability of their disks' gas:

\begin{equation}
\rm Q_{gas}=\frac{\sigma_{gas} \kappa}{\pi G \Sigma_{gas}}
\end{equation}

\noindent where $\rm \kappa = \sqrt{2}V_{rot}/R$ for a flat rotation
curve, and $\rm \Sigma_{gas}$ is the gas mass surface density. For a
disk to be stable, $\rm Q \gtrsim 1$. More accurate measures of Q
incorporate the disks' stellar components as well \citep{D09}, though
we limit our present analysis to the LCBGs' H \textsc{I} since we are
probing the galaxies' kinematics beyond the extent of their stellar
disks, and are particularly interested in whether the kinematics of
their gas are conducive to star formation. In general, incorporating a
stellar component will increase a galaxy's value of Q.

Keeping in mind that a galaxy's total mass is represented by $\rm
M_{dyn} = V_{rot}^{2}R/G$, we can solve for $\rm V_{rot}$ in the $\rm
M_{dyn}$ equation and solve for $\sigma_{gas}$ in Equation 5 with $\rm
Q_{gas} = 1$, and divide the two relationships following \citet{D09}
to highlight the criterion for disk stability in terms of gas
properties we can measure:

\begin{equation}
\rm \frac{V_{rot}}{\sigma_{gas}} < \frac{\sqrt{2}\Sigma_{total}}{\Sigma_{gas}} ,
\end{equation}

\noindent or

\begin{equation}
\rm \frac{V_{rot}}{\sigma_{gas}} < \frac{\sqrt{2}}{f_{gas}} .
\end{equation}

\noindent In this relationship, $\rm \Sigma_{total}$ is the galaxy's
total mass surface density, which is calculated using its $ \rm
M_{dyn}$ within a given radius ($ \rm \Sigma_{total} = M_{dyn}/\pi
R^{2}$). We note that for this discussion, we make the approximation
that $\rm f_{gas} = (M_{HI} + M_{H_{2}})/M_{dyn}$ where $\rm
M_{H_{2}}$ is either the published value for each galaxy if it exists
(these are stated in the Appendix), or $10\%$ of $\rm M_{HI}$ if no
published $\rm M_{H_{2}}$ exists for the galaxy. $\rm M_{dyn}$ is
measured within $\rm R_{HI}$. \citet{CG05} found that LCBGs' molecular
gas mass is typically $\lesssim 10\%$ of their $\rm M_{HI}$ (assuming
a Galactic X factor), so contributions to $\rm f_{gas}$ from molecular
gas are likely to be small for the galaxies in our sample that do not
have published values of $\rm M_{H_{2}}$.

\begin{figure*}[htb!]

\begin{center}
\includegraphics[angle=0, width=0.85\textwidth]{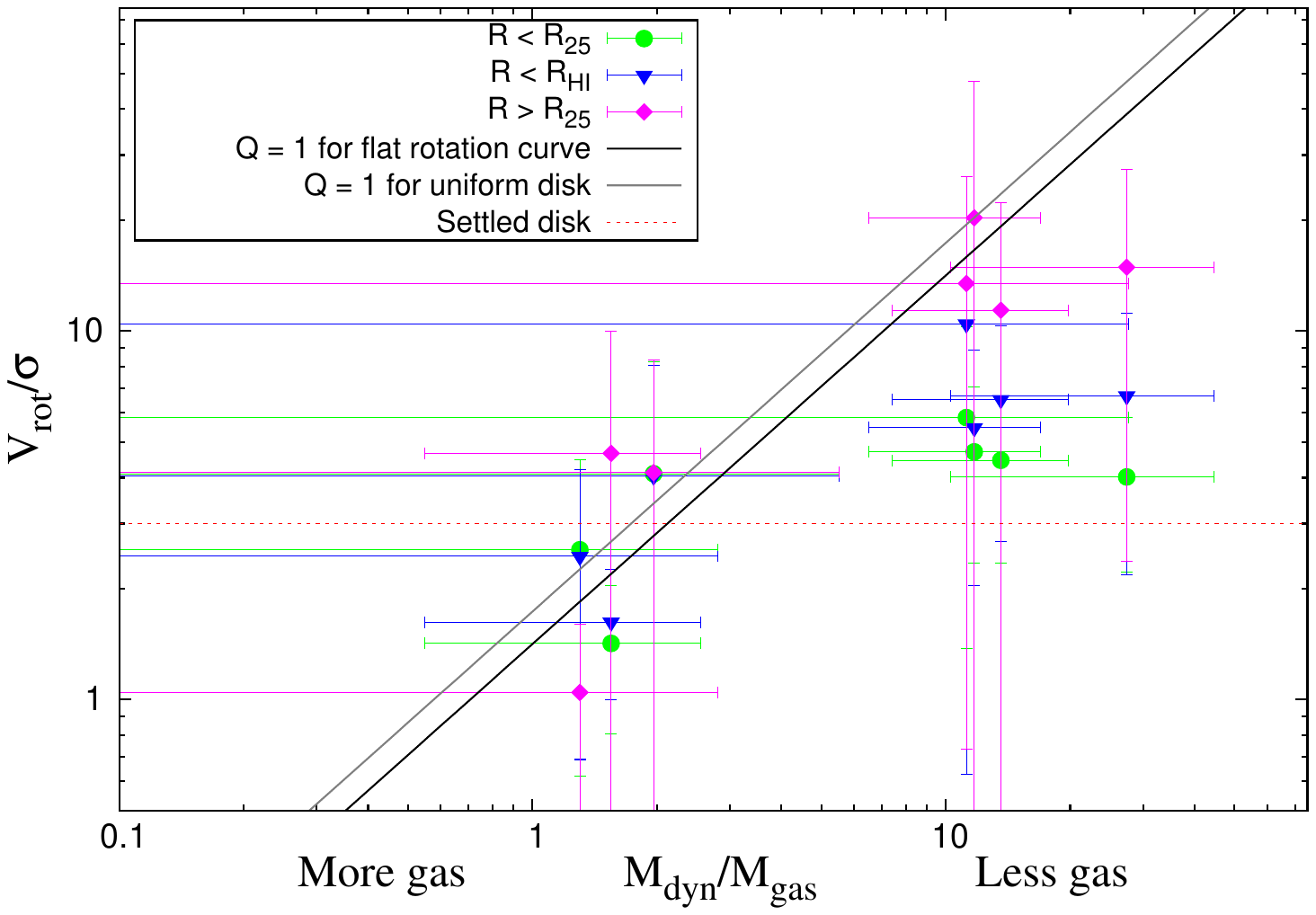}
\end{center}

\caption{$\rm V_{rot}/\sigma$ within $\rm R_{25}$ (green circles),
  $\rm R_{HI}$ (blue triangles) and outside of $\rm R_{25}$ (purple
  diamonds) for the LCBGs in our sample that do not share a common H
  \textsc{I} envelope with another galaxy. The solid black line is the
  Toomre criterion for disk instability for a gas disk with a flat
  rotation curve, and the solid gray line is the Toomre instability
  criterion for a uniform disk \citep{T64}. Above the line, galaxies'
  disks can develop local instabilities. Below the line, turbulence
  prevents gas clumps from forming. The LCBGs in our sample have H
  \textsc{I} disks that are consistent with being stable on average,
  though many of them have large enough error bars on $\rm
  M_{dyn}/M_{gas}$ and $\rm V_{rot}/\sigma$ that we can not rule out
  gravitational instabilities. The horizontal dotted red line is the
  \citet{K12} criterion for disk settling. Above the line ($\rm
  V/\sigma = 3$), galaxies' disks are considered to be settled. \label{fig:vsigma}}

\end{figure*}

We plot the measured values at several radii for each of the LCBGs in
our sample in Figure \ref{fig:vsigma}. This plot excludes
SDSS0728+3532 and SDSS1319+5253B, which are contained within a larger
H \textsc{I} envelope. For these two sources, $\rm M_{HI}$ contains
multiple galaxies, whereas $\rm M_{dyn}$ only reflects the mass of the
target galaxy. All of the LCBGs in our sample except Mrk 325 have
stable gas disks with respect to perturbations over a range of radii
(for $\rm Q<1$, data points would lie above the curve in Figure
\ref{fig:vsigma}, and for $\rm Q>1$, data points lie below the curve),
which means that their $\rm f_{gas}$ would need to be higher for their
disks to be unstable given their present values of $ \rm
V_{rot}/\sigma$. Since the error bars on $\rm V_{rot}/\sigma $ and
$ \rm f_{gas}$ are large, we can not rule out the potential to form
local instabilities at at least one radius for any of the LCBGs. We
note that we assume that $\rm f_{gas}$ is constant at all radii, which
is unlikely to be the case given that the LCBGs' H \textsc{I} emission
is more intense in their centers than at their edges. This effect
would result in a greater likelihood of disk instabilities at smaller
radii than what we plot in Figure \ref{fig:vsigma} (lower-radii data
points will move to the left in Figure \ref{fig:vsigma} if $\rm
f_{gas}$ increases with decreasing radius).

In a study modeling gas infall onto galaxies, \citet{D09} found that
if the cold gas streams that are feeding infall onto the disk are
clumpy, the clumps more easily merge toward the center of the galaxy
and form a spheroid shape, keeping the disk's average $\rm f_{gas} <
0.3$. In this scenario, the disk is usually stable. They found that
conversely, if the streams are smooth, the disk can support clumps for
a longer period of time. \citet{N00} found that requiring the local
gas density to rise above a certain threshold before star-forming
clumps could form yielded simulations consistent with observations of
early- and late- type disks, and that for smaller galaxies, clumps may
not be able to form at all and instead the infalling gas is fed into
the center of the galaxy to form a bar. To estimate the likelihood of
these scenarios, we can compare the $\rm M_{HI}$ values that we
measure for the LCBGs in our sample to the clump masses that have been
observed and modeled in other studies. \citet{E09} found that for
spiral galaxies, each clump contains an average of $0.3\%$ of its
galaxy's stellar mass, while for clump cluster galaxies (galaxies
dominated by several bright clumps) each clump contains about $2\%$ of
the galaxy's stellar mass. If we make the assumption that these ratios
are also approximately valid for $\rm M_{HI}$, we can multiply these
percentages by the average $\rm M_{HI}$ of the LCBGs in our sample to
find the expected clump masses. On average, a clump would contain
about $\rm 1 \times 10^{7}\ M_{\sun}$ of H \textsc{I} in a spiral
LCBG, or about $\rm 7 \times 10^{7}\ M_{\sun}$ of H \textsc{I} in a
clump cluster LCBG given these ratios. \citet{N00} found that in
simulations, clumps were more likely to survive migration toward the
galaxies' centers in galaxies with clumps larger than $\rm \sim
10^{7}\ M_{\sun}$, while in galaxies with smaller clumps, the clumps
were more likely to be disrupted and instead form a short bar from
their gas.

To determine whether any star-forming clumps due to gravitational
instabilities in LCBGs are detectable over long timescales, we
calculated the inspiral time for clumps, $\rm t_{ins} =
(V_{rot}/\sigma)^{2} \times t_{dyn}$, where $\rm t_{dyn} =
R_{HI}/V_{rot}$ \citep[see, for example, ][]{D09, G14} using $\rm
V_{rot}$ measured at $\rm R_{HI}$ and the average $\sigma$ measured
within $\rm R_{HI}$ for each LCBG. For the LCBGs in our sample, $ \rm
t_{ins}$ is longer than 1 Gyr \citep[approximately the lifetime of a
$\rm \sim 2.5\ M_{\sun}$, or late-B to early A-type star; ][]{H88,
M89, R05} for all but two LCBGs. This implies that if clumps can form
in LCBGs, they can persist for several Gyr before they finally sink to
LCBGs' centers if they are not disrupted. When compared with less
compact and less dynamically hot disk galaxies, however, the expected
$\rm t_{ins}$ for LCBGs is relatively short. Since disk galaxies tend
to have higher values of $\rm V_{rot}/\sigma$ at lower redshifts than
at higher redshifts \citep{K12}, the appearances of clumpy galaxies at
$\rm z \sim 1$ may evolve more rapidly than most disk galaxies with
star-forming clumps in the local universe due to their lower $\rm
t_{ins}$. Thus, since local LCBGs have relatively high values of $\rm
V_{rot}/\sigma$ and compact $\rm R_{HI}$, their $\rm t_{ins}$ are
likely more comparable to those of LCBGs at higher redshifts.

Though we do not have the sensitivity in our current study to measure
the masses of gas clumps, nor the resolution to measure local
overdensities in the LCBGs in our sample, future studies at higher
resolutions may be able to identify density variations in LCBGs' disks
and determine the likely course of future evolution of the galaxies'
clumps. As both low-redshift LCBGs and their high-redshift analogs are
often clumpy, further understanding of this phenomenon will be useful
in predicting these galaxies' evolutionary paths. Future resolved
observations of LCBGs will help better measure local velocity
dispersions and disk inclinations, which will better constrain
inspiral times for clumps in their disks. In addition, resolving
individual clumps and measuring their properties will determine
whether feedback within the clumps due to radiation pressure will
disperse the clumps on shorter timescales than $\rm t_{ins}$.

\subsection{Comparison with higher-redshift galaxies}

\citet{FS09} measured $\rm H_{\alpha}$ velocity maps for a sample of
$\rm z \sim 2$ star-forming galaxies from the Spectroscopic Imaging
survey in the Near-Infrared with SINFONI (SINS) survey. Most of the
galaxies in their sample had clumpy $\rm H_{\alpha}$
morphologies. They found that the galaxies in their sample fell into
three groups based on their kinematics, with approximately a third of
their sample falling into each category: rotation-dominated disks,
compact dispersion-dominated galaxies, and merging systems. The H
\textsc{I} velocity fields of the LCBGs in our sample resemble the
$\rm H_{\alpha}$ velocity fields of the rotation-supported galaxies in
their sample, as shown in Figure 17 of \citet{FS09}. The LCBGs in our
sample do not appear to be similar to star-forming galaxies supported
by disordered motions; the dispersion-dominated galaxies in the
\citet{FS09} sample do not show a clear axis of rotation, while even
the least rotationally-supported LCBGs in our sample show a clear
velocity gradient.

\citet{G14} found that for a sample of 19 rotationally-supported,
star-forming disk galaxies from the SINS survey with smooth velocity
gradients and centrally-peaked velocity dispersions at $\rm z \sim 2$,
Q is also centrally-peaked. We can see from Equation 5 and Figure
\ref{fig:vsigma} that this is true for the LCBGs in our sample as
well. The solid line in Figure \ref{fig:vsigma} represents $\rm Q =
1$, with values of Q increasing to the bottom right. We see that for
the LCBGs in our sample, the measurements at smaller radii tend to
have larger values of Q than at larger radii. As mentioned in Section
4.4.1, we assume in Figure \ref{fig:vsigma} that $\rm f_{gas}$ is
constant at all radii, which is unlikely to be true since the H
\textsc{I} emission in the LCBGs in our sample decreases in intensity
at larger radii. If $\rm \Sigma_{gas}$ increases with decreasing
radius in the LCBGs in our sample, then Q will not be as
centrally-peaked as is implied by Figure \ref{fig:vsigma}. \citet{G14}
argue that increased values of Q in the centers of galaxies could lead
to the formation of a central bulge and the quenching of star
formation where $\rm Q > 1$. While our measurements of rotation
velocities, velocity dispersions, and gas masses suggest that a
similar central bulge is possible in the LCBGs in our sample,
higher-resolution measurements of H \textsc{I} are needed for these
galaxies to more conclusively identify central bulges that could lead
to the quenching of star formation in LCBGs.

\citet{K12} measured the gas kinematics in blue star-forming galaxies
from $z \sim 0.2$ to $z \sim 1.2$ and found that the average $\rm
V_{rot}/\sigma$ tends to increase with decreasing redshift. They
interpret this trend as the result of the galaxies' disks settling
over time due to fewer mergers and decreased gas fractions due to gas
depletion from past star formation. They defined a settled disk as
having $\rm V_{rot}/\sigma = 3$. In Figure \ref{fig:vsigma}, we see
that the three galaxies with $\rm V_{rot}/ \sigma < 3$ at $\rm R_{HI}$
also have the highest gas fractions. While this relationship could be
due to underestimating these galaxies' $\rm V_{rot}$ due to
inclination uncertainties, it is also notable that these three
galaxies have the least-ordered velocity contours of the galaxies in
our sample that are not in larger H \textsc{I} envelopes. These
galaxies' disks may be in the process of settling, while the remaining
four galaxies' disks, in addition to being stable on large scales
relative to gravitational instabilities, may have already settled.

Recently, \citet{C16} observed LCBGs in intermediate-redshift
clusters. They had previously found that cluster LCBGs at that
redshift are likely experiencing their first pass through a cluster,
and will likely have their gas stripped and star formation quenched
once they pass through the cluster \citep{C11, C14}. In their most
recent study, \citet{C16} find that the LCBGs in their sample have
sizes, masses, and metallicities that are consistent with dwarf
elliptical galaxies. Because of this, they conclude that
intermediate-redshift cluster LCBGs are likely to rapidly evolve into
dE galaxies with quenched star formation once they pass through the
cluster potential and have their gas stripped. Our results show that
the local LCBGs that we have observed are unlikely to passively evolve
into dE galaxies, since they have significant disk
components. However, we note that the LCBGs in our sample, unlike
those studied by \citet{C16}, are not members of massive clusters. It
is possible that the evolutionary paths of cluster and non-cluster
LCBGs are strongly influenced by their environments (\citet{CGletter}
found that this is likely to be true for $\rm z \sim 0$ LCBGs), and
thus their future morphologies may diverge.

\section{Conclusions}

In this study, we have measured the H \textsc{I} properties of nine
LCBGs from resolved maps of H \textsc{I} emission. We conclude that
\begin{itemize}
\item The LCBGs in our sample are rotating disk galaxies with H
  \textsc{I} intensities and velocity dispersions that decrease with
  increasing radii;
\item The H \textsc{I} linewidths measured for these galaxies by
  single dishes tend to overestimate their rotation velocities, likely
  due to the inclusion of companions or tidal features in the beam;
\item The LCBGs in our sample have H \textsc{I} $\rm
  V_{rot}/\sigma$ values that are consistent with the gas
  kinematics of disk galaxies when measured at $\rm R_{HI}$ and $\rm
  R_{25}$, though some LCBGs have H \textsc{I} $\rm
  V_{rot}/\sigma$ within $\rm R_{eff}$ that are consistent with
  bulgelike behavior; and
\item The disks of the LCBGs in our sample are stable on average with
  respect to local perturbations, though they have the potential to
  form local instabilities at large radii, and three LCBGs' disks have
  $\rm V_{rot}/\sigma$ values at $\rm R_{HI}$ consistent with disks
  that have not yet settled.
\end{itemize}

We have found that the LCBGs exhibit a variety of gas morphologies,
from regular, symmetric rotation to asymmetric, disturbed rotation to
multiple galaxies in a common H \textsc{I} envelope. All of the LCBGs
in our sample appear to be dominated by rotation at large radii ($\rm
V_{rot}/\sigma(R_{HI}) > 1$), and have significant rotation components
at smaller radii. Because we were unable to robustly fit rotation
curves to the LCBGs, we do not have enough information to comment
extensively on the shapes of LCBGs' rotation curves (or their mass
distributions), though we have been able to produce position-velocity
plots from cuts along their major axes that display the shapes of
their velocity profiles. The LCBGs in our sample tend to have
asymmetric velocity profiles, which supports a scenario where LCBGs'
star formation is the result of gas disturbance. We cannot
conclusively distinguish whether this disturbance is externally or
internally triggered, but the presence of companions near most of the
LCBGs in our sample are not inconsistent with external
mechanisms. Even so, even the most disturbed LCBGs have identifiable H
\textsc{I} rotation axes.

When compared to previous single-dish results \citep{CG04}, we measure
lower values of $\rm V_{rot}$ for all of the LCBGs in our sample
except for SDSS1507+5511. Six of the nine LCBGs have smaller measured
$\rm R_{HI}$ values than what was estimated for them using $ \rm
R_{HI} \sim 2\times R_{25}(B)$ (two of the exceptions, SDSS0728+3532
and SDSS1319+5203, have H \textsc{I} envelopes that include other
galaxies, while the other exception, SDSS0936+0106, has a measured
$\rm R_{HI}$ that is only 6$\%$ larger than its estimated $\rm R_{HI}$). These discrepancies tend to result in smaller values of $\rm
M_{dyn}$, and larger values of $\rm f_{HI}$, for LCBGs than those
calculated from single dish measurements.

Though most of them have disturbed kinematics, the LCBGs in our sample
do not appear to be exclusively the result of ongoing major
mergers. While seven ($78\%$) of them have companions at comparable
systemic velocities, only two ($22\%$) of them have other galaxies
within their H \textsc{I} envelopes. This implies that the star
formation in LCBGs is not required to be solely triggered by a major
merger event. Instead, some LCBGs' star formation may be the result of
intrinsic bulge-building, which could be enhanced by interactions or
minor mergers but does not require interactions to proceed (we note,
for example, that for the two LCBGs, SDSS0728+3532 and SDSS1319+5203,
that are in three-galaxy interacting systems, only one of the galaxies
in each system is an LCBG).

We found that the LCBGs in our sample either already follow the
Tully-Fisher relation or have the potential to evolve onto it once
their current elevated star formation, and thus B-band luminosities,
fade. While the $\rm V_{rot}/\sigma$ values of the LCBGs' H \textsc{I}
look like stellar $\rm V_{rot}/\sigma$ of dwarf elliptical galaxies at
the smallest radii ($\rm R < R_{eff}$), they resemble other types of
spiral galaxies more closely at $\rm R_{25}(B)$. From this, we infer
that the LCBGs in our sample are likely not dispersion-dominated
bulges, and that they will not likely be able to passively evolve into
dwarf ellipticals once their star formation is quenched. However, some
of the LCBGs in our sample may be building small central bulges, as
three of the galaxies have $\rm V_{rot}/ \sigma$ values at large radii
($\rm R > R_{25}$) that are above the threshold of disk instability
necessary for infalling gas to create star-forming clumps that may
later merge to form a nucleus or bar. (We note that since we used
inclinations calculated from optical SDSS \textit{i}-band axis ratios
to calculate $\rm V_{rot}$, there are large uncertainties in the
values of $\rm V_{rot}/\sigma$ that we report.)

The variety of optical and H \textsc{I} morphologies, environments and
kinematics of the LCBGs in our sample lends support to the picture of
LCBGs being a heterogeneous class of galaxies undergoing a common,
short-lived evolutionary phase in their star formation
histories. Since the LCBGs in our sample do not have common H
\textsc{I} properties, we cannot predict a single future scenario for
their evolution based on their gas morphologies. This is not
surprising, as the relative abundance of LCBGs at $\rm z \sim 1$ would
suggest that, if LCBGs follow a single common evolutionary path, their
end products would be similarly common in the local universe.

Since LCBGs appear to have such a variety of gas morphologies, future
studies of their gas properties focusing on high-resolution mapping to
further probe their internal dynamics, including the presence of
star-forming gas clumps, bars, or bulges related to local
gravitational instabilities, will better illustrate whether they have
any common intrinsic star formation triggers. Such mapping would
provide data useful for modeling LCBGs' gas evolution to predict their
timescales for future quenching of their star formation. We are
currently studying star formation tracers in a larger local sample of
LCBGs to calculate their current star formation properties. This
information, coupled with LCBGs' $\rm M_{HI}$ values, will help us
understand whether LCBGs' gas depletion timescales given their current
star formation rates are shorter than their expected timescales for
star formation quenching. This will provide better understanding for
the evolutionary paths of this formerly common, currently rare, class
of galaxies.

\acknowledgments
We thank the staff of the GMRT who have made these observations
possible. GMRT is run by the National Centre for Radio Astrophysics of
the Tata Institute of Fundamental Research.

K.R. acknowledges partial support from a Wisconsin Space Grant
Consortium Research Infrastructure Program grant. D.J.P. and
K.R. acknowledge partial support from NSF CAREER grant AST-1149491.

The National Radio Astronomy Observatory is a facility of the National
Science Foundation opereated under cooperative agreement by Associated
Universities, Inc.

This research has made use of the NASA/IPAC Extragalactic Database
(NED) which is operated by the Jet Propulsion Laboratory, California
Institute of Technology, under contract with the National Aeronautics
and Space Administration. We acknowledge the use of NASA's SkyView
facility (http://skyview.gsfc.nasa.gov) located at NASA Goddard Space
Flight Center. 

This publication makes use of data products from the Two Micron All
Sky Survey, which is a joint project of the University of
Massachusetts and the Infrared Processing and Analysis
Center/California Institute of Technology, funded by the National
Aeronautics and Space Administration and the National Science
Foundation.

Funding for SDSS-III has been provided by the Alfred P. Sloan
Foundation, the Participating Institutions, the National Science
Foundation, and the U.S. Department of Energy Office of Science. The
SDSS-III web site is http://www.sdss3.org/.

SDSS-III is managed by the Astrophysical Research Consortium for the
Participating Institutions of the SDSS-III Collaboration including the
University of Arizona, the Brazilian Participation Group, Brookhaven
National Laboratory, Carnegie Mellon University, University of
Florida, the French Participation Group, the German Participation
Group, Harvard University, the Instituto de Astrofisica de Canarias,
the Michigan State/Notre Dame/JINA Participation Group, Johns Hopkins
University, Lawrence Berkeley National Laboratory, Max Planck
Institute for Astrophysics, Max Planck Institute for Extraterrestrial
Physics, New Mexico State University, New York University, Ohio State
University, Pennsylvania State University, University of Portsmouth,
Princeton University, the Spanish Participation Group, University of
Tokyo, University of Utah, Vanderbilt University, University of
Virginia, University of Washington, and Yale University.

\appendix
\section{Individual Galaxies}

\subsection{SDSS0119+1452}

SDSS0119+1452 (NGC 469; Figure \ref{fig:0119}) has asymmetric
H \textsc{I} emission that has its highest column density coinciding
with its optical center. Its velocity contours are disturbed, and its
H \textsc{I} major axis is nearly perpendicular to its optical major
axis. This morphology is similar to polar bulge galaxies \citep{C12},
and thus may be a result of an acquisition of gas after the nuclear
portion of the galaxy was formed. It has a blob of H \textsc{I}
emission to its southwest that does not have an optical counterpart
and is either not connected to the main galaxy, or connected at a low
column density. In low-resolution H \textsc{I} maps, the H \textsc{I}
blob is not separate from the main galaxy. Its area of highest
H \textsc{I} velocity dispersion coincides with its optical emission.

In SDSS optical images, SDSS0119+1452 does not have discernable spiral
arms. Galaxy Zoo 1 \citep{L11} classifies it as ``uncertain'', with a
$31\%$ probability of being a spiral galaxy ($18\%$ probability of
being an edge-on spiral), a $28\%$ probability of being an elliptical
galaxy, and a $24\%$ probability of being a merger. It has the
elongated, clumpy appearance of a ``bent chain'' galaxy similar to
those identified by \citet{EE06}, although \citet{CGletter} classified
it as not being clumpy since it has fewer than three distinct
clumps. Outside of the bright central clumpy emission, it has low
surface brightness optical emission.

\citet{PG11} mapped SDSS0119+1452 in H $\alpha$ emission and identified it as a preturbed rotator following the classification system of \citet{Y08}. This implies that its velocity contours show ordered rotation, including a major axis aligned with the optical major axis, but the peak of the velocity dispersion occurs away from the galaxy's center. In our H \textsc{I} maps, its H \textsc{I} contours do show rotation, and the peak of its H \textsc{I} velocity dispersion coincides with its optical center, which would suggest that its H \textsc{I} should be classified as a rotating disk. However, since its H \textsc{I} major axis is misaligned with its optical major axis, SDSS0119+1452 may be considered to have complex kinematics according to the \citet{Y08} classification system.

SDSS0119+1452 was detected in H \textsc{I} with Arecibo in the ALFALFA
survey \citep{H11}, with an H \textsc{I} mass of $\rm 1.38 \times
10^{9}\ M_{\sun}$.  \citet{CG05} observed SDSS0119+1452 in the J=2-1
transition of CO and did not detect any CO above the $3\sigma$
level. They reported an upper limit for molecular gas mass of $\rm
1.1 \time 10^{8}\ M_{\sun}$. \citet{J15} did detect SDSS0119+1452 in
the CO J=2-1 transition; they derived a molecular gas mass of $\rm
M_{H_{2}} 1.78 \times 10^{8}\ M_{\sun}$.

SDSS0119+1452 does not have any companions that we detect in
H \textsc{I} (the H \textsc{I} clump to the galaxy's southwest does
not coincide with any optical source, and has similar velocity to the
receding edge of the galaxy's main H \textsc{I} emission, so it
appears to be associated with SDSS0119+1452). The nearby galaxies seen
in optical images are members of the higher-redshift cluster Abell 175
(SDSS0119+1452 is not a member of that cluster).

\subsection{SDSS0125+0110}

SDSS0125+0110 (ARK 044; Figure \ref{fig:0125}) has velocity contours
representative of a rotating disk, and has an easily identifiable
H \textsc{I} major axis that is coincident with its optical major
axis. Its H \textsc{I} emission is much more extended than its optical
emission, and its region of highest column density is offset from
its optical emission.

Optical images of SDSS0125+0110 show visible spiral arms. Galaxy Zoo
classifies this galaxy as a spiral ($96\%$ probability) with
anticlockwise arms ($89\%$ probability). Its center is redder than the
centers of other galaxies in our sample, and it does not have large
star-forming clumps, though it does appear to have smaller
star-forming clumps near the outer regions of its disk.

While SDSS0125+0110 does not appear to be interacting with any other
galaxies, it isn't necessarily isolated. \citet{ZM98} include
SDSS0125+0110 in the NGC 533 group using velocity analysis derived
from optical spectroscopy, though it is far enough away from NGC 533
(the group's central brightest cluster galaxy) that we do not detect
other galaxies in H \textsc{I} in our observed data. A significant
fraction of intermediate-redshift LCBGs have been found to be on the
outskirts of groups and clusters \citep{C16}, where their star
formation is enhanced as they interact with the cluster potential and
then quenched as their gas is stripped once they have passed through
the cluster. It is possible that this is the reason for SDSS0125+0110
experiencing an LCBG phase even though it does not have evidence of a
recent interaction or clumpy star formation. We note that following
SDSS-III photometry, SDSS0125+0110 is redder and has a lower B-band
surface brightness than the LCBG criteria defined by \citet{W04}.

\subsection{SDSS0728+3532}

SDSS0728+3532 (ARK 134; Figure \ref{fig:0728}) has H \textsc{I}
emission that is concentrated near its optical center, velocity
contours that are representative of a rotating disk, and velocity
dispersion contours that trace the H \textsc{I} and optical
emission. It also has an H \textsc{I} tidal tail that encompasses a
companion galaxy to its south and extends to the south and east. In
addition, it has a companion to its southwest that we detect in
H \textsc{I} and is rotating in the opposite direction as
SDSS0728+3532.

In SDSS images, the three galaxies do not appear to be obviously
interacting. The H \textsc{I} tidal tail does not have a corresponding
feature visible in SDSS optical images, with the exception of the
companion galaxy that is embedded in it. SDSS0728+3532 is symmetric in
SDSS images without visible extended low surface brightness
features. The companion galaxy also does not have any visible stellar
streams or other obvious optical signs of interactions. SDSS0728+3532
has visible spiral arms and many regions of clumpy star
formation. Galaxy Zoo 1 classifies it as a spiral galaxy ($76\%$
probability) with anticlockwise arms ($70\%$ probability). Galaxy Zoo
2 classifies it as most likely to be a barred spiral galaxy with an
unknown number of arms of medium tightness. Both companion galaxies
are blue and appear to be edge-on disk galaxies (Galaxy Zoo 1
classifies both of them as most likely to be edge-on spirals).

SDSS0728+3532 was observed in the J=2-1 transition of CO
by \citet{CG05} but was not detected. It has a $3\sigma$ upper limit
of molecular gas mass of $\rm M_{H_{2}} = 0.91 \times 10^{8}\
M_{\sun}$ (the beam size of the CO observation is not large enough to
include its companions).

\subsection{SDSS0934+0014}

SDSS0934+0014 (UGC 05097; Figure \ref{fig:0934}) has asymmetric
H \textsc{I} emission that is not much more extended than the optical
emission. The peak of its Moment 0 and Moment 2 maps coincide with the
optical galaxy, although the H \textsc{I} emission is more extended on
the galaxy's east side than its west side. Its velocity contours show
rotation, and its major axis aligns with its optical major axis. The
velocity contours near the optical center are not as closely spaced as
is typical for galaxies that are not close to face-on, though its
optical emission does not appear to be face-on. It has two companions:
SDSS J093410.47+001528.5 to the north, and UGC 05099 to the
southeast. UGC 05099 is strongly detected in H \textsc{I} emission in
the low-resolution Moment 0 map, while SDSS J093410.47+001528.5 is not
detected in H \textsc{I} in the low-resolution map shown in
Figure \ref{fig:0934}. In low-resolution maps made with less flagging
of radio interference, there is H \textsc{I} emission that coincides
with the optical position of SDSS J093410.47+001528.5, though the
column density of this emission is not high enough to distinguish it
from noise.

In optical SDSS images, SDSS0934+0014 is a blue, clumpy disk galaxy
with spiral arms. Its companions are also both disk galaxies, with its
northern companion being a blue disk galaxy fainter than
SDSS0934+0014, and its southeastern companion being a spiral galaxy
with a redder appearance. Galaxy Zoo 1 classifies SDSS0934+0014 as
uncertain, with $74\%$ of voters indicating that they don't know
whether it is a spiral or elliptical galaxy. 

SDSS0934+0014 was detected in the J=1-0, J=2-1, and J=3-2 transitions
of CO by \citet{CG05}. They find its molecular gas mass to be $\rm
M_{H_{2}} = 5.3 \times 10^{8}\ M_{\sun}$. 

\subsection{SDSS0936+0106}

SDSS0936+0106 (CGCG 007-009; Figure \ref{fig:0936}) has velocity
contours indicative of ordered rotation. The peak of its H \textsc{I}
emission is offset from the peak of its optical emission, though the
two peaks are within a beam size of each other. Its H \textsc{I} major
axis aligns with its optical major axis, and its Moment 1 map does not
show evidence of interactions. Its Moment 2 map is centrally peaked,
which is consistent with a rotating disk galaxy.

In SDSS images of SDSS0936+0106, its center is redder than the rest of
the galaxy, which is primarily made up of blue spiral arms. The spiral
arms are much brighter close to the center of the galaxy, but low
surface brightness spiral featues are visible outside of the
center. The spiral arms do not have large clumps. Galaxy Zoo 1
classifies SDSS0936+0106 as a spiral galaxy ($97\%$ probability) with
anticlockwise arms ($91\%$ probability). Galaxy Zoo 2 classifies
SDSS0936+0106 as having evidence of a disk or spiral features ($97\%$
probability) with 3-4 tightly wound spiral arms, a bulge that is just
noticeable, and no bar. Its companion appears to be a blue edge-on
disk galaxy, and is detected in H \textsc{I} in our low-resolution
map.

SDSS0936+0106 was detected in the J=1-0 transition of CO and not
detected in the J=2-1 transition by \citet{CG05}. They measured a
molecular gas mass of $\rm M_{H_{2}} = 2.0 \times 10^{8}\
M_{\sun}$. \citet{P03} identified the companion galaxy that we detect
in H \textsc{I} as being a satellite galaxy of SDSS0936+0106.

\subsection{SDSS1319+5203}

SDSS1319+5203 (SBS 1317+523B; Figure \ref{fig:1319}) is the
second-brightest in a group of three star-forming galaxies that share
a common H \textsc{I} envelope. It also contains the second-most
H \textsc{I} of the three galaxies in the envelope. Both it and the
brightest galaxy in the group, SBS 1317+523A, show evidence of
independent rotation in the Moment 1 map, while the faintest of the
three galaxies, Mrk 251, does not show clear rotation. In their common
H \textsc{I} envelope, the areas with the highest velocity dispersions
appear to be where SDSS1319+5203 interacts with each of its
companions.

SDSS1319+5203 is a compact, blue, fuzzy, relatively featureless galaxy
that appears to have a central star-forming clump. Its SDSS images do
not show signs of past interactions like stellar streams or asymmetric
features. Its brighter companion, SBS 1317+523A, appears more disklike
with clumpy star formation, and its fainter companion, Mrk 251, is
also blue, compact, and fuzzy. Galaxy Zoo 1 classifies SDSS1319+5203
as having $44\%$ probability of being elliptical, and $29\%$
probability of being a spiral galaxy. Galaxy Zoo 2 gives it a higher
probablity of being smooth ($55\%$) than having a disk or features
($45\%$).

Past studies of groups of star-forming galaxies have also identified
SDSS1319+5203 as being in a trio of field
galaxies \citep[e.g.][]{P02}. \citet{P2001} classified SDSS1319+5203
and SBS 1317+523A as a binary pair of galaxies with similar masses
that are likely triggering star formation in each other even if they
are not a merging pair. \citet{P02} identified the brightest galaxy in
the group, SBS 1317+523A, as having experienced a merger in the
past. \citet{CG05} did not detect the J=2-1 transition of CO in
SDSS1319+5203, and reported an upper limit of molecular gas of $\rm
M_{H_{2}} = 1.1 \times 10^{8}\ M_{\sun}$.

\subsection{SDSS1402+0955}

SDSS1402+0955 (NGC 5414; Figure \ref{fig:1402}) has asymmetric,
centrally-concentrated H \textsc{I} emission. Its Moment 1 map shows
that it is a rotating galaxy, though in contrast to other undisturbed
rotating galaxies like SDSS0125+0110 and SDSS0936+0106, its velocity
contours are closer together at the edges of the galaxy at each end of
its minor axis than they are in the center. This results in the
galaxy's velocity dispersion being highest at its western edge,
although the broad distribution of its velocity dispersion follows a
pattern of being lower at larger radii and higher closer to the center
of the galaxy.

SDSS1402+0955 is blue and clumpy in SDSS images. Its optical emission
is elongated and asymmetric, with SDSS images showing fainter optical
emission that extends to the northwest and southeast. The major axis
along its bright optical emission aligns with its H \textsc{I} major
axis. It is irregularly-shaped, with little obvious spiral
structure. Galaxy Zoo 1 classifies this galaxy as uncertain, with a
$12\%$ probability of having an elliptical morphology, a $47\%$
probability of having a spiral morphology, and an $18\%$ probability
of being a merger. $23\%$ of Galaxy Zoo 1 classifiers chose ``don't
know'' as its most likely morphology. Galaxy Zoo 2 classifies this
galaxy as most likely having ``features or a disk'' ($56\%$), with the
most common ``odd'' feature being its irregular shape ($92\%$ of
Galaxy Zoo 2 classifiers indicated an odd feature, and $78\%$ of
classifiers who indicated an odd feature specified an irregular
shape).

While we do not detect other galaxies in H \textsc{I} emission in the
same data cube as SDSS1402+0955, \citet{KG02} identified this galaxy
as possibly being part of the cluster MKW 12. Their velocity analysis
suggested that SDSS1402+0955 is likely part of a foreground
substructure of this cluster. This substructure is not conclusively
kinematically linked to the main cluster, and \citet{KG02}
hypothesized that it may be in the process of joining the main
cluster. This is consistent with the conclusions of \citet{C11}
and \citet{C14} that some LCBGs are making their first pass through a
cluster. SDSS1402+0955 was detected in the J=1-0 and the J=2-1
transitions of CO, with a molecular gas mass of $\rm M_{H_{2}} =
2.6 \times 10^{8}\ M_{\sun}$ \citep{CG05}.

\subsection{SDSS1507+5511}

SDSS1507+5511 (UGC 09737; Figure \ref{fig:1507}) has H \textsc{I}
emission that is elongated near the galaxy's optical emission, and
then extends asymmetrically to the northeast at a lower column
density. While this galaxy looks like it may have H \textsc{I}
emission with a morphology similar to the optical ``cometary'' or
``tadpole'' emission sometimes seen in blue compact star-forming
galaxies \citep[e.g. ][]{LT86, vdB96}, its optical emission does not
show this asymmetric morphology, and its Moment 1 map shows evidence
of rotation throughout its H \textsc{I} emission (its velocity
contours run perpendicular to the elongated concentration of
H \textsc{I} in its Moment 0 map). The galaxy's velocity dispersion
contours roughly follow its H \textsc{I} content, with a few small
``hot spots'' in its lower-column density emission to the northeast. A
companion is visible in the low-resolution Moment 0 map.

SDSS images of SDSS1507+5511 show a relatively featureless galaxy with
a central bright knot, and a stellar disk that appears to have its
light arranged in shell-like structures. Galaxy Zoo 1 classifies this
galaxy as ``uncertain'', with a $37\%$ probability of being
elliptical, and a $48\%$ probability of being a spiral galaxy. Galaxy
Zoo 2 classifies this object as likely being a disk galaxy ($67\%$)
rather than being smooth ($31\%$). It is likely to be barred (a $90\%$
probability from the classifiers who identified it as a disk galaxy),
and does not have obvious spiral arms ($83\%$ of classifiers who
identified it as a disk galaxy voted that they could not tell how many
spiral arms the galaxy had).

The mismatch between the optical and H \textsc{I} morphology in
SDSS1507+5511 looks similar to the H \textsc{I} morphology of another
galaxy, NGC 4522, that was mapped by \citet{K04}. That galaxy is not
an LCBG, but its H \textsc{I} Moment 0 map shows the elongated
emission along the galaxy's major axis with the extra emission above
the plane of the galaxy that SDSS1507+5511 also shows, and its Moment
1 map shows velocity contours that resemble a rotating disk, similar
to those of SDSS1507+5511. \citet{K04} identified the H \textsc{I}
maps of NGC 4522 as showing evidence of the H \textsc{I} undergoing
ram pressure stripping due to interactions between the galaxy's
interstellar medium and the intracluster medium of the group to which
NGC 4522 belongs. \citet{G00} identified SDSS1507+5511 as being part
of a group containing five galaxies. We detect one of the galaxies in
this group in the low-resolution Moment 0 map. The H \textsc{I} in
SDSS1507+5511 does not appear to be as strongly stripped as that of
NGC 4522 (the H \textsc{I} emission does not extend throughout NGC
4522's entire disk, while it extends far beyond the optical disk of
SDSS1507+5511 on all sides), nor is SDSS1507+5511 deficient in
H \textsc{I} as other galaxies that appear to have undergone stripping
are. \citet{V01} simulated H \textsc{I} stripping for galaxies
interacting with clusters, and found that a stripped galaxy's
H \textsc{I} becomes less extended along its disk and more elongated
in the direction of the stripping as the time after the initial
interaction with the cluster increases. SDSS1507+5511's H \textsc{I}
morphology resembles that of a galaxy viewed edge-on in their
simulation that is in the early stages of an interaction with its
cluster. If SDSS1507+5511's H \textsc{I} morphology is due to an
interaction between its ISM and the ICM of its group, this interaction
could also be the cause of its current star formation rate, consistent
with \citet{C11} and \citet{C14}.

Previous studies of LCBGs have detected atomic and molecular gas in
this galaxy. \citet{PG11} mapped the $\rm H\alpha$ emission in
SDSS1507+5511 and identified it as a rotating disk from its velocity
field and velocity dispersion. They found a dynamical mass of $\rm
M_{dyn} = 5.09 \times 10^{9}\ M_{\sun}$. \citet{CG05} detected
SDSS1507+5511 in the J=2-1 transition of CO. They found a molecular
gas mass of $\rm M_{H_{2}} = 0.29 \times 10^{8}\ M_{\sun}$.

\subsection{Mrk 325}

Mrk 325 (NGC 7673; Figure \ref{fig:mrk325}) has H \textsc{I} emission
that does not extend much farther than its optical emission. The
highest column density H \textsc{I} emission traces the galaxy's
optical barlike feature. The velocity contours in its Moment 1 map are
warped, though this galaxy shows a velocity gradient like all of the
other LCBGs in our sample. When selecting a major axis for this
galaxy, there were two possible position angles for which a major axis
could be drawn perpendicular to the velocity contours; we chose the
position angle that passed along the highest H \textsc{I} column
density and the optical bar. The other possible major axis is in the
direction of the bright clump to the northeast of the galaxy. The
galaxy's velocity dispersion distribution is similar to that of its
H \textsc{I} emission, with higher velocity dispersions coinciding
with its optical bar, though in addition it has clumps of higher
velocity dispersion that roughly coincide with bright optical
clumps. It has a larger companion, Mrk 326 (NGC 7677) that we also
detect in H \textsc{I}. The H \textsc{I} emission in Mrk 326 is
symmetric, disk-shaped, and centrally-concentrated. Its velocity
contours show evidence of Mrk 326 being dominated by ordered
rotation, though \citet{N97} found its H \textsc{I} emission shows evidence of warping.

In SDSS images, Mrk 325 is very blue, clumpy, asymmetric, and has
visible irregular spiral structure. It does not have any Galaxy Zoo 1
or 2 classifications. It has a few large, bright, blue clumps, with a
few smaller clumps along its spiral arms. It also appears to have a
bar (or line of central clumps that are not as blue). Its spiral arms
are asymmetric, with no obvious spiral structure southwest of the
bar. Mrk 325 also has lower surface brightness optical emission
surrounding the galaxy. The elongated, redder galaxy to the immediate
north of Mrk 325 in optical images is at a higher redshift (SDSS
calculates a photometric redshift of $\rm z = 0.441 \pm 0.0251$). In
contrast to the morphology of Mrk 325, its nearest companion, Mrk 326,
is a redder spiral galaxy with a bright center, two well-defined
spiral arms, and no visible clumps.

Mrk 325 is a well-studied galaxy that \citet{PG10} and \citet{CM11}
called a ``prototypical LCBG'' due to its clumpy, irregular appearance
and high star formation rate. Past studies have found that its star
formation is concentrated in 5-6 large clumps that are comprised of
smaller star clusters. \citet{PG03} measured 87 star-forming clusters
that show $H\alpha$ emission, and found that only $9\%$ of the
galaxy's ionized gas is diffuse rather than concentrated in the
clumps. \citet{H02} found that one of the large clumps, Clump F, is
likely one large object rather than a collection of clusters, and that
the clumps only extend to half of the galaxy's optical radius. Beyond
the clumps, the galaxy's disk is smooth, faint, and
circular. Connected to this outer disk is an optical shell \citep{D84}
that resembles those found around galaxies that are candidates for
being in the late stages of a past major
merger \citep{H02}. \citet{HG99} argue that an ongoing interaction
with the most obvious candidate, Mrk 326, is unlikely due to the lack
of a tidal tail associated with Mrk 325, as well as the symmetric and
undisturbed morphology of Mrk 326. The most likely progenitor of both
the low surface brightness shell and the star formation activity is
likely to be a past minor merger with a dwarf galaxy \citep{HG99, H02,
CM11}. The brightest clump in Mrk 325, Clump B, has a decoupled
kinematic component \citep[][we also observe a kinematic component
in the direction of Clump B]{PG10}. Clump B also has the highest
$H\alpha$ content of all of the galaxy's clumps \citep[including the
central clump ][]{CM11}. The authors of those studies suggest that
Clump B may be associated with the dwarf galaxy merger candidate.

Past studies of Mrk 325 have found small velocity
gradients \citep{DA82, N97}, though the galaxy also appears to be
face-on. \citet{PG11} classified Mrk 325 as a rotating disk from maps
of $\rm H\alpha$ emission. \citet{P01} found an H \textsc{I} rotation
velocity of $\rm 116\ km\ s^{-1}$ when corrected for inclination, with
an H \textsc{I} radius of 8.3 kpc and a $\rm M_{dyn}$ of $2.5 \times
10^{10}\ M_{\sun}$. \citet{N97} mapped Mrk 325 with the VLA in its
most compact configuration and found that its H \textsc{I} is offset
to the west of the centroid of its low-luminosity optical emission and
has a rotation axis whose position angle changes with increasing
radius. They measured $\rm M_{HI} = 3.6 \times 10^{9}\
M_{\sun}$ \citep{N97}. \citet{CG05} detected the J = 1-0 and J = 2-1
transitions of CO in Mrk 325, and measured a molecular gas mass of
$\rm M_{H_2} = 1.6 \times 10^{8}\ M_{\sun}$. \citet{CG07} mapped Mrk
325 in H \textsc{I} with the VLA with a beam size of $20\arcsec$ in C
configuration, as well as in CO J = 1-0. They measured $\rm M_{HI} =
6.3 \times 10^{9}\ M_{\sun}$, and found that the molecular gas is
concentrated near the optical bar. Our H \textsc{I} map of Mrk 325
combines the \citet{CG07} map with new B configuration data and
archival D configuration data that was publised by \citet{N97}.

\begin{longrotatetable}
\begin{deluxetable*}{lllllllll}

\tablecolumns{9}
\tablewidth{0pt}
\tabletypesize{\small}
\tablecaption{Optical properties \label{fig:obssummary}}
\tablehead{\colhead{Source\tablenotemark{a}} & \colhead{Common} &
  \colhead{D\tablenotemark{b}} &
  \colhead{$\rm{R_{eff}(B)}$\tablenotemark{c}} &
  \colhead{$\rm{B-V}$\tablenotemark{d}} &
  \colhead{$\rm{m_{B}}$\tablenotemark{e}} &
  \colhead{$\rm{M_{B}}$\tablenotemark{f}} &
  \colhead{$\rm{SB_{e}(B)}$\tablenotemark{g}} &
  \colhead{Hubble\tablenotemark{h}} \\ \colhead{} & \colhead{Name} &
  \colhead{(Mpc)} & \colhead{(kpc)} & \colhead{} & \colhead{} &
  \colhead{} & \colhead{($\rm{B\ mag\ \cdot arcsec^{-2}}$)} &
  \colhead{Type}}

\startdata 
SDSS J011932.82+145220.7 & NGC 469 & 54.6 & 1.4 $\pm$ 0.1 & 0.37 $\pm$
0.02 & 14.86 $\pm$ 0.02 & -18.8 $\pm$ 0.2 & 20.48 $\pm$ 0.03 & S? \\
SDSS J012539.72+011041.1\tablenotemark{i} & ARK 044 & 80.4 & 2.4 $\pm$
0.2 & 0.62 $\pm$ 0.01 & 15.46 $\pm$ 0.01 & -19.0 $\pm$ 0.2 & 21.29
$\pm$ 0.01 & Sbc \\
SDSS J072849.75+353255.1 & ARK 134 & 58.6 & 1.4 $\pm$ 0.1 & 0.41 $\pm$
0.01 & 14.71 $\pm$ 0.01 & -19.1 $\pm$ 0.2 & 20.19 $\pm$ 0.01 & SBbc \\
SDSS J093410.54+001430.3 & UGC 05097 & 75.8 & 2.0 $\pm$ 0.2 & 0.46
$\pm$ 0.01 & 14.21 $\pm$ 0.01 & -20.2 $\pm$ 0.2 & 19.78 $\pm$ 0.01 &
Sa \\
SDSS J093635.36+010659.8 & CGCG 007-009 & 76.2 & 2.5 $\pm$ 0.3 & 0.51
$\pm$ 0.01 & 14.71 $\pm$ 0.01 & -19.7 $\pm$ 0.2 & 20.83 $\pm$ 0.02 &
Sb \\
SDSS J131949.94+520341.2 & SBS1317+523B & 69.8 & 1.2 $\pm$ 0.1 & 0.30
$\pm$ 0.01 & 15.50 $\pm$ 0.01 & -18.7 $\pm$ 0.2 & 20.10 $\pm$ 0.01 &
E \\
SDSS J140203.52+095545.5 & NGC 5414 & 65.0 & 1.7 $\pm$ 0.2 & 0.45
$\pm$ 0.01 & 13.94 $\pm$ 0.01 & -20.2 $\pm$ 0.2 & 19.60 $\pm$ 0.03 &
E? \\
SDSS J150748.34+551108.8 & UGC 09737 & 49.3 & 1.9 $\pm$ 0.2 & 0.44
$\pm$ 0.01 & 14.47 $\pm$ 0.01 & -19.0 $\pm$ 0.2 & 20.87 $\pm$ 0.02 &
SBcd \\
Mrk 325\tablenotemark{j} & NGC 7673 & 44.0 & 1.2 $\pm$ 0.1 & 0.41
$\pm$ 0.01 & 13.59 $\pm$ 0.01 & -19.6 $\pm$ 0.2 & 19.40 $\pm$ 0.02 &
Sc \\
\enddata

\tablenotetext{a}{\footnotesize SDSS source names are of the form SDSS
  JHHMMSS.SS+DDMMSS.S and are hereafter shortened to SDSS HHMM+DDMM.}

\tablenotetext{b}{\footnotesize Distances were taken from NED's
  luminosity distances using $\rm H_{0} =
  70\ km\ s^{-1}\ Mpc^{-1}$. NED assumes a $10\%$ uncertainty
    on its luminosity distances, which we use when propagating
    uncertainties.}

\tablenotetext{c}{\footnotesize Half-light radii in the B band
  calculated using SDSS g and r Petrosian radii using
  SDSS-III photometry.}

\tablenotetext{d}{\footnotesize Colors calculated from SDSS g and r
  magnitudes using SDSS-III photometry. Magnitudes are
  adjusted for extinction using SDSS-III Galactic reddening
  correction values.}

\tablenotetext{e}{\footnotesize B-band apparent magnitudes calculated
  from SDSS g and r magnitudes using SDSS-III photometry and
  corrected for extinction.}

\tablenotetext{f}{\footnotesize B-band absolute magnitudes calculated
  from $\rm m_{B}$ and luminosity distances.}

\tablenotetext{g}{\footnotesize Surface brightnesses in the B band
  calculated from $\rm M_{B}$, $\rm R_{eff}(B)$, and luminosity
  distances.}

\tablenotetext{h}{\footnotesize From Hyperleda.}

\tablenotetext{i}{\footnotesize The color and surface brightness of
  SDSS 0125+0110 are outside of the optical parameters that
  \citet{W04} use to define LCBGs when we use DR9 photometry
  to calculate them. As these properties are within the LCBG optical
  parameters using DR4 photometry, we do not exclude this galaxy from
  our analysis. }

\tablenotetext{j}{\footnotesize Mrk 325 is at the J2000 (RA, Dec)
  position (23:27:41.0, +23:35:21).}

\end{deluxetable*}
\end{longrotatetable}

\begin{longrotatetable}
\begin{deluxetable*}{ccccccccccc}
\tablecolumns{11}
\tablewidth{0pt}
\tabletypesize{\footnotesize}
\tablecaption{Imaging parameters \label{fig:cubes}}
\tablehead{
\colhead{Galaxy} &
\colhead{High-res beam} &
\colhead{Robustness} &
\colhead{UV taper} &
\colhead{UV range} &
\colhead{$\#$ CLEAN} &
\colhead{Low-res beam} &
\colhead{Robustness} &
\colhead{UV taper} &
\colhead{UV range} &
\colhead{$\#$ CLEAN} \\
\colhead{} &
\colhead{($\rm arcsec^{2}$)} &
\colhead{} &
\colhead{($\rm k\lambda$)} &
\colhead{($\rm k\lambda$)} &
\colhead{iterations} &
\colhead{($\rm arcsec^{2}$)} &
\colhead{} &
\colhead{($\rm k\lambda$)} &
\colhead{($\rm k\lambda$)} &
\colhead{iterations} 
}
\startdata
SDSS0119+1452 & 13 $\times$ 13 & 2 & 70 $\times$ 70 & 100 & 50 & 52 $\times$ 47 & 5 & 3 $\times$ 3 & 5 & 50\\
SDSS0125+0110 & 22 $\times$ 13 & 5 & 30 $\times$ 30 & 50 & 6000 & 54 $\times$ 45 & 5 & 3 $\times$ 3 & 5 & 35\\
SDSS0728+3532 & 13 $\times$ 8 & 5 & 70 $\times$ 70 & 100 & 180 & 55 $\times$ 53 & 5 & 3 $\times$ 3 & 5 & 60\\
SDSS0934+0014 & 20 $\times$ 20 & 5 & 30 $\times$ 30 & 40 & 200 & 75 $\times$ 49 & 5 & 3 $\times$ 3 & 5 & 25\\
SDSS0936+0106 & 12 $\times$ 11 & 3 & 70 $\times$ 70 & 100 & 100 & 55 $\times$ 46 & 5 & 3 $\times$ 3 & 5 & 950\\
SDSS1319+5203 & 15 $\times$ 12 & 5 & 30 $\times$ 30 & 50 & 400 & 63 $\times$ 50 & 5 & 3 $\times$ 3 & 5 & 50\\
SDSS1402+0955 & 23 $\times$ 14 & 2 & 30 $\times$ 30 & 50 & 6000 & 53 $\times$ 53 & 5 & 7 $\times$ 7 & 10 & 20\\
SDSS1507+5511 & 11 $\times$ 9 & 5 & 30 $\times$ 30 & 50 & 130 & 52 $\times$ 51 & 5 & 3 $\times$ 3 & 5 & 40\\
Mrk 325\tablenotemark{a} & 6 $\times$ 6 & 0 & \nodata & \nodata & 50 \\
\enddata

\tablenotetext{a}{The data cube for Mrk 325 is from a combination of B, C, and D VLA configuration observations.}
\end{deluxetable*}
\end{longrotatetable}

\begin{longrotatetable}
\begin{deluxetable*}{lllccccc}
\tablecolumns{8}
\tablewidth{0pt}
\tablecaption{Companion sources visible in maps \label{fig:companions}}
\tablehead{
\colhead{LCBG} &
\colhead{Companion Name} &
\colhead{RA} &
\colhead{Dec} &
\colhead{Separation} &
\colhead{Separation\tablenotemark{a}} &
\colhead{$\rm V_{sys}$\tablenotemark{b}} &
\colhead{Detected?} \\
\colhead{} &
\colhead{} &
\colhead{(J2000)} &
\colhead{(J2000)} &
\colhead{(arcminutes)} &
\colhead{($\rm R_{eff}(B)$)} &
\colhead{($\rm km \  s^{-1}$)} &
\colhead{}
}

\startdata
SDSS0119+1452 & NGC 471 & 01:19:59.6 & +14:47:10 & 8.2 & 90 & 4137 & N\\
SDSS0728+3532 & GALEXASCJ072841.30+353206.1\tablenotemark{c} & 07:28:41.3 & +35:32:06 & 2.0 & 24 & 3930 & Y\\
 & SDSS072849.02+353124.6 & 07:28:49.0 & +35:31:24 & 1.5 & 18 & 4010 & Y\\
SDSS0934+0014 & SDSS J093410.47+001528.5 & 09:34:10.5 & +00:15:29 & 1.0 & 11 & 4665 & N\\
 & UGC 05099 & 09:34:34.2 & +00:05:23 & 11 & 123 & 4954 & Y\\
SDSS0936+0106 & SDSS093626.68+011128.8 & 09:36:26.7 & +01:11:28 & 5.0 & 44 & 4900 & Y\\
SDSS1319+5203 & SBS1317+520\tablenotemark{d} & 13:19:46.2 & +51:48:06 & 15.6 & 270 & \nodata & Y\\ 
 & SBS1317+523A & 13:19:47.5 & +52:04:13 & 0.6 & 11 & 4588 & Y\\
 & Mrk 251 & 13:20:01.0 & +52:03:03 & 1.8 & 32 & 4581 & Y\\
SDSS1507+5511 & SDSS150804.21+551954.0 & 15:08:04.2 & +55:19:54 & 9.0 & 69 & 3385 & Y\\ 
Mrk 325 & Mrk 326 & 23:28:06.1 & +23:31:52 & 6.7 & 70 & 3519 & Y\\
\enddata

\tablenotetext{a}{Projected separation from target galaxy in multiples of the target galaxy's $\rm R_{eff}(B)$.}
\tablenotetext{b}{$\rm V_{sys}$ is approximated from moment maps if detected, or taken from NED if not detected.}
\tablenotetext{c}{SDSS classifies this object as a star; NED classifies it as a UV source.}
\tablenotetext{d}{NED classifies this object as a QSO at $\rm z=1.06$.}
\end{deluxetable*}
\end{longrotatetable}

\clearpage
\begin{deluxetable*}{ccccccc}

\tablecolumns{7}
\tablewidth{0pt}
\tabletypesize{\footnotesize}
\tablecaption{LCBG H I Profile Properties \label{fig:spectrum}}
\tablehead{
\colhead{Galaxy} &
\colhead{$\rm V_{sys}$\tablenotemark{a}} &
\colhead{$\rm W_{20}$\tablenotemark{b}} &
\colhead{$\rm \int S dv$} &
\colhead{$\rm M_{HI}$} &
\colhead{$\rm \frac{M_{HI_{GMRT}}}{M_{HI_{GBT}}}$\tablenotemark{c}} &
\colhead{Companion in }\\
\colhead{} &
\colhead{(km $\rm s^{-1}$)} &
\colhead{(km $\rm s^{-1}$)} &
\colhead{(Jy km $\rm s^{-1}$)} &
\colhead{$\rm (10^{9}\ M_{\odot})$} &
\colhead{} &
\colhead{GBT beam?\tablenotemark{d}}
}
\startdata
SDSS0119+1452 & 4123 $\pm$ 7 & 52.1 $\pm$ 13.6 & 1.0 $\pm$ 0.4 & 0.85 $\pm$ 0.29 & 0.4 & Y\\
SDSS0125+0110 & 5875 $\pm$ 7 & 126.8 $\pm$ 13.8 & 3.3 $\pm$ 0.3 & 5.4 $\pm$ 0.5 & \nodata & \nodata\\
SDSS0728+3532\tablenotemark{e} & 3962 $\pm$ 7 & 166.1 $\pm$ 13.6 & 9.8 $\pm$ 1.2 & 7.4 $\pm$ 0.9 & 1.2 & N\\
SDSS0934+0014 & 4903 $\pm$ 7 & 125.8 $\pm$ 13.7 & 2.2 $\pm$ 0.4 & 2.5 $\pm$ 0.5 & 0.5 & Y\\
SDSS0936+0106 & 4909 $\pm$ 7 & 181.2 $\pm$ 13.7 & 2.0 $\pm$ 0.4 & 2.3 $\pm$ 0.5 & 0.6 & Y\\
SDSS1319+5203\tablenotemark{e} & 4607 $\pm$ 7 & 139.4 $\pm$ 13.7 & 11.2 $\pm$ 0.7 & 11.4 $\pm$ 0.7 & 1.4 & Y\\
SDSS1402+0955 & 4251 $\pm$ 7 & 262.3 $\pm$ 13.7 & 4.9 $\pm$ 1.1 & 4.3 $\pm$ 0.9 & 0.7 & Y\\
SDSS1507+5511 & 3358 $\pm$ 7 & 124.2 $\pm$ 13.6 & 3.8 $\pm$ 0.4 & 2.0 $\pm$ 0.2 & 1.0 & N\\
Mrk 325\tablenotemark{f} & 3364 $\pm$ 5 & 54.1 $\pm$ 10.3 & 3.2 $\pm$ 0.4 & 1.7 $\pm$ 0.2 & 0.3 & Y\\
\enddata

\tablenotetext{a}{$\rm V_{sys}$ is measured halfway between the
  channels used to measure $\rm W_{20}$. The reported uncertainty is
  half of a channel width.}

\tablenotetext{b}{$\rm W_{20}$ is corrected for random motions
  following Equation 12 of \citet{TF85}. No correction for inclination
  angle has been made. The reported uncertainty is one channel width.}

\tablenotetext{c}{$\rm M_{HI_{GBT}}$ values are taken from
  \citet{CG04}.}

\tablenotetext{d}{Taken from Table 1 of \citet{CG04}.}

\tablenotetext{e}{Properties listed are for the entire H \textsc{I}
  envelope, which contains multiple galaxies.}

\tablenotetext{f}{Measurements are taken from the high-resolution
  cube. For data measured from a low-resolution cube, see Table 4 of
  \citet{N97}. Those authors found $\rm M_{HI} = 3.6 \times
  10^{9}\ M_{\sun}$, which is $60\%$ of the \citet{CG04} single-dish
  value.}

\end{deluxetable*}

\begin{longrotatetable}
\begin{deluxetable*}{ccccccccccc}
\tablecolumns{10} 
\tablewidth{0pt}
\tabletypesize{\footnotesize} 
\tablecaption{Velocities and
    Dynamical Masses from Cuts Along Major Axes \label{fig:pvcuts}} 
\tablehead{
\colhead{Galaxy} & 
\colhead{$\rm V_{opt}$\tablenotemark{a}} &
\colhead{$\rm V_{sys}$\tablenotemark{b}} &
\colhead{D\tablenotemark{c}}& 
\colhead{$\rm R_{25}$\tablenotemark{d}} & 
\colhead{$\rm V_{rot}(R_{25})$\tablenotemark{e}} & 
\colhead{$\rm R_{H I}$\tablenotemark{f}} & 
\colhead{$\rm V_{rot}(R_{H I})$\tablenotemark{g}} &
\colhead{$i_{opt}$\tablenotemark{h}} & 
\colhead{$\rm M_{dyn}(R_{HI})$} \\
\colhead{} &
\colhead{(km $\rm s^{-1}$)} &
\colhead{(km $\rm s^{-1}$)} &
\colhead{(Mpc)}&
\colhead{(kpc)} &
\colhead{(km $\rm s^{-1}$)} &
\colhead{(kpc)} &
\colhead{(km $\rm s^{-1}$)} &
\colhead{(deg)} &
\colhead{($\rm 10^{10}\ M_{\odot}$)}
}
\startdata
SDSS0119+1452\tablenotemark{i} & 4118.6 $\pm$ 13.6 & 4118.6 $\pm$ 13.6
& 58.8 $\pm$ 0.2 & 7.3 $\pm$ 1.1 & 27.1 $\pm$ 13.6 & 6.0 $\pm$ 0.1 &
27.1 $\pm$ 13.6 & 61.5 $\pm$ 0.2 & 0.13 $\pm$ 0.14\\
SDSS0125+0110 & 5877.1 $\pm$ 13.7 & 5877.1 $\pm$ 13.7 & 84.0 $\pm$ 0.2
& 7.9 $\pm$ 1.4 & 68.6 $\pm$ 13.7 & 20.9 $\pm$ 0.2 & 68.6 $\pm$ 13.7 &
35.9 $\pm$ 0.5 & 6.7 $\pm$ 9.8\\
SDSS0728+3532 & 3944.1 $\pm$ 13.5 & 3964.4 $\pm$ 13.5 & 56.6 $\pm$ 0.2
& 5.3 $\pm$ 1.1 & 54.2 $\pm$ 13.5 & 14.2 $\pm$ 0.2 & 74.5 $\pm$ 13.5 &
42.7 $\pm$ 0.3 & 4.0 $\pm$ 3.2\\
SDSS0934+0014 & 4905.8 $\pm$ 13.6 & 4926.3 $\pm$ 13.6 & 70.4 $\pm$ 0.2
& 6.5 $\pm$ 1.2 & 40.9 $\pm$ 13.6\tablenotemark{j} & 6.1 $\pm$ 0.7 &
47.7 $\pm$ 13.6 & 56.4 $\pm$ 0.2 & 0.47 $\pm$ 0.29\\
SDSS0936+0106 & 4883.5 $\pm$ 13.1 & 4909.7 $\pm$ 13.1 & 70.1 $\pm$ 0.2
& 7.4 $\pm$ 1.2 & 91.7 $\pm$ 13.1 & 15.0 $\pm$ 0.2 & 104.8 $\pm$ 13.1
& 48.5 $\pm$ 0.3 & 6.9 $\pm$ 4.1\\
SDSS1319+5203 & 4657.4 $\pm$ 13.6 & 4623.4 $\pm$ 13.6 & 66.0 $\pm$ 0.2
& 5.4 $\pm$ 1.8 & 47.6 $\pm$ 13.6 & 15.8 $\pm$ 0.4 & 20.4 $\pm$ 13.6 &
43.2 $\pm$ 0.5 & 0.33 $\pm$ 0.56\\
SDSS1402+0955 & 4213.0 $\pm$ 13.6 & 4233.4 $\pm$ 13.6 & 60.5 $\pm$ 0.2
& 8.6 $\pm$ 1.1 & 115.3 $\pm$ 13.6 & 10.5 $\pm$ 0.2 & 115.3 $\pm$ 13.6
& 51.2 $\pm$ 0.2 & 5.3 $\pm$ 2.1\\
SDSS1507+5511 & 3347.9 $\pm$ 13.5 & 3320.9 $\pm$ 13.5 & 47.4 $\pm$ 0.2
& 7.8 $\pm$ 0.8 & 67.4 $\pm$ 13.5 & 12.1 $\pm$ 0.3 & 80.9 $\pm$ 13.5 &
54.8 $\pm$ 0.2 & 2.8 $\pm$ 1.2\\
Mrk 325\tablenotemark{i} & 3373.9 $\pm$ 10.3 & 3368.8 $\pm$ 10.3 &
48.1 $\pm$ 0.1 & 7.2 $\pm$ 0.8 & 25.8 $\pm$ 10.3 & 5.4 $\pm$ 0.2 &
25.8 $\pm$ 10.3 & 28.5 $\pm$ 0.4 & 0.37 $\pm$ 0.66\\
\enddata
\tablenotetext{a}{Velocities are measured at the position of the optical galaxy along the major axis.}
\tablenotetext{b}{Systemic velocities are measured at the halfway point of the major axis along the H \textsc{I} diameter.}
\tablenotetext{c}{Distances are $\rm V_{sys} / 70\ km\ s^{-1}\ Mpc$.}
\tablenotetext{d}{$\rm R_{25}(B)$ taken from Hyperleda.}
\tablenotetext{e}{Uncorrected rotation velocity at $\rm R_{25}(B)$.}
\tablenotetext{f}{$\rm R_{HI}$ is the distance between the optical center of the galaxy and the contour with a column density of 1 $\rm M_{\sun}\ pc^{-2}$.}
\tablenotetext{g}{Uncorrected rotation velocity at $\rm R_{HI}$.}
\tablenotetext{h}{Inclinations calculated using the major and minor axis lengths measured in the SDSS \textit{i} band using an exponential disk profile.}
\tablenotetext{i}{SDSS0119+1452 and Mrk 325 are less extended in H \textsc{I} than in the optical, so $\rm V_{rot}(R_{25})$ is taken to be the velocity at $\rm R_{HI}$.}
\tablenotetext{j}{The H \textsc{I} of SDSS0934+0014 is less extended than the optical galaxy on one side. We took the velocity at the side where the H \textsc{I} is more extended than the optical galaxy.}
\end{deluxetable*}
\end{longrotatetable}

\begin{deluxetable*}{ccccccc}
\tablecolumns{7}
\tablewidth{0pt}
\tabletypesize{\small}
\tablecaption{Velocity dispersions \label{fig:sigmatable}}
\tablehead{
\colhead{Galaxy} &
\colhead{$\rm \sigma_{R_{eff}}$\tablenotemark{a}} &
\colhead{$\rm V_{rot}^{R_{eff}}$\tablenotemark{b}} &
\colhead{$\rm \frac{V_{rot}^{R_{eff}}}{\sigma^{R_{eff}}}$} &
\colhead{$\rm \sigma_{R_{25}}$\tablenotemark{c}} &
\colhead{$\rm V_{rot}^{R_{25}}$\tablenotemark{d}} &
\colhead{$\rm \frac{V_{rot}^{R_{25}}}{\sigma^{R25}}$} \\
\colhead{} &
\colhead{($\rm km\ s^{-1}$)} &
\colhead{($\rm km\ s^{-1}$)} &
\colhead{} &
\colhead{($\rm km\ s^{-1}$)} &
\colhead{($\rm km\ s^{-1}$)} &
\colhead{}\\
\hline
\colhead{ } &
\colhead{$\rm \sigma_{R_{HI}}$\tablenotemark{e}} &
\colhead{$\rm \sigma_{>R_{25}}$\tablenotemark{f}} &
\colhead{$\rm V_{rot}^{R_{HI}}$\tablenotemark{g}} &
\colhead{$\rm \frac{V_{rot}^{R_{HI}}}{\sigma^{R_{HI}}}$} &
\colhead{$\rm \frac{V_{rot}^{R_{HI}}}{\sigma_{>R_{25}}}$} &
\colhead{$\rm t_{ins}$\tablenotemark{h}}\\
\colhead{} &
\colhead{($\rm km\ s^{-1}$)} &
\colhead{($\rm km\ s^{-1}$)} &
\colhead{($\rm km\ s^{-1}$)} &
\colhead{} &
\colhead{} &
\colhead{(Gyr)}
}
\startdata
SDSS0119+1452\tablenotemark{i} & 14.1 $\pm$ 6.5 & 7.7 $\pm$ 15.5 &
0.55 $\pm$ 1.13 & 12.1 $\pm$ 6.7 & 30.8 $\pm$ 15.9 & 2.5 $\pm$ 1.9 \\*
& 12.6 $\pm$ 6.3 & 29.5 $\pm$ 3.4 & 30.8 $\pm$ 15.9 & 2.4 $\pm$ 1.8 &
1.0 $\pm$ 0.6 & 1.2 $\pm$ 1.3\\*
SDSS0125+0110\tablenotemark{j} & $<$ 9.6 & 70.2 $\pm$ 54.7 & \nodata &
20.1 $\pm$ 4.3 & 117.0 $\pm$ 85.7 & 5.8 $\pm$ 4.4 \\* & 11.2 $\pm$ 6.6
& 8.7 $\pm$ 5.2 & 117.0 $\pm$ 85.7 & 10.4 $\pm$ 9.8 & 13.4 $\pm$ 12.7
& 19 $\pm$ 27\\*
SDSS0728+3532 & 30.0 $\pm$ 6.1 & 30.0 $\pm$ 22.7 & 1.0 $\pm$ 0.8 &
33.0 $\pm$ 5.2 & 80.0 $\pm$ 35.1 & 2.4 $\pm$ 1.1 \\* & 20.6 $\pm$ 11.1
& 10.0 $\pm$ 7.3 & 109.9 $\pm$ 44.4 & 5.3 $\pm$ 3.6 & 11.0 $\pm$ 9.1 &
3.7 $\pm$ 4.2\\*
SDSS0934+0014 & 37.5 $\pm$ 5.4 & 8.2 $\pm$ 16.4 & 0.22 $\pm$ 0.44 &
34.6 $\pm$ 8.7 & 49.1 $\pm$ 17.3 & 1.4 $\pm$ 0.6 \\* & 35.3 $\pm$ 8.2
& 12.3 $\pm$ 13.6 & 57.3 $\pm$ 17.6 & 1.6 $\pm$ 0.6 & 4.7 $\pm$ 5.3 &
0.28 $\pm$ 0.16\\*
SDSS0936+0106 & 28.3 $\pm$ 7.3 & 70.0 $\pm$ 25.7 & 2.5 $\pm$ 1.1 &
30.5 $\pm$ 10.0 & 122.5 $\pm$ 37.3 & 4.0 $\pm$ 1.8 \\* & 21.0 $\pm$
12.7 & 9.4 $\pm$ 7.4 & 140.0 $\pm$ 41.5 & 6.7 $\pm$ 4.5 & 14.9 $\pm$
12.5 & 4.8 $\pm$ 5.9\\*
SDSS1319+5203 & 29.2 $\pm$ 13.7 & 9.9 $\pm$ 20.6 & 0.34 $\pm$ 0.72 &
39.5 $\pm$ 7.0 & 69.5 $\pm$ 42.8 & 1.8 $\pm$ 1.1 \\* & 24.7 $\pm$ 14.3
& 16.8 $\pm$ 12.6 & 29.8 $\pm$ 25.7 & 1.2 $\pm$ 1.3 & 1.8 $\pm$ 2.0 &
0.77 $\pm$ 1.11\\*
SDSS1402+0955 & 33.1 $\pm$ 7.2 & 43.5 $\pm$ 18.8 & 1.3 $\pm$ 0.6 &
31.5 $\pm$ 14.5 & 148.0 $\pm$ 29.4 & 4.7 $\pm$ 2.4 \\* & 27.1 $\pm$
16.1 & 7.3 $\pm$ 9.7 & 148.0 $\pm$ 29.4 & 5.5 $\pm$ 3.4 & 20.3 $\pm$
27.2 & 2.1 $\pm$ 2.5\\*
SDSS1507+5511 & 18.3 $\pm$ 3.0 & 8.3 $\pm$ 16.6 & 0.45 $\pm$ 0.91 &
18.5 $\pm$ 7.5 & 82.5 $\pm$ 20.5 & 4.5 $\pm$ 2.1 \\* & 15.2 $\pm$ 8.3
& 8.7 $\pm$ 8.1 & 99.0 $\pm$ 22.0 & 6.5 $\pm$ 3.8 & 11.4 $\pm$ 10.9 &
5.2 $\pm$ 5.8\\*
Mrk 325\tablenotemark{i} & 9.7 $\pm$ 6.2 & 32.5 $\pm$ 34.0 &
3.4 $\pm$ 4.1 & 13.2 $\pm$ 6.0 & 54.1 $\pm$ 48.7 &
4.1 $\pm$ 4.1 \\* & 13.4 $\pm$ 5.9 & 13.1 $\pm$ 6.3 &
54.1 $\pm$ 48.7 & 4.0 $\pm$ 4.0 & 4.1 $\pm$
  4.2 & 1.6 $\pm$ 2.0\\*
\enddata

\tablenotetext{a}{\small $\rm \sigma_{R_{eff}}$ is the average value of the Moment 2 map within $\rm R_{eff}$.}
\tablenotetext{b}{\small $\rm V_{rot}^{R_{eff}}$ is the rotation velocity measured at $\rm R_{eff}$ corrected for inclination.}
\tablenotetext{c}{\small $\rm \sigma_{R_{25}}$ is the average value of the Moment 2 map within $\rm R_{25}$.}
\tablenotetext{d}{\small $\rm V_{rot}^{R_{25}}$ is the rotation velocity measured at $\rm R_{25}$ corrected for inclination.}
\tablenotetext{e}{\small $\rm \sigma_{R_{HI}}$ is the average value of the Moment 2 map within $\rm R_{HI}$.}
\tablenotetext{f}{\small $\rm \sigma_{>R25}$ is the average value of the Moment 2 map outside of $\rm R_{25}$.}
\tablenotetext{g}{\small $\rm V_{rot}^{R_{HI}}$ is the rotation velocity measured at $\rm R_{HI}$ corrected for inclination}
\tablenotetext{h}{\small $\rm t_{ins}$ is the inspiral time for clumps to reach the center of a rotating disk.}
\tablenotetext{i}{\small $\rm R_{25}$ is larger than \textbf{$\rm R_{HI}$} for SDSS0119+1452 and Mrk 325.}
\tablenotetext{j}{\small The correction for beam smearing for SDSS0125+0110 is larger than the measured \textbf{$\rm \sigma_{R_{eff}}$}. The reported value is a $\rm 3\sigma$ upper limit.}

\end{deluxetable*}

\begin{longrotatetable}
\begin{deluxetable*}{cccccccccc}
\tablecolumns{10}
\tablewidth{0pt}
\tabletypesize{\footnotesize}
\tablecaption{Comparison of H \textsc{I} properties to those derived from single dish data \label{fig:compare}}
\tablehead{
\colhead{Galaxy} &
\colhead{$\rm V_{rot}^{G04}$\tablenotemark{a}} &
\colhead{$\rm V_{rot}(R_{HI})$\tablenotemark{b}} &
\colhead{$\rm R_{HI}^{G04}$\tablenotemark{c}} &
\colhead{$\rm R_{HI}$} &
\colhead{$\rm M_{dyn}^{G04}(R_{HI})$\tablenotemark{d}} &
\colhead{$\rm M_{dyn}(R_{HI})$\tablenotemark{e}} &
\colhead{$\rm M_{*}$} &
\colhead{$\rm f_{HI}^{G04}$\tablenotemark{f}} &
\colhead{$\rm f_{HI}$\tablenotemark{g}} \\
\colhead{} &
\colhead{($\rm km \ s^{-1}$)} &
\colhead{($\rm km \ s^{-1}$)} &
\colhead{(kpc)} &
\colhead{(kpc)} &
\colhead{($\rm 10^{10}\ M_{\sun}$)} &
\colhead{($\rm 10^{10}\ M_{\sun}$)} &
\colhead{($\rm 10^{9}\ M_{\sun}$)} &
\colhead{} &
\colhead{} 
}
\startdata
SDSS0119+1452 & 115 & 31 $\pm$ 16 & 11.2 & 6.0 $\pm$ 0.1 & 3.3 & 0.13
$\pm$ 0.14 & 4.1 $\pm$ 0.5 & 0.06 & 0.63 $\pm$ 0.69\\
SDSS0125+0110 & \nodata & 117 $\pm$ 86 & \nodata & 20.9 $\pm$ 0.2 &
\nodata & 6.7 $\pm$ 9.8 & 4.5 $\pm$ 0.4 & \nodata & 0.08 $\pm$ 0.12\\
SDSS0728+3532\tablenotemark{h} & 160 & 110 $\pm$ 44 & 10.8 & 14.2
$\pm$ 0.2 & 6.7 & 4.0 $\pm$ 3.2 & 4.4 $\pm$ 0.3 & 0.09 & 0.19 $\pm$
0.15\\
SDSS0934+0014 & 189 & 57 $\pm$ 18 & 13.6 & 6.1 $\pm$ 0.7 & 10.8 & 0.47
$\pm$ 0.29 & 15.5 $\pm$ 0.9 & 0.05 & 0.53 $\pm$ 0.35\\
SDSS0936+0106 & 146 & 140 $\pm$ 41 & 14.2 & 15.0 $\pm$ 0.2 & 6.3 & 6.9
$\pm$ 4.0 & 7.9 $\pm$ 0.7 & 0.06 & 0.034 $\pm$ 0.021\\
SDSS1319+5203\tablenotemark{h} & 120 & 30 $\pm$ 26 & 10.6 & 15.8 $\pm$
0.4 & 4.0 & 0.33 $\pm$ 0.56 & 2.8 $\pm$ 0.3 & 0.2 & 3.5 $\pm$ 6.0\\
SDSS1402+0955 & 186 & 148 $\pm$ 29 & 17.4 & 10.5 $\pm$ 0.2 &
15.0 & 5.3 $\pm$ 2.1 & 11.6 $\pm$ 0.6 & 0.04 & 0.080
$\pm$ 0.036\\
SDSS1507+5511 & 72 & 99 $\pm$ 22 & 15.8 & 12.1 $\pm$ 0.3 & 2.0 & 2.8
$\pm$ 1.2 & 4.2 $\pm$ 0.3 & 0.1 & 0.072 $\pm$ 0.033\\
Mrk 325 & 121 & 54 $\pm$ 49 & 19.2 & 5.4 $\pm$ 0.2 & 6.3 & 0.37 $\pm$
0.66 & 10.8 $\pm$ 0.3 & 0.1 & 0.47 $\pm$ 0.84\\
\enddata

\tablenotetext{a}{$\rm V_{rot}^{G04}$ is calculated using half of $\rm
  W_{20}$, which is corrected for inclination and random motions and
  taken from Table 3 of \citet{CG04}.}  
\tablenotetext{b}{$\rm
  V_{rot}(R_{HI})$ is measured using a cut along the major axis and is
  taken from Table \ref{fig:pvcuts} and corrected for inclination.}
\tablenotetext{c}{$\rm R_{HI}^{G04}$ is esitmated to be 2 $\rm \times
  R_{25}$ and is taken from \citet{CG04}.}  
\tablenotetext{d}{$\rm
  M_{dyn}^{G04}(R_{HI})$ is calculated using the $\rm M_{HI}$ and $\rm
  M_{HI}M_{dyn}^{-1}$ values in Table 3 and Table 4 of \citet{CG04}.}
\tablenotetext{e}{$\rm M_{dyn}(R_{HI})$ is calculated using $\rm
  V_{rot}(R_{HI})$ and $\rm R_{HI}$ and is taken from Table
  \ref{fig:pvcuts}.}  
\tablenotetext{f}{$\rm f_{HI}^{G04} = (M_{H
    I}/M_{dyn})^{G04}$} 
\tablenotetext{g}{$\rm f_{HI} = M_{H
    I}/M_{dyn}$} 
\tablenotetext{h}{SDSS0728+3532 and SDSS1319+5203 lie
  within H \textsc{I} envelopes that include other galaxies. Their H
  \textsc{I} masses include the entire envelope, while their rotation
  velocities and dynamical masses calculated from a cut along their
  major axes include only the LCBGs.}

\end{deluxetable*}
\end{longrotatetable}



\clearpage
\begin{thebibliography}{}

\bibitem[Ahn et al.(2012)]{DR9} Ahn, C.~P., Alexandroff, R., Allende Prieto, C., et al.\ 2012, \apjs, 203, 21 

\bibitem[Amram \& {\"O}stlin(2001)]{AO01} Amram, P., {\"O}stlin,
  G.\ 2001, The Messenger, 103, 31

\bibitem[Barton \& van Zee(2001)]{BvZ01} Barton, E.~J., \& van Zee,
  L.\ 2001, \apjl, 550, L35

\bibitem[Barton et al.(2006)]{B06} Barton, E.~J., van Zee, L., \&
  Bershady, M.~A.\ 2006, \apj, 649, 129

\bibitem[Bell \& de Jong(2001)]{B01} Bell, E.~F., \& de Jong,
  R.~S.\ 2001, \apj, 550, 212

\bibitem[Bershady et al.(2005)]{B04} Bershady, M.~A., Vils, M., Hoyos,
  C., Guzm{\'a}n, R., \& Koo, D.~C.\ 2005, Starbursts: From 30 Doradus
  to Lyman Break Galaxies, 329, 177

\bibitem[Broeils \& van Woerden(1994)]{B94} Broeils, A.~H., \& van
  Woerden, H.\ 1994, \aaps, 107, 129

\bibitem[Cardamone et al.(2009)]{C09} Cardamone, C., Schawinski, K.,
  Sarzi, M., et al.\ 2009, \mnras, 399, 1191

\bibitem[Castillo-Morales et al.(2011)]{CM11} Castillo-Morales, A., Gallego, J., P{\'e}rez-Gallego, J., et al.\ 2011, \mnras, 411, 1819 

\bibitem[Chandra et al.(2004)]{Chandra04} Chandra, P., Ray,
  A., \& Bhatnagar, S.\ 2004, \apj, 612, 974

\bibitem[Corsini et al.(2012)]{C12} Corsini, E.~M., M{\'e}ndez-Abreu, J., Pastorello, N., et al.\ 2012, \mnras, 423, L79

\bibitem[Crawford et al.(2011)]{C11} Crawford, S.~M., Wirth, G.~D.,
  Bershady, M.~A., \& Hon, K.\ 2011, \apj, 741, 98

\bibitem[Crawford et al.(2014)]{C14} Crawford, S.~M., Wirth, G.~D., \&
  Bershady, M.~A.\ 2014, \apj, 786, 30

\bibitem[Crawford et al.(2016)]{C16} Crawford, S.~M., Wirth,
  G.~D., Bershady, M.~A., \& Randriamampandry, S.~M.\ 2016, \apj, 817,
  87

\bibitem[Cresci et al.(2009)]{Cresci09} Cresci, G., Hicks,
  E.~K.~S., Genzel, R., et al.\ 2009, \apj, 697, 115 

\bibitem[de Blok et al.(2008)]{dB08} de Blok, W.~J.~G.,
  Walter, F., Brinks, E., et al.\ 2008, \aj, 136, 2648-2719

\bibitem[Dekel et al.(2009)]{D09} Dekel, A., Sari, R., \& Ceverino,
  D.\ 2009, \apj, 703, 785

\bibitem[Dettmar et al.(1984)]{D84} Dettmar, R.~J., Klein, U.,
  Wielebinski, R., \& Heidmann, J.\ 1984, \aap, 130, 424

\bibitem[Di Teodoro \& Fraternali(2015)]{DF15} Di Teodoro,
  E.~M., \& Fraternali, F.\ 2015, \mnras, 451, 3021

\bibitem[Duflot-Augarde \& Alloin(1982)]{DA82} Duflot-Augarde,
  R., \& Alloin, D.\ 1982, \aap, 112, 257 

\bibitem[Elmegreen \& Elmegreen(2006)]{EE06} Elmegreen, D.~M.,
  \& Elmegreen, B.~G.\ 2006, \apj, 651, 676 

\bibitem[Elmegreen et al.(2009)]{E09} Elmegreen, B.~G., Elmegreen,
  D.~M., Fernandez, M.~X., \& Lemonias, J.~J.\ 2009, \apj, 692, 12

\bibitem[F{\"o}rster Schreiber et al.(2009)]{FS09} F{\"o}rster
  Schreiber, N.~M., Genzel, R., Bouch{\'e}, N., et al.\ 2009, \apj,
  706, 1364

\bibitem[France et al.(2010)]{F10} France, K., Nell, N., Green, J.~C.,
  \& Leitherer, C.\ 2010, \apjl, 722, L80

\bibitem[Gallego et al.(1996)]{G96} Gallego, J., Zamorano, J., Rego, M., Alonso, O., \& Vitores, A.~G.\ 1996, \aaps, 120, 323

\bibitem[Garland et al.(2004)]{CG04} Garland, C.~A., Pisano, D.~J.,
  Williams, J.~P., Guzm{\'a}n, R., \& Castander, F.~J.\ 2004, \apj,
  615, 689

\bibitem[Garland et al.(2005)]{CG05} Garland, C.~A., Williams, J.~P.,
  Pisano, D.~J., et al.\ 2005, \apj, 624, 714

\bibitem[Garland et al.(2007)]{CG07} Garland, C.~A., Pisano, D.~J.,
  Williams, J.~P., et al.\ 2007, \apj, 671, 310

\bibitem[Garland et al.(2015)]{CGletter} Garland, C.~A.,
  Pisano, D.~J., Mac Low, M.~M., et al.\ 2015, \apj, 807, 134

\bibitem[Garland et al.(in prep)]{CGbigpaper} Garland, C.~A. et al. in
  prep

\bibitem[Geha et al.(2003)]{Geha03} Geha, M., Guhathakurta, P., \& van
  der Marel, R.~P.\ 2003, \aj, 126, 1794

\bibitem[Genzel et al.(2014)]{G14} Genzel, R., F{\"o}rster Schreiber,
  N.~M., Lang, P., et al.\ 2014, \apj, 785, 75

\bibitem[Giuricin et al.(2000)]{G00} Giuricin, G., Marinoni,
  C., Ceriani, L., \& Pisani, A.\ 2000, \apj, 543, 178 

\bibitem[Guzm{\'a}n et al.(1997)]{G97} Guzm{\'a}n, R., Gallego, J.,
  Koo, D.~C., et al.\ 1997, \apj, 489, 559

\bibitem[Guzm{\'a}n(1999)]{G99} Guzm{\`a}n, R.\ 1999, The Evolution of
  Galaxies on Cosmological Timescales, 187, 271

\bibitem[Guzm{\'a}n(2001)]{G01} Guzm{\'a}n, R.\ 2001, Astrophysics and
  Space Science Supplement, 277, 507

\bibitem[Guzm{\'a}n et al.(2003)]{G03} Guzm{\'a}n, R., {\"O}stlin, G.,
  Kunth, D., et al.\ 2003, \apjl, 586, L45

\bibitem[Hammer et al.(2001)]{H01} Hammer, F., Gruel, N., Thuan,
  T.~X., Flores, H., \& Infante, L.\ 2001, \apj, 550, 570

\bibitem[Harmanec(1988)]{H88} Harmanec, P.\ 1988, Bulletin of the
  Astronomical Institutes of Czechoslovakia, 39, 329

\bibitem[Haynes et al.(2011)]{H11} Haynes, M.~P., Giovanelli,
  R., Martin, A.~M., et al.\ 2011, \aj, 142, 170 

\bibitem[Heckman et al.(2005)]{H05} Heckman, T.~M., Hoopes, C.~G.,
  Seibert, M., et al.\ 2005, \apjl, 619, L35

\bibitem[Homeier \& Gallagher(1999)]{HG99} Homeier, N.~L., \& Gallagher, J.~S.\ 1999, \apj, 522, 199 

\bibitem[Homeier et al.(2002)]{H02} Homeier, N., Gallagher, J.~S., III, \& Pasquali, A.\ 2002, \aap, 391, 857 

\bibitem[Hoyos et al.(2004)]{H04} Hoyos, C., Guzm{\'a}n, R., Bershady,
  M.~A., Koo, D.~C., \& D{\'{\i}}az, A.~I.\ 2004, \aj, 128, 1541

\bibitem[Hoyos et al.(2007)]{H07} Hoyos, C., Guzm{\'a}n, R.,
  D{\'{\i}}az, A.~I., Koo, D.~C., \& Bershady, M.~A.\ 2007, \aj, 134,
  2455

\bibitem[Hunt(2017)]{LucasThesis} Hunt, L. \ 2017, Ph.D.~Thesis  

\bibitem[Inoue \& Saitoh(2012)]{I12} Inoue, S., \& Saitoh,
  T.~R.\ 2012, \mnras, 422, 1902

\bibitem[Jiang et al.(2015)]{J15} Jiang, X.-J., Wang, Z., Gu, Q.,
  Wang, J., \& Zhang, Z.-Y.\ 2015, \apj, 799, 92

\bibitem[Kassin et al.(2007)]{K07} Kassin, S.~A., Weiner,
  B.~J., Faber, S.~M., et al.\ 2007, \apjl, 660, L35

\bibitem[Kassin et al.(2012)]{K12} Kassin, S.~A., Weiner, B.~J.,
  Faber, S.~M., et al.\ 2012, \apj, 758, 106

\bibitem[Kenney et al.(2004)]{K04} Kenney, J.~D.~P., van
  Gorkom, J.~H., \& Vollmer, B.\ 2004, \aj, 127, 3361 

\bibitem[Koo et al.(1994)]{K94} Koo, D.~C., Bershady, M.~A., Wirth,
  G.~D., Stanford, S.~A., \& Majewski, S.~R.\ 1994, \apjl, 427, L9

\bibitem[Koranyi \& Geller(2002)]{KG02} Koranyi, D.~M., \& Geller, M.~J.\ 2002, \aj, 123, 100 

\bibitem[Kormendy \& Kennicutt(2004)]{KK04} Kormendy, J., \&
  Kennicutt, R.~C., Jr.\ 2004, \araa, 42, 603

\bibitem[Lintott et al.(2011)]{L11} Lintott, C., Schawinski, K.,
  Bamford, S., et al.\ 2011, \mnras, 410, 166

\bibitem[Loose \& Thuan(1986)]{LT86} Loose, H.-H., \& Thuan, T.~X.\ 1986, Star-forming Dwarf Galaxies and Related Objects, 73 

\bibitem[Maeder \& Meynet(1989)]{M89} Maeder, A., \& Meynet, G.\ 1989,
  \aap, 210, 155

\bibitem[McGaugh \& Schombert(2015)]{MS15} McGaugh, S.~S., \&
  Schombert, J.~M.\ 2015, \apj, 802, 18 

\bibitem[Noeske et al.(2006)]{N06} Noeske, K.~G., Koo, D.~C.,
  Phillips, A.~C., et al.\ 2006, \apjl, 640, L143

\bibitem[Noguchi(1998)]{N98} Noguchi, M.\ 1998, \nat, 392, 253

\bibitem[Noguchi(1999)]{N99} Noguchi, M.\ 1999, \apj, 514, 77

\bibitem[Noguchi(2000)]{N00} Noguchi, M.\ 2000, \mnras, 312, 194

\bibitem[Noguchi(2001)]{N01} Noguchi, M.\ 2001, \apj, 555, 289

\bibitem[Noordermeer et al.(2007)]{N07} Noordermeer, E., van
  der Hulst, J.~M., Sancisi, R., Swaters, R.~S., \& van Albada,
  T.~S.\ 2007, \mnras, 376, 1513

\bibitem[Nordgren et al.(1997)]{N97} Nordgren, T.~E., Chengalur,
  J.~N., Salpeter, E.~E., \& Terzian, Y.\ 1997, \aj, 114, 77

\bibitem[{\"O}stlin et al.(2001)]{O01} {\"O}stlin, G., Amram, P.,
  Boulesteix, J., et al.\ 2001, Astrophysics and Space Science
  Supplement, 277, 433

\bibitem[Overzier et al.(2009)]{O09} Overzier, R.~A., Heckman, T.~M.,
  Tremonti, C., et al.\ 2009, \apj, 706, 203

\bibitem[P{\'e}rez-Gallego et al.(2010)]{PG10} P{\'e}rez-Gallego, J.,
  Guzm{\'a}n, R., Castillo-Morales, A., et al.\ 2010, \mnras, 402,
  1397

\bibitem[P{\'e}rez-Gallego et al.(2011)]{PG11} P{\'e}rez-Gallego, J.,
  Guzm{\'a}n, R., Castillo-Morales, A., et al.\ 2011, \mnras, 418,
  2350

\bibitem[P{\'e}rez-Gonz{\'a}lez et al.(2003)]{PG03} P{\'e}rez-Gonz{\'a}lez, P.~G., Zamorano, J., Gallego, J., Arag{\'o}n-Salamanca, A., \& Gil de Paz, A.\ 2003, \apj, 591, 827 

\bibitem[Petrosian et al.(2002)]{P02} Petrosian, A., McLean, B., Allen, R.~J., et al.\ 2002, \aj, 123, 2280 

\bibitem[Phillips et al.(1997)]{P97} Phillips, A.~C., Guzm{\'a}n, R.,
  Gallego, J., et al.\ 1997, \apj, 489, 543

\bibitem[Pisano et al.(2001)]{P01} Pisano, D.~J., Kobulnicky, H.~A.,
  Guzm{\'a}n, R., Gallego, J., \& Bershady, M.~A.\ 2001, \aj, 122,
  1194

\bibitem[Prada et al.(2003)]{P03} Prada, F., Vitvitska, M.,
  Klypin, A., et al.\ 2003, \apj, 598, 260

\bibitem[Pustilnik et al.(2001)]{P2001} Pustilnik, S.~A., Kniazev, A.~Y., Lipovetsky, V.~A., \& Ugryumov, A.~V.\ 2001, \aap, 373, 24 

\bibitem[Romano et al.(2005)]{R05} Romano, D., Chiappini, C.,
  Matteucci, F., Tosi, M.\ 2005, \aap, 430, 491

\bibitem[Sault et al.(1995)]{S95} Sault, R.~J., Teuben, P.~J.,
  \& Wright, M.~C.~H.\ 1995, Astronomical Data Analysis Software and
  Systems IV, 77, 433

\bibitem[Serra et al.(2016)]{S16} Serra, P., Oosterloo, T.,
  Cappellari, M., den Heijer, M., \& J{\'o}zsa, G.~I.~G.\ 2016,
  \mnras, 460, 1382

\bibitem[Skrutskie et al.(2006)]{S06} Skrutskie, M.~F., Cutri,
  R.~M., Stiening, R., et al.\ 2006, \aj, 131, 1163

\bibitem[Sparke \& Gallagher(2007)]{SG} Sparke, L.~S., \& Gallagher,
  J.~S., III 2007, Galaxies in the Universe: An Introduction.~Second
  Edition.~By Linda S.~Sparke and John S.~Gallagher, III.~ISBN-13
  978-0-521-85593-8 (HB); ISBN-13 978-0-521-67186-6 (PB).~Published by
  Cambridge University Press, Cambridge, UK, 2007.

\bibitem[Stott et al.(2016)]{Stott16} Stott, J.~P., Swinbank,
  A.~M., Johnson, H.~L., et al.\ 2016, \mnras, 457, 1888

\bibitem[Tamburro et al.(2009)]{T09} Tamburro, D., Rix, H.-W., Leroy,
  A.~K., et al.\ 2009, \aj, 137, 4424

\bibitem[Tollerud et al.(2010)]{T10} Tollerud, E.~J., Barton, E.~J.,
  van Zee, L., \& Cooke, J.\ 2010, \apj, 708, 1076

\bibitem[Tonini et al.(2014)]{T14} Tonini, C., Jones, D.~H., Mould,
  J., et al.\ 2014, \mnras, 438, 3332

\bibitem[Toomre(1964)]{T64} Toomre, A.\ 1964, \apj, 139, 1217

\bibitem[Tully \& Fisher(1977)]{TF77} Tully, R.~B., \& Fisher,
  J.~R.\ 1977, \aap, 54, 661

\bibitem[Tully \& Fouque(1985)]{TF85} Tully, R.~B., \& Fouque,
  P.\ 1985, \apjs, 58, 67

\bibitem[Tully \& Pierce(2000)]{TP00} Tully, R.~B., \& Pierce,
  M.~J.\ 2000, \apj, 533, 744

\bibitem[van den Bergh et al.(1996)]{vdB96} van den Bergh, S.,
  Abraham, R.~G., Ellis, R.~S., et al.\ 1996, \aj, 112, 359 

\bibitem[van Zee et al.(2004)]{vZ04} van Zee, L., Skillman, E.~D., \&
  Haynes, M.~P.\ 2004, \aj, 128, 121

\bibitem[Vollmer et al.(2001)]{V01} Vollmer, B., Cayatte, V.,
  Balkowski, C., \& Duschl, W.~J.\ 2001, \apj, 561, 708

\bibitem[Walter et al.(2008)]{W08} Walter, F., Brinks, E., de Blok,
  W.~J.~G., et al.\ 2008, \aj, 136, 2563

\bibitem[Weiner et al.(2006)]{W06} Weiner, B.~J., Willmer,
  C.~N.~A., Faber, S.~M., et al.\ 2006, \apj, 653, 1027

\bibitem[Werk et al.(2004)]{W04} Werk, J.~K., Jangren, A., \& Salzer,
  J.~J.\ 2004, \apj, 617, 1004

\bibitem[Willett et al.(2013)]{W13} Willett, K.~W., Lintott,
  C.~J., Bamford, S.~P., et al.\ 2013, \mnras, 435, 2835

\bibitem[Yang et al.(2008)]{Y08} Yang, Y., Flores, H., Hammer,
  F., et al.\ 2008, \aap, 477, 789 

\bibitem[Zabludoff \& Mulchaey(1998)]{ZM98} Zabludoff, A.~I.,
  \& Mulchaey, J.~S.\ 1998, \apj, 496, 39


\end{thebibliography}
\end{document}